\documentclass[aps,prl,notitlepage,twocolumn,superscriptaddress,longbibliography,nofootinbib]{revtex4-1}
\pdfoutput=1
\usepackage[utf8]{inputenc}

\usepackage{amsmath,amsthm,amssymb}
\usepackage{scalerel}
\usepackage{mathtools}
\usepackage{booktabs}
\usepackage{graphicx}
\usepackage{subfigure}
\usepackage[normalem]{ulem}
\usepackage[colorlinks = true,
            linkcolor = blue,
            urlcolor  = blue,
            citecolor = blue,
            anchorcolor = blue]{hyperref}
\usepackage{mathrsfs}
\usepackage{bbold}
\usepackage[]{units}
\usepackage{bm}
\usepackage{braket}
\usepackage{color}
\usepackage{makecell}
\usepackage{enumitem}
\usepackage{upgreek}
\usepackage{blindtext}
\usepackage{graphics}
\usepackage{verbatim} 
\usepackage{algorithm}
\usepackage[noend]{algpseudocode}
\usepackage[dvipsnames]{xcolor}
\usepackage{bbm}
\usepackage[bb=boondox]{mathalfa}
\usepackage{array}
\usepackage{multirow}
\usepackage{tabularx}
\usepackage{float}
\usepackage{graphicx}
\usepackage{dcolumn}
\usepackage{bm}
\usepackage{amsmath,amsfonts,amssymb}
\usepackage{slashed}
\usepackage{braket,xcolor}
\usepackage{verbatim}
\usepackage{multirow}
\usepackage{dsfont}
\usepackage{amsfonts, mathtools,resizegather}
\usepackage[export]{adjustbox}

\begin{document}

\title{{Unitary $k$-designs without independent Hamiltonian quenches}}

\author{Pratik Nandy\,\,\href{https://orcid.org/0000-0001-5383-2458}
{\includegraphics[scale=0.05]{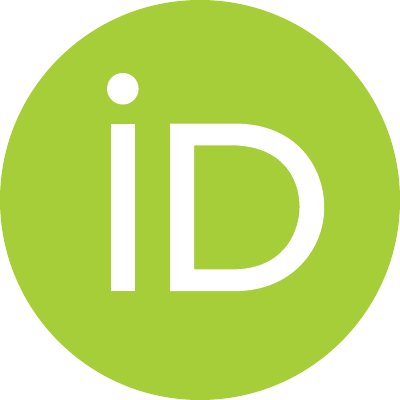}}}
\email{pratik.nandy@vub.be}
\affiliation{Theoretische Natuurkunde, Vrije Universiteit Brussel (VUB) and \\The International Solvay Institutes, Pleinlaan 2, B-1050 Brussels, Belgium}
\affiliation{RIKEN Centre for Interdisciplinary Theoretical and Mathematical Sciences (iTHEMS),
Wako, Saitama 351-0198, Japan}

\begin{abstract}

Unitary $k$-designs provide a resource-efficient framework for emulating Haar randomness up to the $k$-th order. Quench-based protocols have recently been shown to generate such designs, but achieving this typically requires multiple Hamiltonian realizations, even when using temporal ensembles. Here, we show that no independent Hamiltonians are required: a single chaotic Hamiltonian is sufficient to generate approximate unitary $k$-designs, even when that Hamiltonian is spatially local. We introduce a \emph{two-Pauli-kick} (2PK) protocol, in which unitary evolution under a fixed Hamiltonian is interspersed with two Pauli operator insertions (kicks). We find that the resulting frame potential approaches the Haar value at long times, with deviations suppressed by the inverse Hilbert-space dimension. We verify this for single realizations of Gaussian random matrices, the Majorana and Spin Sachdev–Ye–Kitaev models, and deterministic local quantum spin chains. Remarkably, in deterministic spin systems, the 2PK protocol generates approximate unitary designs in regimes where quench-based protocols are either inapplicable or fail to converge. Furthermore, our protocol provides a finite-temperature extension of the frame potential and establishes an analytic bound in terms of the equilibrium partition function. We discuss a holographic perspective on this mechanism.

\end{abstract}

~~~~~~~~~~~~~~~~~~~~~~~~~~~~~~~~~~ RIKEN-iTHEMS-Report-26

\maketitle

\emph{Introduction:} Haar-random unitaries \cite{Mele:2023ojv} provide an ideal benchmark for quantum chaos and play a central role in quantum information protocols. However, their exact realization in many-body quantum systems is exponentially expensive, motivating the search for physically realizable ensembles that reproduce Haar statistics up to a desired order. Unitary $k$-designs provide a powerful alternative, reproducing the first $k$ moments of the Haar measure with significantly fewer resources \cite{Gross:2007xgw, Roberts:2016hpo, Cotler:2017jue, Emerson:2003www, Brandao:2016glh, Brandao:2019sgy, Schuster:2024ajb}. Consequently, the construction of unitary $k$-designs has attracted considerable recent attention, with many directions actively explored \cite{Hunter-Jones:2019lps, Choi:2021npc, Ho:2021dmh, Cotler:2021pbc, Claeys:2022hts, Fava:2023pac, Nakata:2024tla, Dowling:2025cxr, Cui:2025teh}.

A central question is whether a simple dynamical process based solely on Hamiltonian evolution can generate unitary $k$-designs. Recently, Zhou et al.\,\cite{Zhou:2025noh, Zhou:2026ubz} introduced quench protocols that realize such designs. They considered Hamiltonian ensembles, where randomness comes from sampling different Hamiltonians \cite{Chenu:2017qdv, Chenu:2018spm}, and temporal ensembles, where randomness arises from sampling evolution times in multi-step quenches involving at least three different Hamiltonians \cite{Zhou:2026ubz}. A related protocol using two Hamiltonians and a randomly sampled Pauli operator was later proposed in Ref.\,\cite{Sun:2026wrk} (Table \ref{comparison}).

\begin{figure}[t]
    \centering
\includegraphics[width=0.97\linewidth]{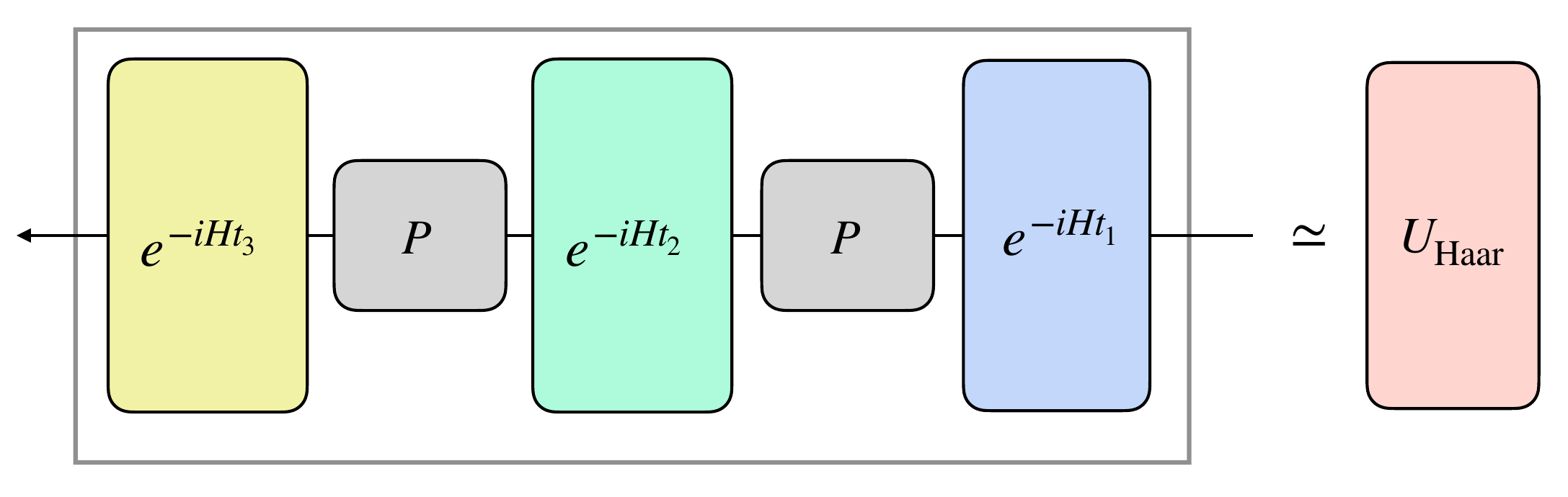}
    \caption{Diagrammatic illustration of the two-Pauli-kick (2PK) protocol. The system evolves under a \emph{fixed} Hamiltonian $H$ for a duration $t_1$, followed by a Pauli kick. It then evolves under the \emph{same} Hamiltonian for a duration $t_2$, after which the \emph{same} Pauli kick is applied again, and the evolution continues for a duration $t_3$. The arrow dictates the direction of the operation. The resulting temporal ensemble approaches the Haar ensemble, characterized by the frame potential \eqref{eq:F2PK}.}
    \label{fig:2PK_diagram}
\end{figure}

However, these constructions rely on independent generators from ensembles at different stages, making them unsuitable for deterministic Hamiltonians. They apply only at infinite temperature and do not naturally incorporate thermal states. This raises a fundamental question: can a temporal ensemble generated by a single, fixed chaotic Hamiltonian produce Haar-like ensembles, particularly in spatially local deterministic systems? Such a construction would be directly relevant for experimentally realizable systems and for finite-temperature dynamics, neither of which is accessible within multi-Hamiltonian constructions of Refs.\,\cite{Zhou:2026ubz, Sun:2026wrk}.

Here, we answer this question affirmatively by introducing a
\emph{two-Pauli-kick} (2PK) protocol, illustrated in
Fig.\,\ref{fig:2PK_diagram}. Starting from a fixed chaotic Hamiltonian $H$,
the system evolves three times, interrupted twice by a fixed Pauli operator $P$. Importantly, once chosen, both $H$ and $P$ remain fixed throughout the protocol, and randomness enters only through the sampled evolution times.
Thus, the protocol eliminates the need for independent Hamiltonians: the two
Pauli kicks provide the scrambling mechanism without requiring any further independent Hamiltonians. By evaluating the $k$-th frame potential in various systems, including Gaussian ensembles (GXE, $\mathrm{X}\in\{\mathrm{O},\mathrm{U}\}$), the Majorana and Spin Sachdev--Ye--Kitaev (SYK) models, and quantum spin chains, we demonstrate that the resulting temporal ensemble generates approximate unitary
$k$-designs. In all cases, our protocol uses only a single Hamiltonian realization, and hence fewer independent random generators. This advantage is particularly striking for deterministic spin chains, where previous protocols are either inapplicable or succeed only under special conditions. Our construction admits a natural finite-temperature generalization by incorporating the thermal density matrix $\rho_{\beta}$, yielding a finite-temperature frame potential that is not naturally accessible within multi-Hamiltonian quenches \cite{Zhou:2026ubz, Sun:2026wrk}. We also provide a gravitational interpretation of our protocol. See Table \ref{comparison} for a comparison of different protocols.

\begin{table}[t]
\centering
\scriptsize
\setlength{\tabcolsep}{0.5pt}
\renewcommand{\arraystretch}{1.15}
\begin{tabular}{@{}lcccc@{}}
\hline\hline
Work & Protocol & Randomness & Local $H$ & Finite $T$ \\
\hline
Zhou et al.\,\cite{Zhou:2026ubz}
 & $3H_{\rm ind}$ (3SP)
 & $H$ + temporal
 & $\times$ & $\times$ \\
Sun et al.\,\cite{Sun:2026wrk}
 & $2H_{\mathrm{ind}}$ + $P_{\mathrm{rand}}$
 & $H$ + $P$ + temporal
 & N/E & $\times$ \\
\textbf{This work}
 & \textbf{$H_{\rm fix}+P_{\rm fix}$ (2PK)}
 & \textbf{Temporal only}
 & $\checkmark$ & $\checkmark$ \\
\hline\hline
\end{tabular}
\caption{Comparison with previous protocols. Here $\times$ implies either not applicable or fails. N/E means not established.} \label{comparison}
\end{table}

\emph{The Two-Pauli-Kick (2PK) protocol:} 
Let $\mathcal{P}_n := \{\mathds{1},X,Y,Z\}^{\otimes n}$ denote the set of Hermitian Pauli strings on $n$ qubits, and let $P \in \mathcal{P}_n\backslash\{\mathds{1}^{\otimes n}\}$ be a non-identity string, drawn once and held fixed thereafter. We require $[P,H]\neq 0$, for chaotic $H$ with no Pauli symmetry. Starting from a fixed Hamiltonian $H$, the unitary evolution in the 2PK protocol is defined as
\begin{align}
U_{\mathrm{2PK}}(t_1,t_2,t_3) :=
e^{-iHt_3}\,P\, e^{-iHt_2}\,P\, e^{-iHt_1}\,,
\label{eq:U2PK}
\end{align}
with the \emph{same} $P$ inserted at both kicks. The central idea is to \emph{reuse} the \emph{fixed} chaotic Hamiltonian and the \emph{fixed} Pauli operator throughout the
evolution. These Pauli insertions mimic an effective quench without requiring separate Hamiltonian quenches (see Supplemental Material (SM) \ref{sec:matrix_elements}). The corresponding temporal ensemble is generated by sampling the evolution times from a uniform probability distribution $p(t)=\frac{1}{T}\mathbf{1}_{[0,T]}(t)$,
\begin{align}
\mathcal{E}_{\mathrm{2PK}}
:= \left\{U_{\mathrm{2PK}}(t_1,t_2,t_3):~ t_i\sim p(t) \right\}\,.
\label{eq:E2PK}
\end{align}
where $\mathbf{1}_{A}(t)$ is the indicator function, $\mathbf{1}_{A}(t) = 1$ for $t\in A$ and zero otherwise. In the $T \rightarrow \infty$ limit, an approximate $k$-design is characterized by the $k$-th frame potential \cite{Gross:2007xgw, Roberts:2016hpo}
\begin{align}
F^{(k)}_{\mathrm{2PK}} :=
\mathds{E}_{U,V\in\mathcal{E}_{\mathrm{2PK}}}\left|\mathrm{Tr}\left(U^\dagger V\right)
\right|^{2k}\,,~~ k \in \mathbb{N}\,.
\label{eq:F2PK}
\end{align}
Here, the expectation is taken with respect to the temporal ensembles. The Haar ensemble provides the
minimal value of the frame potential $F^{(k)}_{\mathcal{E}_{\mathrm{Haar}}} = k!$ for $k \le d$ \cite{Scott_2008}, with $d$ being the Hilbert-space dimension (or that of relevant symmetry sectors) on which the trace is evaluated.

Here we show that the frame potential of the 2PK protocol converges to the corresponding Haar value,
\begin{align}
F^{(k)}_{\mathrm{2PK}}
\longrightarrow k! [1 + O(1/d)]\,, ~~~~~ (T \rightarrow \infty)\,, \label{f2pk}
\end{align}
implying that the protocol realizes an approximate $k$-design in the large-$d$ limit at fixed $k$, with corrections controlled for $k^2 \ll d$. The $k=1$ case is exactly solvable, giving the deviation
$\Delta^{(1)}_{\mathrm{2PK}}:=F^{(1)}_{\mathrm{2PK}}-1=2(\Lambda-1)^{2}/d$
at leading order in $1/d$. Here $\Lambda$ is the kurtosis factor of the
Pauli matrix elements in the energy basis, taking the values $2$ (GUE)
and $3$ (GOE), and fixes the 1PK frame potential at the same order. Proofs are given in SM \ref{sec:k1exact} and \ref{sec:S3}. Numerical results corroborate this behavior.

We also introduce the \emph{one-Pauli-kick} (1PK) protocol:
\begin{align}
U_{\mathrm{1PK}}(t_1,t_2) :=
e^{-iHt_2} P e^{-iHt_1}\,.
\label{eq:U1PK}
\end{align}
As we show below, the resulting temporal ensemble does not converge to the Haar frame potential and therefore fails to generate an approximate $k$-design. Rather, it approaches the value at leading order in $1/d$:
\begin{align}
F^{(k)}_{\mathrm{1PK, GXE}} =
k!\sum_{j=0}^{k} \binom{k}{j}\, !(k-j)\, \Lambda^j\,, \label{dearr}
\end{align}
where $!n$ is the derangement number. For large $k$, Eq.\,\eqref{dearr} gives $F^{(k)}_{\mathrm{1PK, GXE}} \sim e^{\Lambda - 1} (k!)^2$. See SM \ref{sec:S3} for a proof. 
\newline
\textbf{Proposition 1.} For any unitary $P$, the 2PK protocol \eqref{eq:U2PK}
equals, up to a fixed left multiplication by $P^2$, a three-step protocol (3SP)
\cite{Zhou:2026ubz} with three \emph{isospectral} Hamiltonians
$H_1=H$, $H_2=P^\dagger HP$, $H_3=P^{\dagger2}HP^2$. For our protocol, which uses a Hermitian Pauli string, where $P^2=\mathds{I}$ and $P^{\dagger} = P$, the prefactor is trivial and $H_3=H_1=H$, leaving the two
Hamiltonians $\{H,\,P HP\}$.\\
\emph{Proof.} Using $P^\dagger P= \mathds{1}$ and
$P^{\dagger} e^{-iHt}P=e^{-i(P^{\dagger}HP)t}$
\begin{align}
U_{\mathrm{2PK}}
&= e^{-iHt_3}Pe^{-iHt_2}Pe^{-iHt_1}\nonumber\\
&= P^2\,e^{-i(P^{\dagger2}HP^2)t_3}\,e^{-i(P^\dagger HP)t_2}\,e^{-iHt_1}\,.\nonumber
\end{align}
Since $U$ and $V$ carry the same $P$, the global factor $P^2$ cancels in the
frame potential, $\mathrm{Tr}(U^\dagger V)=\mathrm{Tr}(\widetilde U^\dagger
P^{\dagger2}P^2\widetilde V)=\mathrm{Tr}(\widetilde U^\dagger\widetilde V)$,
where $\widetilde U,\widetilde V$ omit the prefactor. $P^2=\mathds{1}$ trivializes both the prefactor and $H_3=H$, giving $U_{\mathrm{2PK}}=e^{-iHt_3}e^{-i(PHP)t_2}e^{-iHt_1}$.

However, there is a crucial difference compared to the 3SP \cite{Zhou:2026ubz}, which employs three statistically independent Hamiltonians. In the 2PK protocol \eqref{eq:U2PK}, the effective Hamiltonian is a unitary conjugate of a \emph{single} $H$ and thus shares the same spectrum. There is thus no a priori guarantee that such isospectral, non-independent Hamiltonians generate unitary designs; that they do is a nontrivial property of the eigenvectors. This distinction becomes important for local spin chains, as we show below. Diagonalizing the Hamiltonian once, $H=WDW^\dagger$, we can write the frame potentials \eqref{eq:F2PK} as
\begin{align}
F_{1\mathrm{PK}}^{(k)}
&=\mathds{E}_{\{t_j,t_j'\}}
\left|\mathrm{Tr} \left(
e^{iD\Delta t_1} P_W^{\dagger}e^{iD\Delta t_2} P_W \right)\right|^{2k}\,,
\label{12PKnum0} \\
F_{2\mathrm{PK}}^{(k)}
&=
\mathds{E}_{\{t_j,t_j'\}} \Bigl|
\mathrm{Tr} \Bigl(e^{iD\Delta t_1} P_W^{\dagger} e^{iDt_2} P_W^{\dagger}
\nonumber \\
&\qquad\qquad\qquad
\times
e^{iD\Delta t_3} P_W
e^{-iDt_2'} P_W \Bigr) \Bigr|^{2k}\,, \label{12PKnum}
\end{align}
where $P_W=W^\dagger P W$ and $\Delta t_j=t_j-t_j'$ for $j=1,2,3$ (also $P_W^{\dagger} = P_W$ for our case). Consequently, the protocol requires only a single diagonalization of the Hamiltonian, substantially reducing the numerical cost compared with protocols involving multiple independent Hamiltonians.

\begin{figure}[t]
    \hspace{-0.5cm}
\includegraphics[width=1.03\linewidth]{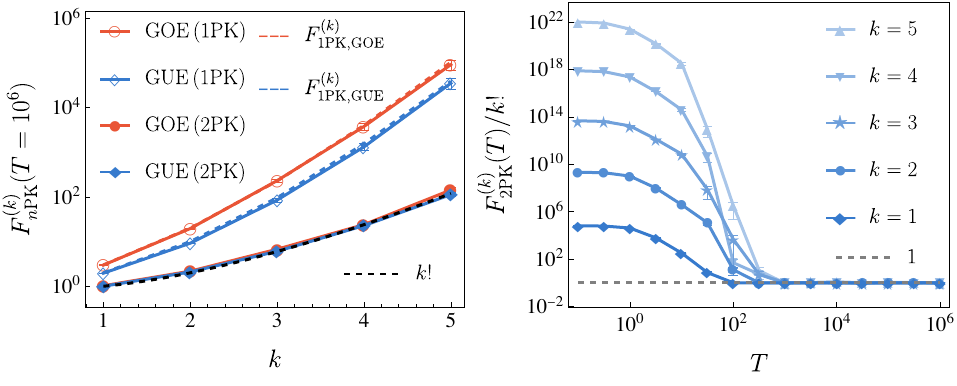}
    \caption{\textbf{Left panel:} Frame potential $F^{(k)}$ for the GXE (single realization) under the 1PK and 2PK protocols. \textbf{Right panel:} The normalized 2PK $F_{\mathrm{2PK}}^{(k)}/k!$ for GUE, which stabilizes to unity for $T \geq 10^4$. The chosen Pauli operator has a weight (number of non-identities) of $6$. The red/blue and black dashed lines denote the analytical prediction in \eqref{dearr} with $\Lambda = 3$ (GOE), $\Lambda = 2$ (GUE), and the Haar value $k!$, respectively. The system size is $d=2^8$, and the results are averaged over $10^4$ independent temporal realizations.}
    \label{fig:FP12PKGUEplot}
\end{figure}

\emph{$k$-design in GXE:} As a benchmark, we consider the Gaussian ensembles (GXE) (traceless), whose spectral statistics provide a paradigmatic model of quantum chaos. Since GXE has no additional symmetries, the Haar benchmark is simply $k!$. Figure \ref{fig:FP12PKGUEplot} compares the frame potentials of 1PK and 2PK protocols in GXE. For 1PK, the frame potential converges to Eq.\,\eqref{dearr}, which remains above $k!$ for all $k$. In contrast, the 2PK frame potential approaches $k!$, indicating the emergence of an approximate unitary $k$-design. In SM \ref{sec:pauli_independence}, we have verified that this behavior is insensitive to the particular choice of $P$. The long-time limit $T = 10^6$ is justified in the right panel, where $F_{\mathrm{2PK}}^{(k)}/k!$ stabilizes to unity after $T \geq 10^4$. The finite-size analysis in the SM \ref{sec:finitesize} shows that the
deviation from Haar, $\Delta^{(k)}_{\mathrm{2PK}}:=F^{(k)}_{\mathrm{2PK}}-k!$,
scales as $1/d$, so the design becomes exact in the large-$d$ limit.

\emph{$k$-design in Majorana and Spin SYK:} We next apply the protocol to all-to-all chaotic systems, focusing on Majorana and Spin Sachdev--Ye--Kitaev (SYK) \cite{PhysRevLett.70.3339, Kittu} models. The $q$-body Hamiltonians are compactly written as
\begin{align}
H_{q}
&= \sqrt{\frac{(q-1)!}{M^{q-1}}} 
\sum_{I} i^{\eta_I} J_I O_I\,,~ (\mathrm{Maj.~\&~Spin~SYK})\,
\end{align}
where $I=(i_1,\ldots,i_q)$ denotes a collective index and $J_I$ are Gaussian random couplings with zero mean and unit variance. For Majorana SYK, $O_I=O_{i_1}\cdots O_{i_q}$ are products of Majorana operators $O_I \equiv \chi_i$ satisfying $\{\chi_i,\chi_j\}=2\delta_{ij}$. For Spin SYK, $O_I$ are Pauli strings constructed from local spin operators \cite{Hanada:2023rkf, Basu:2025ubf}. The factors $i^{\eta_I}$ are introduced to ensure Hermiticity. Here, $M$ denotes the number of degrees of freedom, with $M=N_{\mathrm{Maj}}$ for Majorana and $M=N_{\mathrm{Spin}}=N_{\mathrm{Maj}}/2$ for Spin SYK. We focus on Majorana SYK$_4$, Spin SYK$_4$, and Spin SYK$_2$ models. Due to the hard-core boson constraints \cite{Rosa:2019jin}, even Spin SYK$_2$ retains signatures of quantum chaos \cite{Basu:2025ubf}. These models therefore provide a controlled setting to investigate whether the emergence of unitary designs depends on the microscopic structure of SYK interactions.

\begin{figure}[t]
\hspace{-0.5cm}
\includegraphics[width=0.49\linewidth]{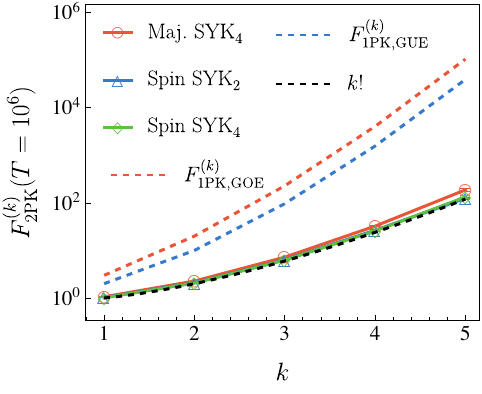}
\includegraphics[width=0.49\linewidth]{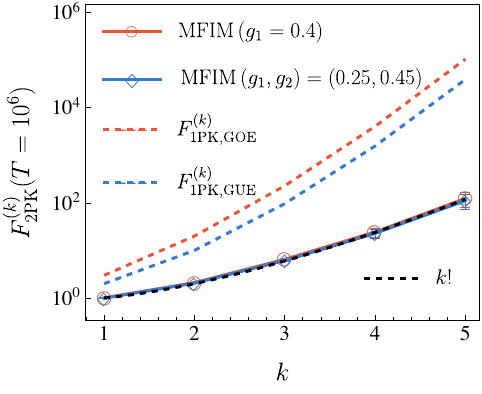}
\caption{\textbf{Left panel:} Frame potential for Majorana ($N_{\mathrm{Maj}} = 16$ (GOE) \cite{Ludwig}) and Spin SYK ($N_{\mathrm{Spin}} = 8$ (GUE) \cite{Hanada:2023rkf, Basu:2025ubf}) under 2PK protocol. The chosen Pauli operator is of weight $6$. \textbf{Right panel:} Corresponding results for the MFIM models with $L = 10$. The Pauli operator is $P = ZXYZYXYXYY$. The dashed lines denote \eqref{dearr} and the Haar value $k!$, respectively. In all cases, the results are averaged over $10^5$ (for SYK) and  $10^4$ (for MFIM) independent temporal realizations.}
\label{fig:FP2PKallSYKMFIMplot}
\end{figure}

Figure \ref{fig:FP2PKallSYKMFIMplot} (left) shows that the frame potentials for the Majorana and Spin SYK models approach the Haar value, demonstrating the emergence of approximate unitary $k$-designs. Remarkably, this includes the Spin SYK$_2$ model, whose interaction structure is much simpler than that of SYK$_4$. Since no averaging over the SYK ensemble is performed, these results show that Haar-like randomness arises from the chaotic dynamics of a single Hamiltonian realization rather than from random couplings.

\emph{Mixed-Field Ising Models:} In previous examples, the protocol uses a single Hamiltonian, but with randomly drawn couplings. This motivates the question of whether Hamiltonian randomness is necessary for the success of the 2PK protocol. To answer this, we consider deterministic mixed-field Ising models (MFIM):
\begin{align}
    H_{\mathrm{MFIM}} = \sum_{i} (Z_i Z_{i+1} + h_x  X_i + h_z  Z_i) + \sum_i g_i X_i\,, \label{MFIM}
\end{align}
where $g_i$ are fixed site-dependent defects introduced to break the translation and reflection symmetries of the clean chain. We consider an open chain (OBC) with a single defect, $g_1=0.4, g_{i>1}=0$, and a periodic chain (PBC) with two defects, $(g_1, g_2)= (0.25, 0.45)$, $g_{i>2}=0$. In both cases, with $(h_x,h_z)=(-1.05,0.5)$ \cite{Banuls:2010zki}, the level-spacing ratio $\langle r\rangle\simeq0.53$ confirms GOE statistics \cite{Oganesyan:2007wpd, Atas2013distribution}.

We take the full-weight kick $P=ZXYZYXYXYY$. Unlike the GXE, locality makes this choice crucial: the protocol is typically effective for weights $\omega\gtrsim L/2$ and fails for low-weight strings. Geometrically, the kick rotates $H$ into the effective Hamiltonian $PHP$ (cf.\ Proposition~1), with overlap ratio $c=\mathrm{Tr}(H\,PHP)/\mathrm{Tr}(H^{2})=1-2f$, where $f$ is the fraction of the Frobenius weight of $H$ whose sign the kick flips. SM~\ref{supp:MFIM} shows that an extensive fraction must be rotated, so we restrict to $|c|\ll1$, which is necessary but not a sufficient condition.

\begin{figure}[t]
\hspace{-0.8cm}
\includegraphics[width=1\linewidth]{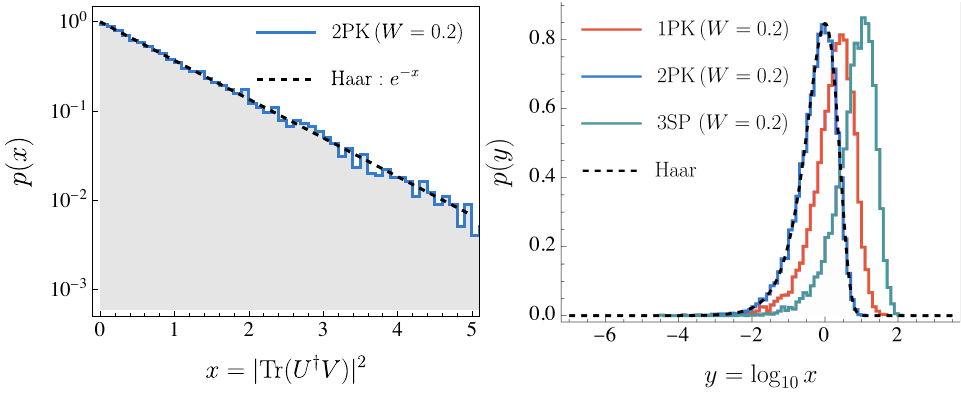}
\caption{Weakly disordered MFIM with $g_i = W\mathcal{N}(0,1)$ and $W=0.2$
(Cases 6--7 of Table~\ref{tab:3sp}). \textbf{Left panel:} Distribution of
$x=|\mathrm{Tr}(U^{\dagger}V)|^2$ for 2PK against the Haar prediction
$e^{-x}$. \textbf{Right panel:} The same data for $y=\log_{10}x$, with 1PK
and 3SP for comparison. Densities are normalized in $y$, so
$p(y)=p(x)|dx/dy|=\ln(10)\,10^{y}e^{-10^{y}}$ (a scaled Gumbel density),
peaking at $y=0$ with height $\ln 10/e = 0.847$. The system size is $L=10$,
and $10^5$ temporal samples are taken.}
    \label{fig:momentMFIM}
\end{figure}

Figure \ref{fig:FP2PKallSYKMFIMplot} (right) shows that 2PK forms an approximate $k$-design with a completely deterministic local Hamiltonian. This provides, to the best of our knowledge, the first realization of designs in deterministic settings, with randomness entering only through the sampled evolution times. It lies outside the scope of Refs.\,\cite{Zhou:2026ubz, Sun:2026wrk}, which require independent Hamiltonian realizations and, in the latter case, a resampled Pauli operator. For a direct comparison with 3SP, we introduce weak disorder with $g_i\sim W\mathcal{N}(0,1)$. While 2PK generates approximate designs, 3SP fails because weak disorder does not sufficiently rotate the eigenbases of independent local Hamiltonians ($|c|\approx1$). In contrast, Pauli conjugation induces a controlled orthogonal rotation in operator space while preserving the Hamiltonian norm ($|c|\ll1$) (Table \ref{tab:3sp}), yielding the overlap statistics with complex Gaussian $z$ of unit variance. Thus, 2PK satisfies the criterion \eqref{eq:criterion0}, whereas 1PK and 3SP fail, as shown in Fig.\,\ref{fig:momentMFIM}. 

In SM \ref{suffcond}, for any ensemble $\mathcal{E}$, we give a criterion in terms of
$\Delta^{(1)}_{\mathcal{E}}:=F^{(1)}_{\mathcal{E}}-1$ and
$\kappa_{\mathcal{E}}:=F^{(2)}_{\mathcal{E}}/[2(F^{(1)}_{\mathcal{E}})^{2}]-1$:
\begin{align}
\mathcal{E}\ \text{approximate }k\text{-design}\,\Rightarrow\,
\Delta^{(1)}_{\mathcal{E}}\to0\ \text{\&}\  \kappa_{\mathcal{E}}\to0\,.
\label{eq:criterion0}
\end{align}
Here $\kappa_{\mathcal{E}}$ constrains the Gaussian shape of the distribution of
$z=\mathrm{Tr}(U^{\dagger}V)$, while $\Delta^{(1)}_{\mathcal{E}}$ constrains its scale. The converse holds given the factorization \eqref{eq:S3-2PKresult} but
not in general. For 2PK, $F^{(1)}_{\mathrm{2PK}}=\mathrm{Tr}(w^{4})$ is exact for any
non-degenerate $H$, with $w_{ab}:=|\langle E_a|P|E_b\rangle|^{2}$, so that $\Delta^{(1)}_{\mathrm{2PK}}\to 0$ once $P$ is delocalized in the energy eigenbasis. It therefore provides a design diagnostic requiring one diagonalization and no temporal sampling.

\begin{figure}[t]
    \hspace{-0.5cm}
\includegraphics[width=1.03\linewidth]{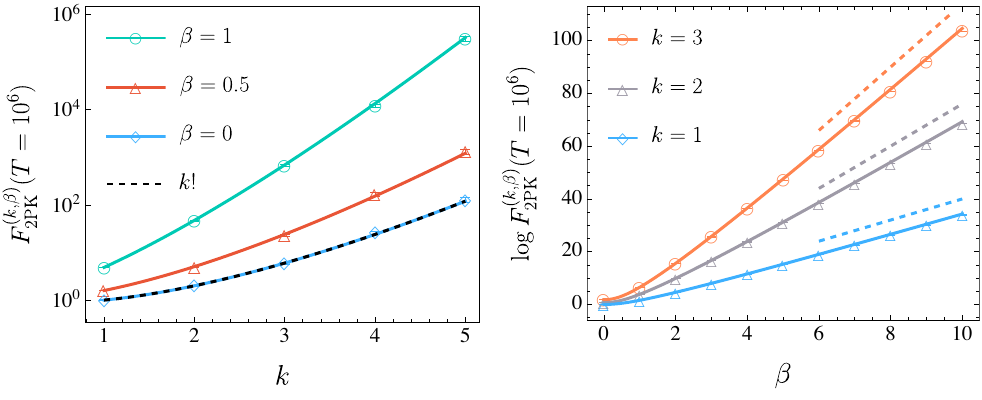}
    \caption{\textbf{Left panel:} Finite-temperature 2PK frame potential \eqref{eq:F2PKbeta} for GUE for different $\beta$. \textbf{Right panel:} Logarithm of the frame potential with $\beta$ for different $k$. In both plots, the point markers and the solid lines are the numerical and analytic results \eqref{ineq}, respectively. The dashed lines indicate the slope $\alpha_k = 4k$, from \eqref{gs}. System size is $d=2^8$, and the results are averaged over $10^4$ independent temporal realizations.}
    \label{fig:FP2PKGUEwithbetaplot}
\end{figure}

\emph{2PK protocol and frame potential at finite temperature:} Since our 2PK protocol is generated by a single Hamiltonian, it naturally admits a finite-temperature generalization \cite{Roberts:2016hpo, Cotler:2017jue, Tian-GangZhou:2026hbi} to thermal ensembles. We define the finite-temperature frame potential as
\begin{align}
F^{(k, \beta)}_{\mathrm{2PK}} :=
\mathds{E}_{U,V\in\mathcal{E}_{\mathrm{2PK}}}\left|\mathrm{Tr}\left(\rho_{\beta}U^\dagger V\right)
\right|^{2k}\,,
\label{eq:F2PKbeta}
\end{align}
where $\rho_{\beta} = e^{-\beta H}$ is the unnormalized Gibbs density matrix. To evaluate \eqref{eq:F2PKbeta}, we recast it into the form of \eqref{12PKnum} and make the replacement $\Delta t_1 \rightarrow \Delta t_1 + i \beta$. Figure \ref{fig:FP2PKGUEwithbetaplot} (left) shows the finite-temperature frame potential for the 2PK protocol at different values of $\beta$. As $\beta$ increases, the frame potential deviates from $k!$, reflecting the dominance of low-energy states in the unnormalized thermal weighting factor $e^{-\beta H}$. More generally, we can show that
\begin{align}
    F^{(k, \beta)}_{\mathrm{2PK}} = k! \bigg(\frac{\mathcal{Z}(2\beta)}{d}\bigg)^k [1 + O(1/d)] \,, \label{ineq}
\end{align}
for all $\beta$, and becomes $k!$ for $\beta = 0$. Here $\mathcal{Z}(\beta)=\mathrm{Tr}(e^{-\beta H})$ is the partition function and $d$ denotes the dimension of the Hilbert space (or of the symmetry-resolved sector). The proof is given in the SM \ref{sec:SM-bound}. The left panel of Fig.\,\ref{fig:FP2PKGUEwithbetaplot} numerically confirms our result.

More generally, \eqref{ineq} suggests a universal lower bound. Indeed, evaluating the finite-temperature frame potential $F^{(k,\beta)}_{\mathcal{E}_{\mathrm{Haar}}}$ for independent Haar-random unitaries $U$ and $V$ yields the same value as $F^{(k, \beta)}_{\mathrm{2PK}}$. We prove (see SM \ref{sec:SM-bound}) that for an arbitrary ensemble of unitaries $\mathcal{E}$
\begin{align}
F^{(k,\beta)}_{\mathcal{E}} \;\ge\; F^{(k,\beta)}_{\mathcal{E}_{\mathrm{Haar}}}
= k!\left(\frac{\mathcal{Z}(2\beta)}{d}\right)^{k}\big[1+O(1/d)\big].
\label{lowerbound}
\end{align}
Since $\mathcal{Z}(2\beta)\ge d$ for traceless $H$ by the arithmetic--geometric mean inequality, the right-hand side is bounded below by $k!\,[1+O(1/d)]$, with equality at $\beta=0$, where the bound reduces to $F^{(k,0)}_{\mathcal{E}}\ge k!$
exactly. The first inequality is exact at finite $d$ and is saturated iff $\mathcal{E}$ forms a $k$-design; the $O(1/d)$ enters only through the closed-form evaluation of the Haar value. Thus \eqref{lowerbound} establishes a relationship between the frame potential and the partition function, or equivalently the unnormalized thermal purity, complementing earlier observations in Brownian circuits \cite{Jian:2022pvj} and a spin model on a triangular lattice \cite{Hunter-Jones:2019lps}. 

The right panel of Fig.\,\ref{fig:FP2PKGUEwithbetaplot} illustrates the dependence of the frame potential on the inverse temperature $\beta$ for several values of $k$. The numerical data are in excellent agreement with the analytic expression in \eqref{ineq}. In the low-temperature (large-$\beta$) regime, the partition function is dominated by the ground-state contribution of $\mathcal{Z}(2\beta) = \sum_n e^{-2 \beta E_n} \simeq e^{-2 \beta E_0}$, leading to
\begin{align}
   F^{(k, \beta)}_{\mathrm{2PK}} \sim \frac{k!}{d^k} \, e^{2 k \beta |E_0|}\,,~~~ (\beta \rightarrow \infty)\,, \label{gs}
\end{align}
where $E_0 = - |E_0| < 0$ is the ground-state energy. Consequently, the asymptotic growth exponent is $\alpha_k = 2 k |E_0|$. For the GUE with spectral support in $[-2,2]$, one has $E_0 \simeq -2$ in the large-$d$ limit, yielding $\alpha_k = 4 k$. This is in agreement with the numerical results in  Fig.\,\ref{fig:FP2PKGUEwithbetaplot} (right panel, dashed lines). In SM \ref{sec:Wightman}, we consider symmetric insertions of $\rho_{\beta}^{1/2}$, \emph{i.e.}, the Wightman regularization \cite{Maldacena:2015waa, Roberts:2016hpo, Cotler:2017jue}, and compare with the asymmetric version \eqref{eq:F2PKbeta}.

\emph{Gravity interpretation:} Interestingly, the 2PK protocol admits a holographic interpretation. In the interaction picture, \eqref{eq:U2PK} can be written as
\begin{align}
U_{\mathrm{2PK}}(t_1,t_2,t_3) :=
e^{-iH(t_1+t_2+t_3)}P(t_1 + t_2) P(t_1)\,. \label{Ush}
\end{align}
Hence, $U_{\mathrm{2PK}}$ is a single global evolution of total duration $T := t_1+t_2+t_3$ dressed by two precursors $P(\tau)\equiv e^{iH\tau}Pe^{-iH\tau}$ at cumulative times $t_1$ and $t_1+t_2$. Holographically, each precursor is a bulk shockwave \cite{Stanford:2014complexity}, so $U_{\mathrm{2PK}}$ creates two such shockwaves. Therefore, the frame potential, which counts the overlap, quantifies how effectively repeated shockwaves interact, randomizing quantum information and producing Haar-like dynamics.

\emph{Conclusion:} Unlike multi-Hamiltonian constructions
\cite{Zhou:2025noh, Zhou:2026ubz, Sun:2026wrk}, our 2PK protocol generates
approximate unitary $k$-designs from a single chaotic Hamiltonian interspersed
with two Pauli kicks. It requires neither exactly Haar-random eigenstates nor an ensemble of Hamiltonians, only non-resonant spectral phases from temporal sampling and a kick that rotates $H$ into a nearly orthogonal operator direction ($|c|\ll 1$). These ingredients are sufficient to drive the overlap $z:=\mathrm{Tr}(U^\dagger V)$ towards the centered complex-Gaussian statistics of unit variance characteristic of the Haar ensemble. Consequently, 2PK remains effective for deterministic spin chains, where the 3SP \cite{Zhou:2026ubz} is either not applicable or fails. Built on a single Hamiltonian, it admits a canonical finite-temperature extension and a lower bound on the frame potential in terms of the equilibrium partition function.

Future directions include chaos-to-integrability transitions in spin chains, SYK models \cite{Garcia-Garcia:2017bkg}, and Rosenzweig--Porter type ensembles \cite{RPmodel, Amini:2015mvz, Cadez:2024mjc, Jahnke:2025exd}, where the latter may provide a quantitative connection between fractality and $k$-design formation. It is also interesting to consider continuous symmetries. In the presence of a conserved charge, the kick must lie within the corresponding commuting subalgebra, raising the question of when such symmetry-restricted kicks can still generate designs. Finally, it remains to be understood whether $k$-design formation is governed by $k$-freeness \cite{Fava:2023pac}; the 2PK protocol provides a natural setting in which to investigate this connection.

A rigorous study of the holographic interpretation of the 2PK protocol may shed light on the shockwave geometries. In fact, generalizing \eqref{Ush} to the $n$PK protocol:
\begin{align}
    U_{n\mathrm{PK}}(t_1,\cdots,t_{n+1}) := e^{-iHT}\prod_{j=n}^{1} P(\tau_j)\,,~~ \tau_j=\sum_{i = 1}^j t_i\,, \nonumber
\end{align}
where $\prod_{j=n}^{1}P(\tau_j) := P(\tau_n)P(\tau_{n-1})\cdots P(\tau_1)$ and $T := \sum_{i=1}^{n+1} t_i$ suggests that dressed Paulis may yield multi-shockwave \cite{Shenker:2013yza} or caterpillar \cite{Magan:2024aet, Magan:2025hce, Bavaro:2025ooe} geometries in the bulk gravity. We leave a detailed study for future work.

\emph{Note added:} While completing this work, we became aware of Ref.\,\cite{Hou:2026cyf}. Their unitary construction requires Hamiltonian randomness, whereas our 2PK protocol is fully deterministic. After this work, Ref.\,\cite{Yang:2026opw} appeared and confirmed our GUE results for 1PK and 2PK protocols. However, deterministic and spatially local chaotic Hamiltonians are not addressed in their study.

\emph{Acknowledgments:} We would like to thank V. Jahnke, I. M. Khaymovich, J. Magan, D. Rosa, J. Sonner, S. Sur, and Y-N. Zhou for related discussions. The final stages of the work were carried out at the University of Geneva, supported by the ``COST Action CA22113: Short Term Scientific Mission (STSM) 2026''. The work is also supported by FWO-Vlaanderen projects G012222N and G0A2226N, and by the VUB Research Council through the Strategic Research Program High-Energy Physics.

\bibliography{references}    

\begin{thebibliography}{49}%
\makeatletter
\providecommand \@ifxundefined [1]{%
 \@ifx{#1\undefined}
}%
\providecommand \@ifnum [1]{%
 \ifnum #1\expandafter \@firstoftwo
 \else \expandafter \@secondoftwo
 \fi
}%
\providecommand \@ifx [1]{%
 \ifx #1\expandafter \@firstoftwo
 \else \expandafter \@secondoftwo
 \fi
}%
\providecommand \natexlab [1]{#1}%
\providecommand \enquote  [1]{``#1''}%
\providecommand \bibnamefont  [1]{#1}%
\providecommand \bibfnamefont [1]{#1}%
\providecommand \citenamefont [1]{#1}%
\providecommand \href@noop [0]{\@secondoftwo}%
\providecommand \href [0]{\begingroup \@sanitize@url \@href}%
\providecommand \@href[1]{\@@startlink{#1}\@@href}%
\providecommand \@@href[1]{\endgroup#1\@@endlink}%
\providecommand \@sanitize@url [0]{\catcode `\\12\catcode `\$12\catcode
  `\&12\catcode `\#12\catcode `\^12\catcode `\_12\catcode `\%12\relax}%
\providecommand \@@startlink[1]{}%
\providecommand \@@endlink[0]{}%
\providecommand \url  [0]{\begingroup\@sanitize@url \@url }%
\providecommand \@url [1]{\endgroup\@href {#1}{\urlprefix }}%
\providecommand \urlprefix  [0]{URL }%
\providecommand \Eprint [0]{\href }%
\providecommand \doibase [0]{http://dx.doi.org/}%
\providecommand \selectlanguage [0]{\@gobble}%
\providecommand \bibinfo  [0]{\@secondoftwo}%
\providecommand \bibfield  [0]{\@secondoftwo}%
\providecommand \translation [1]{[#1]}%
\providecommand \BibitemOpen [0]{}%
\providecommand \bibitemStop [0]{}%
\providecommand \bibitemNoStop [0]{.\EOS\space}%
\providecommand \EOS [0]{\spacefactor3000\relax}%
\providecommand \BibitemShut  [1]{\csname bibitem#1\endcsname}%
\let\auto@bib@innerbib\@empty
\bibitem [{\citenamefont {Mele}(2024)}]{Mele:2023ojv}%
  \BibitemOpen
  \bibfield  {author} {\bibinfo {author} {\bibfnamefont {Antonio~Anna}\
  \bibnamefont {Mele}},\ }\bibfield  {title} {\enquote {\bibinfo {title}
  {{Introduction to Haar Measure Tools in Quantum Information: A Beginner's
  Tutorial}},}\ }\href {\doibase 10.22331/q-2024-05-08-1340} {\bibfield
  {journal} {\bibinfo  {journal} {Quantum}\ }\textbf {\bibinfo {volume} {8}},\
  \bibinfo {pages} {1340} (\bibinfo {year} {2024})}\BibitemShut {NoStop}%
\bibitem [{\citenamefont {Gross}\ \emph {et~al.}(2007)\citenamefont {Gross},
  \citenamefont {Audenaert},\ and\ \citenamefont {Eisert}}]{Gross:2007xgw}%
  \BibitemOpen
  \bibfield  {author} {\bibinfo {author} {\bibfnamefont {D.}~\bibnamefont
  {Gross}}, \bibinfo {author} {\bibfnamefont {K.}~\bibnamefont {Audenaert}}, \
  and\ \bibinfo {author} {\bibfnamefont {J.}~\bibnamefont {Eisert}},\
  }\bibfield  {title} {\enquote {\bibinfo {title} {{Evenly distributed
  unitaries: On the structure of unitary designs}},}\ }\href {\doibase
  10.1063/1.2716992} {\bibfield  {journal} {\bibinfo  {journal} {J. Math.
  Phys.}\ }\textbf {\bibinfo {volume} {48}},\ \bibinfo {pages} {052104}
  (\bibinfo {year} {2007})}\BibitemShut {NoStop}%
\bibitem [{\citenamefont {Roberts}\ and\ \citenamefont
  {Yoshida}(2017)}]{Roberts:2016hpo}%
  \BibitemOpen
  \bibfield  {author} {\bibinfo {author} {\bibfnamefont {Daniel~A.}\
  \bibnamefont {Roberts}}\ and\ \bibinfo {author} {\bibfnamefont {Beni}\
  \bibnamefont {Yoshida}},\ }\bibfield  {title} {\enquote {\bibinfo {title}
  {Chaos and complexity by design},}\ }\href {\doibase 10.1007/JHEP04(2017)121}
  {\bibfield  {journal} {\bibinfo  {journal} {JHEP}\ }\textbf {\bibinfo
  {volume} {04}},\ \bibinfo {pages} {121} (\bibinfo {year} {2017})}\BibitemShut
  {NoStop}%
\bibitem [{\citenamefont {Cotler}\ \emph {et~al.}(2017)\citenamefont {Cotler},
  \citenamefont {Hunter-Jones}, \citenamefont {Liu},\ and\ \citenamefont
  {Yoshida}}]{Cotler:2017jue}%
  \BibitemOpen
  \bibfield  {author} {\bibinfo {author} {\bibfnamefont {Jordan}\ \bibnamefont
  {Cotler}}, \bibinfo {author} {\bibfnamefont {Nicholas}\ \bibnamefont
  {Hunter-Jones}}, \bibinfo {author} {\bibfnamefont {Junyu}\ \bibnamefont
  {Liu}}, \ and\ \bibinfo {author} {\bibfnamefont {Beni}\ \bibnamefont
  {Yoshida}},\ }\bibfield  {title} {\enquote {\bibinfo {title} {Chaos,
  complexity, and random matrices},}\ }\href {\doibase 10.1007/JHEP11(2017)048}
  {\bibfield  {journal} {\bibinfo  {journal} {JHEP}\ }\textbf {\bibinfo
  {volume} {11}},\ \bibinfo {pages} {048} (\bibinfo {year} {2017})}\BibitemShut
  {NoStop}%
\bibitem [{\citenamefont {Emerson}\ \emph {et~al.}(2003)\citenamefont
  {Emerson}, \citenamefont {Weinstein}, \citenamefont {Saraceno}, \citenamefont
  {Lloyd},\ and\ \citenamefont {Cory}}]{Emerson:2003www}%
  \BibitemOpen
  \bibfield  {author} {\bibinfo {author} {\bibfnamefont {Joseph}\ \bibnamefont
  {Emerson}}, \bibinfo {author} {\bibfnamefont {Yaakov~S.}\ \bibnamefont
  {Weinstein}}, \bibinfo {author} {\bibfnamefont {Marcos}\ \bibnamefont
  {Saraceno}}, \bibinfo {author} {\bibfnamefont {Seth}\ \bibnamefont {Lloyd}},
  \ and\ \bibinfo {author} {\bibfnamefont {David~G.}\ \bibnamefont {Cory}},\
  }\bibfield  {title} {\enquote {\bibinfo {title} {{Pseudo-Random Unitary
  Operators for Quantum Information Processing}},}\ }\href {\doibase
  10.1126/science.1090790} {\bibfield  {journal} {\bibinfo  {journal}
  {Science}\ }\textbf {\bibinfo {volume} {302}},\ \bibinfo {pages} {2098--2100}
  (\bibinfo {year} {2003})}\BibitemShut {NoStop}%
\bibitem [{\citenamefont {Brand{\~a}o}\ \emph {et~al.}(2016)\citenamefont
  {Brand{\~a}o}, \citenamefont {Harrow},\ and\ \citenamefont
  {Horodecki}}]{Brandao:2016glh}%
  \BibitemOpen
  \bibfield  {author} {\bibinfo {author} {\bibfnamefont {Fernando G.
  {\,}S.~{\,}L.}\ \bibnamefont {Brand{\~a}o}}, \bibinfo {author} {\bibfnamefont
  {Aram~W.}\ \bibnamefont {Harrow}}, \ and\ \bibinfo {author} {\bibfnamefont
  {Micha{\l}}\ \bibnamefont {Horodecki}},\ }\bibfield  {title} {\enquote
  {\bibinfo {title} {{Efficient Quantum Pseudorandomness}},}\ }\href {\doibase
  10.1103/PhysRevLett.116.170502} {\bibfield  {journal} {\bibinfo  {journal}
  {Phys. Rev. Lett.}\ }\textbf {\bibinfo {volume} {116}},\ \bibinfo {pages}
  {170502} (\bibinfo {year} {2016})}\BibitemShut {NoStop}%
\bibitem [{\citenamefont {Brand{\~a}o}\ \emph {et~al.}(2021)\citenamefont
  {Brand{\~a}o}, \citenamefont {Chemissany}, \citenamefont {Hunter-Jones},
  \citenamefont {Kueng},\ and\ \citenamefont {Preskill}}]{Brandao:2019sgy}%
  \BibitemOpen
  \bibfield  {author} {\bibinfo {author} {\bibfnamefont {Fernando G. S.~L.}\
  \bibnamefont {Brand{\~a}o}}, \bibinfo {author} {\bibfnamefont {Wissam}\
  \bibnamefont {Chemissany}}, \bibinfo {author} {\bibfnamefont {Nicholas}\
  \bibnamefont {Hunter-Jones}}, \bibinfo {author} {\bibfnamefont {Richard}\
  \bibnamefont {Kueng}}, \ and\ \bibinfo {author} {\bibfnamefont {John}\
  \bibnamefont {Preskill}},\ }\bibfield  {title} {\enquote {\bibinfo {title}
  {{Models of Quantum Complexity Growth}},}\ }\href {\doibase
  10.1103/PRXQuantum.2.030316} {\bibfield  {journal} {\bibinfo  {journal} {PRX
  Quantum}\ }\textbf {\bibinfo {volume} {2}},\ \bibinfo {pages} {030316}
  (\bibinfo {year} {2021})}\BibitemShut {NoStop}%
\bibitem [{\citenamefont {Schuster}\ \emph {et~al.}(2025)\citenamefont
  {Schuster}, \citenamefont {Haferkamp},\ and\ \citenamefont
  {Huang}}]{Schuster:2024ajb}%
  \BibitemOpen
  \bibfield  {author} {\bibinfo {author} {\bibfnamefont {Thomas}\ \bibnamefont
  {Schuster}}, \bibinfo {author} {\bibfnamefont {Jonas}\ \bibnamefont
  {Haferkamp}}, \ and\ \bibinfo {author} {\bibfnamefont {Hsin-Yuan}\
  \bibnamefont {Huang}},\ }\bibfield  {title} {\enquote {\bibinfo {title}
  {{Random unitaries in extremely low depth}},}\ }\href {\doibase
  10.1126/science.adv8590} {\bibfield  {journal} {\bibinfo  {journal}
  {Science}\ }\textbf {\bibinfo {volume} {389}},\ \bibinfo {pages} {adv8590}
  (\bibinfo {year} {2025})}\BibitemShut {NoStop}%
\bibitem [{\citenamefont {Hunter-Jones}(2019)}]{Hunter-Jones:2019lps}%
  \BibitemOpen
  \bibfield  {author} {\bibinfo {author} {\bibfnamefont {Nicholas}\
  \bibnamefont {Hunter-Jones}},\ }\bibfield  {title} {\enquote {\bibinfo
  {title} {{Unitary designs from statistical mechanics in random quantum
  circuits}},}\ }\href@noop {} {\  (\bibinfo {year} {2019})},\ \Eprint
  {http://arxiv.org/abs/1905.12053} {arXiv:1905.12053 [quant-ph]} \BibitemShut
  {NoStop}%
\bibitem [{\citenamefont {Choi}\ \emph {et~al.}(2023)\citenamefont {Choi} \emph
  {et~al.}}]{Choi:2021npc}%
  \BibitemOpen
  \bibfield  {author} {\bibinfo {author} {\bibfnamefont {Joonhee}\ \bibnamefont
  {Choi}} \emph {et~al.},\ }\bibfield  {title} {\enquote {\bibinfo {title}
  {{Preparing random states and benchmarking with many-body quantum chaos}},}\
  }\href {\doibase 10.1038/s41586-022-05442-1} {\bibfield  {journal} {\bibinfo
  {journal} {Nature}\ }\textbf {\bibinfo {volume} {613}},\ \bibinfo {pages}
  {468--473} (\bibinfo {year} {2023})}\BibitemShut {NoStop}%
\bibitem [{\citenamefont {Ho}\ and\ \citenamefont {Choi}(2022)}]{Ho:2021dmh}%
  \BibitemOpen
  \bibfield  {author} {\bibinfo {author} {\bibfnamefont {Wen~Wei}\ \bibnamefont
  {Ho}}\ and\ \bibinfo {author} {\bibfnamefont {Soonwon}\ \bibnamefont
  {Choi}},\ }\bibfield  {title} {\enquote {\bibinfo {title} {{Exact emergent
  quantum state designs from quantum chaotic dynamics}},}\ }\href {\doibase
  10.1103/PhysRevLett.128.060601} {\bibfield  {journal} {\bibinfo  {journal}
  {Phys. Rev. Lett.}\ }\textbf {\bibinfo {volume} {128}},\ \bibinfo {pages}
  {060601} (\bibinfo {year} {2022})}\BibitemShut {NoStop}%
\bibitem [{\citenamefont {Cotler}\ \emph {et~al.}(2023)\citenamefont {Cotler},
  \citenamefont {Mark}, \citenamefont {Huang}, \citenamefont {Hernandez},
  \citenamefont {Choi}, \citenamefont {Shaw}, \citenamefont {Endres},\ and\
  \citenamefont {Choi}}]{Cotler:2021pbc}%
  \BibitemOpen
  \bibfield  {author} {\bibinfo {author} {\bibfnamefont {Jordan~S.}\
  \bibnamefont {Cotler}}, \bibinfo {author} {\bibfnamefont {Daniel~K.}\
  \bibnamefont {Mark}}, \bibinfo {author} {\bibfnamefont {Hsin-Yuan}\
  \bibnamefont {Huang}}, \bibinfo {author} {\bibfnamefont {Felipe}\
  \bibnamefont {Hernandez}}, \bibinfo {author} {\bibfnamefont {Joonhee}\
  \bibnamefont {Choi}}, \bibinfo {author} {\bibfnamefont {Adam~L.}\
  \bibnamefont {Shaw}}, \bibinfo {author} {\bibfnamefont {Manuel}\ \bibnamefont
  {Endres}}, \ and\ \bibinfo {author} {\bibfnamefont {Soonwon}\ \bibnamefont
  {Choi}},\ }\bibfield  {title} {\enquote {\bibinfo {title} {{Emergent Quantum
  State Designs from Individual Many-Body Wave Functions}},}\ }\href {\doibase
  10.1103/PRXQuantum.4.010311} {\bibfield  {journal} {\bibinfo  {journal} {PRX
  Quantum}\ }\textbf {\bibinfo {volume} {4}},\ \bibinfo {pages} {010311}
  (\bibinfo {year} {2023})}\BibitemShut {NoStop}%
\bibitem [{\citenamefont {Claeys}\ and\ \citenamefont
  {Lamacraft}(2022)}]{Claeys:2022hts}%
  \BibitemOpen
  \bibfield  {author} {\bibinfo {author} {\bibfnamefont {Pieter~W.}\
  \bibnamefont {Claeys}}\ and\ \bibinfo {author} {\bibfnamefont {Austen}\
  \bibnamefont {Lamacraft}},\ }\bibfield  {title} {\enquote {\bibinfo {title}
  {{Emergent quantum state designs and biunitarity in dual-unitary circuit
  dynamics}},}\ }\href {\doibase 10.22331/q-2022-06-15-738} {\bibfield
  {journal} {\bibinfo  {journal} {Quantum}\ }\textbf {\bibinfo {volume} {6}},\
  \bibinfo {pages} {738} (\bibinfo {year} {2022})}\BibitemShut {NoStop}%
\bibitem [{\citenamefont {Fava}\ \emph {et~al.}(2025)\citenamefont {Fava},
  \citenamefont {Kurchan},\ and\ \citenamefont {Pappalardi}}]{Fava:2023pac}%
  \BibitemOpen
  \bibfield  {author} {\bibinfo {author} {\bibfnamefont {Michele}\ \bibnamefont
  {Fava}}, \bibinfo {author} {\bibfnamefont {Jorge}\ \bibnamefont {Kurchan}}, \
  and\ \bibinfo {author} {\bibfnamefont {Silvia}\ \bibnamefont {Pappalardi}},\
  }\bibfield  {title} {\enquote {\bibinfo {title} {{Designs via Free
  Probability}},}\ }\href {\doibase 10.1103/PhysRevX.15.011031} {\bibfield
  {journal} {\bibinfo  {journal} {Phys. Rev. X}\ }\textbf {\bibinfo {volume}
  {15}},\ \bibinfo {pages} {011031} (\bibinfo {year} {2025})}\BibitemShut
  {NoStop}%
\bibitem [{\citenamefont {Nakata}\ \emph {et~al.}(2025)\citenamefont {Nakata},
  \citenamefont {Takeuchi}, \citenamefont {Kliesch},\ and\ \citenamefont
  {Darmawan}}]{Nakata:2024tla}%
  \BibitemOpen
  \bibfield  {author} {\bibinfo {author} {\bibfnamefont {Yoshifumi}\
  \bibnamefont {Nakata}}, \bibinfo {author} {\bibfnamefont {Yuki}\ \bibnamefont
  {Takeuchi}}, \bibinfo {author} {\bibfnamefont {Martin}\ \bibnamefont
  {Kliesch}}, \ and\ \bibinfo {author} {\bibfnamefont {Andrew}\ \bibnamefont
  {Darmawan}},\ }\bibfield  {title} {\enquote {\bibinfo {title} {{Computational
  Complexity of Unitary and State Design Properties}},}\ }\href {\doibase
  10.1103/21vm-bz3t} {\bibfield  {journal} {\bibinfo  {journal} {PRX Quantum}\
  }\textbf {\bibinfo {volume} {6}},\ \bibinfo {pages} {030345} (\bibinfo {year}
  {2025})}\BibitemShut {NoStop}%
\bibitem [{\citenamefont {Dowling}\ \emph {et~al.}(2025)\citenamefont
  {Dowling}, \citenamefont {De~Nardis}, \citenamefont {Heinrich}, \citenamefont
  {Turkeshi},\ and\ \citenamefont {Pappalardi}}]{Dowling:2025cxr}%
  \BibitemOpen
  \bibfield  {author} {\bibinfo {author} {\bibfnamefont {Neil}\ \bibnamefont
  {Dowling}}, \bibinfo {author} {\bibfnamefont {Jacopo}\ \bibnamefont
  {De~Nardis}}, \bibinfo {author} {\bibfnamefont {Markus}\ \bibnamefont
  {Heinrich}}, \bibinfo {author} {\bibfnamefont {Xhek}\ \bibnamefont
  {Turkeshi}}, \ and\ \bibinfo {author} {\bibfnamefont {Silvia}\ \bibnamefont
  {Pappalardi}},\ }\bibfield  {title} {\enquote {\bibinfo {title} {{Free
  Independence and Unitary Design from Random Matrix Product Unitaries}},}\
  }\href@noop {} {\  (\bibinfo {year} {2025})},\ \Eprint
  {http://arxiv.org/abs/2508.00051} {arXiv:2508.00051 [quant-ph]} \BibitemShut
  {NoStop}%
\bibitem [{\citenamefont {Cui}\ \emph {et~al.}(2025)\citenamefont {Cui},
  \citenamefont {Schuster}, \citenamefont {Mao}, \citenamefont {Huang},\ and\
  \citenamefont {Brandao}}]{Cui:2025teh}%
  \BibitemOpen
  \bibfield  {author} {\bibinfo {author} {\bibfnamefont {Laura}\ \bibnamefont
  {Cui}}, \bibinfo {author} {\bibfnamefont {Thomas}\ \bibnamefont {Schuster}},
  \bibinfo {author} {\bibfnamefont {Liang}\ \bibnamefont {Mao}}, \bibinfo
  {author} {\bibfnamefont {Hsin-Yuan}\ \bibnamefont {Huang}}, \ and\ \bibinfo
  {author} {\bibfnamefont {Fernando}\ \bibnamefont {Brandao}},\ }\bibfield
  {title} {\enquote {\bibinfo {title} {{Random unitaries from Hamiltonian
  dynamics}},}\ }\href@noop {} {\  (\bibinfo {year} {2025})},\ \Eprint
  {http://arxiv.org/abs/2510.08434} {arXiv:2510.08434 [quant-ph]} \BibitemShut
  {NoStop}%
\bibitem [{\citenamefont {Zhou}\ \emph
  {et~al.}(2026{\natexlab{a}})\citenamefont {Zhou}, \citenamefont
  {L{\"o}wenberg},\ and\ \citenamefont {Sonner}}]{Zhou:2025noh}%
  \BibitemOpen
  \bibfield  {author} {\bibinfo {author} {\bibfnamefont {Yi-Neng}\ \bibnamefont
  {Zhou}}, \bibinfo {author} {\bibfnamefont {Robin}\ \bibnamefont
  {L{\"o}wenberg}}, \ and\ \bibinfo {author} {\bibfnamefont {Julian}\
  \bibnamefont {Sonner}},\ }\bibfield  {title} {\enquote {\bibinfo {title}
  {{Realizing Unitary k-Designs with a Single Quench}},}\ }\href {\doibase
  10.1103/rvwb-r9lv} {\bibfield  {journal} {\bibinfo  {journal} {Phys. Rev.
  Lett.}\ }\textbf {\bibinfo {volume} {136}},\ \bibinfo {pages} {220403}
  (\bibinfo {year} {2026}{\natexlab{a}})}\BibitemShut {NoStop}%
\bibitem [{\citenamefont {Zhou}\ \emph
  {et~al.}(2026{\natexlab{b}})\citenamefont {Zhou}, \citenamefont {Zhou},\ and\
  \citenamefont {Sonner}}]{Zhou:2026ubz}%
  \BibitemOpen
  \bibfield  {author} {\bibinfo {author} {\bibfnamefont {Yi-Neng}\ \bibnamefont
  {Zhou}}, \bibinfo {author} {\bibfnamefont {Tian-Gang}\ \bibnamefont {Zhou}},
  \ and\ \bibinfo {author} {\bibfnamefont {Julian}\ \bibnamefont {Sonner}},\
  }\bibfield  {title} {\enquote {\bibinfo {title} {{Three Hamiltonians are
  Sufficient for Unitary $k$-Design in Temporal Ensemble}},}\ }\href@noop {} {\
   (\bibinfo {year} {2026}{\natexlab{b}})},\ \Eprint
  {http://arxiv.org/abs/2604.04205} {arXiv:2604.04205 [quant-ph]} \BibitemShut
  {NoStop}%
\bibitem [{\citenamefont {Chenu}\ \emph {et~al.}(2018)\citenamefont {Chenu},
  \citenamefont {Egusquiza}, \citenamefont {Molina-Vilaplana},\ and\
  \citenamefont {del Campo}}]{Chenu:2017qdv}%
  \BibitemOpen
  \bibfield  {author} {\bibinfo {author} {\bibfnamefont {A.}~\bibnamefont
  {Chenu}}, \bibinfo {author} {\bibfnamefont {I.~L.}\ \bibnamefont
  {Egusquiza}}, \bibinfo {author} {\bibfnamefont {J.}~\bibnamefont
  {Molina-Vilaplana}}, \ and\ \bibinfo {author} {\bibfnamefont
  {A.}~\bibnamefont {del Campo}},\ }\bibfield  {title} {\enquote {\bibinfo
  {title} {{Quantum work statistics, Loschmidt echo and information
  scrambling}},}\ }\href {\doibase 10.1038/s41598-018-30982-w} {\bibfield
  {journal} {\bibinfo  {journal} {Sci. Rep.}\ }\textbf {\bibinfo {volume}
  {8}},\ \bibinfo {pages} {12634} (\bibinfo {year} {2018})}\BibitemShut
  {NoStop}%
\bibitem [{\citenamefont {Chenu}\ \emph {et~al.}(2019)\citenamefont {Chenu},
  \citenamefont {Molina-Vilaplana},\ and\ \citenamefont
  {Del~Campo}}]{Chenu:2018spm}%
  \BibitemOpen
  \bibfield  {author} {\bibinfo {author} {\bibfnamefont {Aur{\'e}lia}\
  \bibnamefont {Chenu}}, \bibinfo {author} {\bibfnamefont {Javier}\
  \bibnamefont {Molina-Vilaplana}}, \ and\ \bibinfo {author} {\bibfnamefont
  {Adolfo}\ \bibnamefont {Del~Campo}},\ }\bibfield  {title} {\enquote {\bibinfo
  {title} {{Work Statistics, Loschmidt Echo and Information Scrambling in
  Chaotic Quantum Systems}},}\ }\href {\doibase 10.22331/q-2019-03-04-127}
  {\bibfield  {journal} {\bibinfo  {journal} {Quantum}\ }\textbf {\bibinfo
  {volume} {3}},\ \bibinfo {pages} {127} (\bibinfo {year} {2019})}\BibitemShut
  {NoStop}%
\bibitem [{\citenamefont {Sun}\ and\ \citenamefont
  {Zhang}(2026)}]{Sun:2026wrk}%
  \BibitemOpen
  \bibfield  {author} {\bibinfo {author} {\bibfnamefont {Ning}\ \bibnamefont
  {Sun}}\ and\ \bibinfo {author} {\bibfnamefont {Pengfei}\ \bibnamefont
  {Zhang}},\ }\bibfield  {title} {\enquote {\bibinfo {title} {{Unitary Designs
  from Two Chaotic Hamiltonians and a Random Pauli Operation}},}\ }\href@noop
  {} {\  (\bibinfo {year} {2026})},\ \Eprint {http://arxiv.org/abs/2604.10122}
  {arXiv:2604.10122 [quant-ph]} \BibitemShut {NoStop}%
\bibitem [{\citenamefont {Scott}(2008)}]{Scott_2008}%
  \BibitemOpen
  \bibfield  {author} {\bibinfo {author} {\bibfnamefont {A~J}\ \bibnamefont
  {Scott}},\ }\bibfield  {title} {\enquote {\bibinfo {title} {Optimizing
  quantum process tomography with unitary 2-designs},}\ }\href {\doibase
  10.1088/1751-8113/41/5/055308} {\bibfield  {journal} {\bibinfo  {journal}
  {Journal of Physics A: Mathematical and Theoretical}\ }\textbf {\bibinfo
  {volume} {41}},\ \bibinfo {pages} {055308} (\bibinfo {year}
  {2008})}\BibitemShut {NoStop}%
\bibitem [{\citenamefont {Sachdev}\ and\ \citenamefont
  {Ye}(1993)}]{PhysRevLett.70.3339}%
  \BibitemOpen
  \bibfield  {author} {\bibinfo {author} {\bibfnamefont {Subir}\ \bibnamefont
  {Sachdev}}\ and\ \bibinfo {author} {\bibfnamefont {Jinwu}\ \bibnamefont
  {Ye}},\ }\bibfield  {title} {\enquote {\bibinfo {title} {Gapless spin-fluid
  ground state in a random quantum heisenberg magnet},}\ }\href {\doibase
  10.1103/PhysRevLett.70.3339} {\bibfield  {journal} {\bibinfo  {journal}
  {Phys. Rev. Lett.}\ }\textbf {\bibinfo {volume} {70}},\ \bibinfo {pages}
  {3339--3342} (\bibinfo {year} {1993})}\BibitemShut {NoStop}%
\bibitem [{\citenamefont {Kitaev}(2015)}]{Kittu}%
  \BibitemOpen
  \bibfield  {author} {\bibinfo {author} {\bibfnamefont {A.}~\bibnamefont
  {Kitaev}},\ }\href@noop {} {\enquote {\bibinfo {title} {A simple model of
  quantum holography (part 1) and (part 2)},}\ }\bibinfo {howpublished}
  {\url{https://online.kitp.ucsb.edu/online/joint98/kitaev/},
  \url{https://online.kitp.ucsb.edu/online/entangled15/kitaev2/}} (\bibinfo
  {year} {2015}),\ \bibinfo {note} {talk given at KITP}\BibitemShut {NoStop}%
\bibitem [{\citenamefont {Hanada}\ \emph {et~al.}(2024)\citenamefont {Hanada},
  \citenamefont {Jevicki}, \citenamefont {Liu}, \citenamefont {Rinaldi},\ and\
  \citenamefont {Tezuka}}]{Hanada:2023rkf}%
  \BibitemOpen
  \bibfield  {author} {\bibinfo {author} {\bibfnamefont {Masanori}\
  \bibnamefont {Hanada}}, \bibinfo {author} {\bibfnamefont {Antal}\
  \bibnamefont {Jevicki}}, \bibinfo {author} {\bibfnamefont {Xianlong}\
  \bibnamefont {Liu}}, \bibinfo {author} {\bibfnamefont {Enrico}\ \bibnamefont
  {Rinaldi}}, \ and\ \bibinfo {author} {\bibfnamefont {Masaki}\ \bibnamefont
  {Tezuka}},\ }\bibfield  {title} {\enquote {\bibinfo {title} {{A model of
  randomly-coupled Pauli spins}},}\ }\href {\doibase 10.1007/JHEP05(2024)280}
  {\bibfield  {journal} {\bibinfo  {journal} {JHEP}\ }\textbf {\bibinfo
  {volume} {05}},\ \bibinfo {pages} {280} (\bibinfo {year} {2024})}\BibitemShut
  {NoStop}%
\bibitem [{\citenamefont {Basu}\ \emph {et~al.}(2026)\citenamefont {Basu},
  \citenamefont {Das},\ and\ \citenamefont {Nandy}}]{Basu:2025ubf}%
  \BibitemOpen
  \bibfield  {author} {\bibinfo {author} {\bibfnamefont {Pallab}\ \bibnamefont
  {Basu}}, \bibinfo {author} {\bibfnamefont {Suman}\ \bibnamefont {Das}}, \
  and\ \bibinfo {author} {\bibfnamefont {Pratik}\ \bibnamefont {Nandy}},\
  }\bibfield  {title} {\enquote {\bibinfo {title} {{Complexity of quadratic
  quantum chaos}},}\ }\href {\doibase 10.1007/JHEP04(2026)081} {\bibfield
  {journal} {\bibinfo  {journal} {JHEP}\ }\textbf {\bibinfo {volume} {04}},\
  \bibinfo {pages} {081} (\bibinfo {year} {2026})}\BibitemShut {NoStop}%
\bibitem [{\citenamefont {Rosa}\ \emph {et~al.}(2020)\citenamefont {Rosa},
  \citenamefont {Rossini}, \citenamefont {Andolina}, \citenamefont {Polini},\
  and\ \citenamefont {Carrega}}]{Rosa:2019jin}%
  \BibitemOpen
  \bibfield  {author} {\bibinfo {author} {\bibfnamefont {Dario}\ \bibnamefont
  {Rosa}}, \bibinfo {author} {\bibfnamefont {Davide}\ \bibnamefont {Rossini}},
  \bibinfo {author} {\bibfnamefont {Gian~Marcello}\ \bibnamefont {Andolina}},
  \bibinfo {author} {\bibfnamefont {Marco}\ \bibnamefont {Polini}}, \ and\
  \bibinfo {author} {\bibfnamefont {Matteo}\ \bibnamefont {Carrega}},\
  }\bibfield  {title} {\enquote {\bibinfo {title} {{Ultra-stable charging of
  fast-scrambling SYK quantum batteries}},}\ }\href {\doibase
  10.1007/JHEP11(2020)067} {\bibfield  {journal} {\bibinfo  {journal} {JHEP}\
  }\textbf {\bibinfo {volume} {11}},\ \bibinfo {pages} {067} (\bibinfo {year}
  {2020})}\BibitemShut {NoStop}%
\bibitem [{\citenamefont {You}\ \emph {et~al.}(2017)\citenamefont {You},
  \citenamefont {Ludwig},\ and\ \citenamefont {Xu}}]{Ludwig}%
  \BibitemOpen
  \bibfield  {author} {\bibinfo {author} {\bibfnamefont {Yi-Zhuang}\
  \bibnamefont {You}}, \bibinfo {author} {\bibfnamefont {Andreas W.~W.}\
  \bibnamefont {Ludwig}}, \ and\ \bibinfo {author} {\bibfnamefont {Cenke}\
  \bibnamefont {Xu}},\ }\bibfield  {title} {\enquote {\bibinfo {title}
  {{Sachdev-Ye-Kitaev model and thermalization on the boundary of many-body
  localized fermionic symmetry-protected topological states}},}\ }\href
  {\doibase 10.1103/PhysRevB.95.115150} {\bibfield  {journal} {\bibinfo
  {journal} {Phys. Rev. B}\ }\textbf {\bibinfo {volume} {95}},\ \bibinfo
  {pages} {115150} (\bibinfo {year} {2017})}\BibitemShut {NoStop}%
\bibitem [{\citenamefont {Ba{\~n}uls}\ \emph {et~al.}(2011)\citenamefont
  {Ba{\~n}uls}, \citenamefont {Cirac},\ and\ \citenamefont
  {Hastings}}]{Banuls:2010zki}%
  \BibitemOpen
  \bibfield  {author} {\bibinfo {author} {\bibfnamefont {Mari~Carmen}\
  \bibnamefont {Ba{\~n}uls}}, \bibinfo {author} {\bibfnamefont {J.~Ignacio}\
  \bibnamefont {Cirac}}, \ and\ \bibinfo {author} {\bibfnamefont {Matthew~B.}\
  \bibnamefont {Hastings}},\ }\bibfield  {title} {\enquote {\bibinfo {title}
  {{Strong and Weak Thermalization of Infinite Nonintegrable Quantum
  Systems}},}\ }\href {\doibase 10.1103/PhysRevLett.106.050405} {\bibfield
  {journal} {\bibinfo  {journal} {Phys. Rev. Lett.}\ }\textbf {\bibinfo
  {volume} {106}},\ \bibinfo {pages} {050405} (\bibinfo {year}
  {2011})}\BibitemShut {NoStop}%
\bibitem [{\citenamefont {Oganesyan}\ and\ \citenamefont
  {Huse}(2007)}]{Oganesyan:2007wpd}%
  \BibitemOpen
  \bibfield  {author} {\bibinfo {author} {\bibfnamefont {Vadim}\ \bibnamefont
  {Oganesyan}}\ and\ \bibinfo {author} {\bibfnamefont {David~A.}\ \bibnamefont
  {Huse}},\ }\bibfield  {title} {\enquote {\bibinfo {title} {{Localization of
  interacting fermions at high temperature}},}\ }\href {\doibase
  10.1103/PhysRevB.75.155111} {\bibfield  {journal} {\bibinfo  {journal} {Phys.
  Rev. B}\ }\textbf {\bibinfo {volume} {75}},\ \bibinfo {pages} {155111}
  (\bibinfo {year} {2007})}\BibitemShut {NoStop}%
\bibitem [{\citenamefont {Atas}\ \emph {et~al.}(2013)\citenamefont {Atas},
  \citenamefont {Bogomolny}, \citenamefont {Giraud},\ and\ \citenamefont
  {Roux}}]{Atas2013distribution}%
  \BibitemOpen
  \bibfield  {author} {\bibinfo {author} {\bibfnamefont {Y.~Y.}\ \bibnamefont
  {Atas}}, \bibinfo {author} {\bibfnamefont {E.}~\bibnamefont {Bogomolny}},
  \bibinfo {author} {\bibfnamefont {O.}~\bibnamefont {Giraud}}, \ and\ \bibinfo
  {author} {\bibfnamefont {G.}~\bibnamefont {Roux}},\ }\bibfield  {title}
  {\enquote {\bibinfo {title} {Distribution of the ratio of consecutive level
  spacings in random matrix ensembles},}\ }\href {\doibase
  10.1103/PhysRevLett.110.084101} {\bibfield  {journal} {\bibinfo  {journal}
  {Phys. Rev. Lett.}\ }\textbf {\bibinfo {volume} {110}},\ \bibinfo {pages}
  {084101} (\bibinfo {year} {2013})}\BibitemShut {NoStop}%
\bibitem [{\citenamefont {Zhou}\ and\ \citenamefont
  {Giamarchi}(2026)}]{Tian-GangZhou:2026hbi}%
  \BibitemOpen
  \bibfield  {author} {\bibinfo {author} {\bibfnamefont {Tian-Gang}\
  \bibnamefont {Zhou}}\ and\ \bibinfo {author} {\bibfnamefont {Thierry}\
  \bibnamefont {Giamarchi}},\ }\bibfield  {title} {\enquote {\bibinfo {title}
  {{Solvable Random Unitary Dynamics in a Disordered Tomonaga-Luttinger
  Liquid}},}\ }\href@noop {} {\  (\bibinfo {year} {2026})},\ \Eprint
  {http://arxiv.org/abs/2604.25995} {arXiv:2604.25995 [quant-ph]} \BibitemShut
  {NoStop}%
\bibitem [{\citenamefont {Jian}\ \emph {et~al.}(2023)\citenamefont {Jian},
  \citenamefont {Bentsen},\ and\ \citenamefont {Swingle}}]{Jian:2022pvj}%
  \BibitemOpen
  \bibfield  {author} {\bibinfo {author} {\bibfnamefont {Shao-Kai}\
  \bibnamefont {Jian}}, \bibinfo {author} {\bibfnamefont {Gregory}\
  \bibnamefont {Bentsen}}, \ and\ \bibinfo {author} {\bibfnamefont {Brian}\
  \bibnamefont {Swingle}},\ }\bibfield  {title} {\enquote {\bibinfo {title}
  {{Linear growth of circuit complexity from Brownian dynamics}},}\ }\href
  {\doibase 10.1007/JHEP08(2023)190} {\bibfield  {journal} {\bibinfo  {journal}
  {JHEP}\ }\textbf {\bibinfo {volume} {08}},\ \bibinfo {pages} {190} (\bibinfo
  {year} {2023})}\BibitemShut {NoStop}%
\bibitem [{\citenamefont {Maldacena}\ \emph {et~al.}(2016)\citenamefont
  {Maldacena}, \citenamefont {Shenker},\ and\ \citenamefont
  {Stanford}}]{Maldacena:2015waa}%
  \BibitemOpen
  \bibfield  {author} {\bibinfo {author} {\bibfnamefont {Juan}\ \bibnamefont
  {Maldacena}}, \bibinfo {author} {\bibfnamefont {Stephen~H.}\ \bibnamefont
  {Shenker}}, \ and\ \bibinfo {author} {\bibfnamefont {Douglas}\ \bibnamefont
  {Stanford}},\ }\bibfield  {title} {\enquote {\bibinfo {title} {A bound on
  chaos},}\ }\href {\doibase 10.1007/JHEP08(2016)106} {\bibfield  {journal}
  {\bibinfo  {journal} {JHEP}\ }\textbf {\bibinfo {volume} {08}},\ \bibinfo
  {pages} {106} (\bibinfo {year} {2016})}\BibitemShut {NoStop}%
\bibitem [{\citenamefont {Stanford}\ and\ \citenamefont
  {Susskind}(2014)}]{Stanford:2014complexity}%
  \BibitemOpen
  \bibfield  {author} {\bibinfo {author} {\bibfnamefont {Douglas}\ \bibnamefont
  {Stanford}}\ and\ \bibinfo {author} {\bibfnamefont {Leonard}\ \bibnamefont
  {Susskind}},\ }\bibfield  {title} {\enquote {\bibinfo {title} {Complexity and
  shock wave geometries},}\ }\href {\doibase 10.1103/PhysRevD.90.126007}
  {\bibfield  {journal} {\bibinfo  {journal} {Phys. Rev. D}\ }\textbf {\bibinfo
  {volume} {90}},\ \bibinfo {pages} {126007} (\bibinfo {year}
  {2014})}\BibitemShut {NoStop}%
\bibitem [{\citenamefont {Garc{\'\i}a-Garc{\'\i}a}\ \emph
  {et~al.}(2018)\citenamefont {Garc{\'\i}a-Garc{\'\i}a}, \citenamefont
  {Loureiro}, \citenamefont {Romero-Berm{\'u}dez},\ and\ \citenamefont
  {Tezuka}}]{Garcia-Garcia:2017bkg}%
  \BibitemOpen
  \bibfield  {author} {\bibinfo {author} {\bibfnamefont {Antonio~M.}\
  \bibnamefont {Garc{\'\i}a-Garc{\'\i}a}}, \bibinfo {author} {\bibfnamefont
  {Bruno}\ \bibnamefont {Loureiro}}, \bibinfo {author} {\bibfnamefont
  {Aurelio}\ \bibnamefont {Romero-Berm{\'u}dez}}, \ and\ \bibinfo {author}
  {\bibfnamefont {Masaki}\ \bibnamefont {Tezuka}},\ }\bibfield  {title}
  {\enquote {\bibinfo {title} {{Chaotic-Integrable Transition in the
  Sachdev-Ye-Kitaev Model}},}\ }\href {\doibase 10.1103/PhysRevLett.120.241603}
  {\bibfield  {journal} {\bibinfo  {journal} {Phys. Rev. Lett.}\ }\textbf
  {\bibinfo {volume} {120}},\ \bibinfo {pages} {241603} (\bibinfo {year}
  {2018})}\BibitemShut {NoStop}%
\bibitem [{\citenamefont {Rosenzweig}\ and\ \citenamefont
  {Porter}(1960)}]{RPmodel}%
  \BibitemOpen
  \bibfield  {author} {\bibinfo {author} {\bibfnamefont {Norbert}\ \bibnamefont
  {Rosenzweig}}\ and\ \bibinfo {author} {\bibfnamefont {Charles~E.}\
  \bibnamefont {Porter}},\ }\bibfield  {title} {\enquote {\bibinfo {title}
  {{``Repulsion of Energy Levels'' in Complex Atomic Spectra}},}\ }\href
  {\doibase 10.1103/PhysRev.120.1698} {\bibfield  {journal} {\bibinfo
  {journal} {Phys. Rev.}\ }\textbf {\bibinfo {volume} {120}},\ \bibinfo {pages}
  {1698--1714} (\bibinfo {year} {1960})}\BibitemShut {NoStop}%
\bibitem [{\citenamefont {Kravtsov}\ \emph {et~al.}(2015)\citenamefont
  {Kravtsov}, \citenamefont {Khaymovich}, \citenamefont {Cuevas},\ and\
  \citenamefont {Amini}}]{Amini:2015mvz}%
  \BibitemOpen
  \bibfield  {author} {\bibinfo {author} {\bibfnamefont {V.~E.}\ \bibnamefont
  {Kravtsov}}, \bibinfo {author} {\bibfnamefont {I.~M.}\ \bibnamefont
  {Khaymovich}}, \bibinfo {author} {\bibfnamefont {E.}~\bibnamefont {Cuevas}},
  \ and\ \bibinfo {author} {\bibfnamefont {M.}~\bibnamefont {Amini}},\
  }\bibfield  {title} {\enquote {\bibinfo {title} {{A random matrix model with
  localization and ergodic transitions}},}\ }\href {\doibase
  10.1088/1367-2630/17/12/122002} {\bibfield  {journal} {\bibinfo  {journal}
  {New J. Phys.}\ }\textbf {\bibinfo {volume} {17}},\ \bibinfo {pages} {122002}
  (\bibinfo {year} {2015})}\BibitemShut {NoStop}%
\bibitem [{\citenamefont {\v{C}ade\v{z}}\ \emph {et~al.}(2024)\citenamefont
  {\v{C}ade\v{z}}, \citenamefont {Kumar~Nandy}, \citenamefont {Rosa},
  \citenamefont {Andreanov},\ and\ \citenamefont {Dietz}}]{Cadez:2024mjc}%
  \BibitemOpen
  \bibfield  {author} {\bibinfo {author} {\bibfnamefont {Tilen}\ \bibnamefont
  {\v{C}ade\v{z}}}, \bibinfo {author} {\bibfnamefont {Dillip}\ \bibnamefont
  {Kumar~Nandy}}, \bibinfo {author} {\bibfnamefont {Dario}\ \bibnamefont
  {Rosa}}, \bibinfo {author} {\bibfnamefont {Alexei}\ \bibnamefont
  {Andreanov}}, \ and\ \bibinfo {author} {\bibfnamefont {Barbara}\ \bibnamefont
  {Dietz}},\ }\bibfield  {title} {\enquote {\bibinfo {title} {{The
  Rosenzweig\textendash{}Porter model revisited for the three
  Wigner\textendash{}Dyson symmetry classes}},}\ }\href {\doibase
  10.1088/1367-2630/ad5d86} {\bibfield  {journal} {\bibinfo  {journal} {New J.
  Phys.}\ }\textbf {\bibinfo {volume} {26}},\ \bibinfo {pages} {083018}
  (\bibinfo {year} {2024})}\BibitemShut {NoStop}%
\bibitem [{\citenamefont {Jahnke}\ \emph {et~al.}(2025)\citenamefont {Jahnke},
  \citenamefont {Nandy}, \citenamefont {Pal}, \citenamefont {Camargo},\ and\
  \citenamefont {Kim}}]{Jahnke:2025exd}%
  \BibitemOpen
  \bibfield  {author} {\bibinfo {author} {\bibfnamefont {Viktor}\ \bibnamefont
  {Jahnke}}, \bibinfo {author} {\bibfnamefont {Pratik}\ \bibnamefont {Nandy}},
  \bibinfo {author} {\bibfnamefont {Kuntal}\ \bibnamefont {Pal}}, \bibinfo
  {author} {\bibfnamefont {Hugo~A.}\ \bibnamefont {Camargo}}, \ and\ \bibinfo
  {author} {\bibfnamefont {Keun-Young}\ \bibnamefont {Kim}},\ }\bibfield
  {title} {\enquote {\bibinfo {title} {{Free probability approach to spectral
  and operator statistics in Rosenzweig-Porter random matrix ensembles}},}\
  }\href {\doibase 10.1007/JHEP12(2025)002} {\bibfield  {journal} {\bibinfo
  {journal} {JHEP}\ }\textbf {\bibinfo {volume} {12}},\ \bibinfo {pages} {002}
  (\bibinfo {year} {2025})}\BibitemShut {NoStop}%
\bibitem [{\citenamefont {Shenker}\ and\ \citenamefont
  {Stanford}(2014)}]{Shenker:2013yza}%
  \BibitemOpen
  \bibfield  {author} {\bibinfo {author} {\bibfnamefont {Stephen~H.}\
  \bibnamefont {Shenker}}\ and\ \bibinfo {author} {\bibfnamefont {Douglas}\
  \bibnamefont {Stanford}},\ }\bibfield  {title} {\enquote {\bibinfo {title}
  {Multiple shocks},}\ }\href {\doibase 10.1007/JHEP12(2014)046} {\bibfield
  {journal} {\bibinfo  {journal} {JHEP}\ }\textbf {\bibinfo {volume} {12}},\
  \bibinfo {pages} {046} (\bibinfo {year} {2014})}\BibitemShut {NoStop}%
\bibitem [{\citenamefont {Magan}\ \emph {et~al.}(2025)\citenamefont {Magan},
  \citenamefont {Sasieta},\ and\ \citenamefont {Swingle}}]{Magan:2024aet}%
  \BibitemOpen
  \bibfield  {author} {\bibinfo {author} {\bibfnamefont {Javier~M.}\
  \bibnamefont {Magan}}, \bibinfo {author} {\bibfnamefont {Martin}\
  \bibnamefont {Sasieta}}, \ and\ \bibinfo {author} {\bibfnamefont {Brian}\
  \bibnamefont {Swingle}},\ }\bibfield  {title} {\enquote {\bibinfo {title}
  {{Random circuits in the black hole interior}},}\ }\href {\doibase
  10.21468/SciPostPhys.19.1.007} {\bibfield  {journal} {\bibinfo  {journal}
  {SciPost Phys.}\ }\textbf {\bibinfo {volume} {19}},\ \bibinfo {pages} {007}
  (\bibinfo {year} {2025})}\BibitemShut {NoStop}%
\bibitem [{\citenamefont {Mag{\'a}n}\ \emph {et~al.}(2025)\citenamefont
  {Mag{\'a}n}, \citenamefont {Sasieta},\ and\ \citenamefont
  {Swingle}}]{Magan:2025hce}%
  \BibitemOpen
  \bibfield  {author} {\bibinfo {author} {\bibfnamefont {Javier~M.}\
  \bibnamefont {Mag{\'a}n}}, \bibinfo {author} {\bibfnamefont {Martin}\
  \bibnamefont {Sasieta}}, \ and\ \bibinfo {author} {\bibfnamefont {Brian}\
  \bibnamefont {Swingle}},\ }\bibfield  {title} {\enquote {\bibinfo {title}
  {{ER for typical EPR}},}\ }\href {\doibase 10.1103/btw6-44ry} {\bibfield
  {journal} {\bibinfo  {journal} {Phys. Rev. Lett.}\ }\textbf {\bibinfo
  {volume} {135}},\ \bibinfo {pages} {161601} (\bibinfo {year}
  {2025})}\BibitemShut {NoStop}%
\bibitem [{\citenamefont {Bavaro}\ \emph {et~al.}(2025)\citenamefont {Bavaro},
  \citenamefont {Magan},\ and\ \citenamefont {Martinek}}]{Bavaro:2025ooe}%
  \BibitemOpen
  \bibfield  {author} {\bibinfo {author} {\bibfnamefont {Enzo}\ \bibnamefont
  {Bavaro}}, \bibinfo {author} {\bibfnamefont {Javier~M.}\ \bibnamefont
  {Magan}}, \ and\ \bibinfo {author} {\bibfnamefont {Leandro}\ \bibnamefont
  {Martinek}},\ }\bibfield  {title} {\enquote {\bibinfo {title} {{Quantum
  microstate counting from Brownian motion: from many-body systems to black
  holes}},}\ }\href@noop {} {\  (\bibinfo {year} {2025})},\ \Eprint
  {http://arxiv.org/abs/2512.15854} {arXiv:2512.15854 [hep-th]} \BibitemShut
  {NoStop}%
\bibitem [{\citenamefont {Hou}\ \emph {et~al.}(2026)\citenamefont {Hou},
  \citenamefont {Hou},\ and\ \citenamefont {Yang}}]{Hou:2026cyf}%
  \BibitemOpen
  \bibfield  {author} {\bibinfo {author} {\bibfnamefont {Shengxian}\
  \bibnamefont {Hou}}, \bibinfo {author} {\bibfnamefont {Zong-Yue}\
  \bibnamefont {Hou}}, \ and\ \bibinfo {author} {\bibfnamefont {Zhi-Cheng}\
  \bibnamefont {Yang}},\ }\bibfield  {title} {\enquote {\bibinfo {title}
  {{State $k$-designs from Hamiltonian evolution}},}\ }\href@noop {} {\
  (\bibinfo {year} {2026})},\ \Eprint {http://arxiv.org/abs/2607.18537}
  {arXiv:2607.18537 [quant-ph]} \BibitemShut {NoStop}%
\bibitem [{\citenamefont {Yang}\ and\ \citenamefont {Wu}(2026)}]{Yang:2026opw}%
  \BibitemOpen
  \bibfield  {author} {\bibinfo {author} {\bibfnamefont {Zhongyi}\ \bibnamefont
  {Yang}}\ and\ \bibinfo {author} {\bibfnamefont {Biao}\ \bibnamefont {Wu}},\
  }\bibfield  {title} {\enquote {\bibinfo {title} {{Unitary designs from
  perturbed time evolutions of a chaotic Hamiltonian}},}\ }\href@noop {} {\
  (\bibinfo {year} {2026})},\ \Eprint {http://arxiv.org/abs/2607.24851}
  {arXiv:2607.24851 [quant-ph]} \BibitemShut {NoStop}%
\bibitem [{\citenamefont {Camargo}\ \emph {et~al.}(2025)\citenamefont
  {Camargo}, \citenamefont {Fu}, \citenamefont {Jahnke}, \citenamefont {Kim},\
  and\ \citenamefont {Pal}}]{Camargo:2025zxr}%
  \BibitemOpen
  \bibfield  {author} {\bibinfo {author} {\bibfnamefont {Hugo~A.}\ \bibnamefont
  {Camargo}}, \bibinfo {author} {\bibfnamefont {Yichao}\ \bibnamefont {Fu}},
  \bibinfo {author} {\bibfnamefont {Viktor}\ \bibnamefont {Jahnke}}, \bibinfo
  {author} {\bibfnamefont {Keun-Young}\ \bibnamefont {Kim}}, \ and\ \bibinfo
  {author} {\bibfnamefont {Kuntal}\ \bibnamefont {Pal}},\ }\bibfield  {title}
  {\enquote {\bibinfo {title} {{Quantum signatures of chaos from free
  probability}},}\ }\href {\doibase 10.1007/JHEP10(2025)138} {\bibfield
  {journal} {\bibinfo  {journal} {JHEP}\ }\textbf {\bibinfo {volume} {10}},\
  \bibinfo {pages} {138} (\bibinfo {year} {2025})}\BibitemShut {NoStop}%
\bibitem [{\citenamefont {Parker}\ \emph {et~al.}(2019)\citenamefont {Parker},
  \citenamefont {Cao}, \citenamefont {Avdoshkin}, \citenamefont {Scaffidi},\
  and\ \citenamefont {Altman}}]{Parker:2018yvk}%
  \BibitemOpen
  \bibfield  {author} {\bibinfo {author} {\bibfnamefont {Daniel~E.}\
  \bibnamefont {Parker}}, \bibinfo {author} {\bibfnamefont {Xiangyu}\
  \bibnamefont {Cao}}, \bibinfo {author} {\bibfnamefont {Alexander}\
  \bibnamefont {Avdoshkin}}, \bibinfo {author} {\bibfnamefont {Thomas}\
  \bibnamefont {Scaffidi}}, \ and\ \bibinfo {author} {\bibfnamefont {Ehud}\
  \bibnamefont {Altman}},\ }\bibfield  {title} {\enquote {\bibinfo {title} {{A
  Universal Operator Growth Hypothesis}},}\ }\href {\doibase
  10.1103/PhysRevX.9.041017} {\bibfield  {journal} {\bibinfo  {journal} {Phys.
  Rev. X}\ }\textbf {\bibinfo {volume} {9}},\ \bibinfo {pages} {041017}
  (\bibinfo {year} {2019})}\BibitemShut {NoStop}%
\end{thebibliography}%

\clearpage
\onecolumngrid            

\begin{center}
  \textbf{\large Supplemental Material:\\[2pt]
  Unitary $k$-designs without independent Hamiltonian quenches}\\[8pt]
  Pratik Nandy$^{1,2}$\\[5pt]
  {\itshape\small $^{1}$Theoretische Natuurkunde, Vrije Universiteit Brussel (VUB) and\\
  The International Solvay Institutes, Pleinlaan 2, B-1050 Brussels, Belgium}\\[3pt]
  {\itshape\small $^{2}$RIKEN Centre for Interdisciplinary Theoretical and Mathematical Sciences (iTHEMS), Wako, Saitama 351-0198, Japan}
\end{center}
\vspace{1em}

\setcounter{secnumdepth}{3}
\setcounter{section}{0}
\setcounter{equation}{0}
\setcounter{figure}{0}
\setcounter{table}{0}
\renewcommand{\thesection}{S\arabic{section}}
\renewcommand{\theequation}{S\arabic{equation}}
\renewcommand{\thefigure}{S\arabic{figure}}
\renewcommand{\thetable}{S\arabic{table}}

In this Supplemental Material (SM), we provide detailed derivations of the results presented in the main text and a comprehensive analysis of their properties. Unless otherwise specified, all analyses in the RMT are carried out for the Gaussian Unitary Ensemble (GUE) along with a few analytic cases in the Gaussian Orthogonal Ensemble (GOE). Section \ref{supp:MFIM} is solely dedicated to the Mixed Field Ising Model (MFIM).

\section{Matrix-element structure of the 2PK protocol}
\label{sec:matrix_elements}

In this section, we analyze the matrix-element structure underlying the 2PK
protocol. Our aim is to identify, at the level of exact operator identities and
standard random-matrix statistics, the mechanism by which a single fixed
chaotic Hamiltonian dressed by a fixed Pauli operator reproduces spectral
quantities that would otherwise be associated with statistically independent
Hamiltonians. Throughout, $H \equiv H_1$ is a fixed traceless GUE Hamiltonian
with eigenbasis $H\ket{E_n} = E_n\ket{E_n}$, and $P$ is a fixed non-identity Hermitian Pauli string, $P^2 = \mathds{1}$, $P^\dagger = P$. The second
effective Hamiltonian generated by the protocol is $H_2 = P H P$
(cf.\ Proposition 1 of the main text).

\subsection{Exact structure}

Because $P$ is both unitary and Hermitian, $H_2 = P H P = P H P^\dagger$ is a unitary conjugation of $H$. Two exact consequences follow. First, $H_2$ is
isospectral to $H$, with eigenstates $\ket{E_m^{(2)}} = P\ket{E_m}$,
\begin{align}
    H_2\big(P\ket{E_m}\big) = P H P P\ket{E_m} = P H \ket{E_m}
    = E_m \big(P\ket{E_m}\big)\,.
\end{align}
Second, since adjacent factors satisfy $P^2 = \mathds{1}$, every power
collapses to
\begin{align}
    H_2^{\alpha} = (PHP)^{\alpha} = P H^{\alpha} P \,,
    \label{eq:power_collapse}
\end{align}
and more generally $f(H_2) = P f(H) P$ for any function $f$, \emph{e.g.}, $e^{-iH_2 t} = P e^{-iH t} P$.

Using \eqref{eq:power_collapse} and inserting the resolution of identity
$\mathds{1} = \sum_m \ket{E_m}\bra{E_m}$, the diagonal matrix elements of
$H_2^{\alpha}$ in the $H$ eigenbasis are
\begin{align}
    \bra{E_n} H_2^{\alpha} \ket{E_n}
    = \bra{E_n} P H^{\alpha} P \ket{E_n}
    = \sum_{m} |P_{nm}|^2\, E_m^{\alpha}\,,
    \qquad P_{nm} := \bra{E_n} P \ket{E_m},
    \label{eq:matrix_element}
\end{align}
where we used $\bra{E_m}P\ket{E_n} = P_{mn} = P_{nm}^{*}$. The quantities $P_{nm}$ are
precisely the entries of $P_W = W^\dagger P W$ appearing in Eqs.\,\eqref{12PKnum0}-\eqref{12PKnum} of the main text, with $H = W D W^\dagger$. Equation \eqref{eq:matrix_element}
expresses the matrix element as a weighted average of the spectral powers
$E_m^{\alpha}$, with weights $|P_{nm}|^2$ which form a normalized probability
distribution over $m$ for each fixed $n$,
\begin{align}
    \sum_{m} |P_{nm}|^2 = \bra{E_n} P^2 \ket{E_n} = 1\,.
    \label{eq:weight_norm}
\end{align}
These statements are exact for the single pair $(H, P)$; no averaging over
any ensemble has been performed. 

\subsection{Exact spectral identity}

Averaging \eqref{eq:matrix_element} over the eigenstates, or equivalently using
cyclicity of the trace together with $P^2 = \mathds{1}$, gives the exact
identity
\begin{align}
    \frac{1}{d}\sum_{n} \bra{E_n} H_2^{\alpha} \ket{E_n}
    = \frac{1}{d}\,\mathrm{Tr}\big(H_2^{\alpha}\big)
    = \frac{1}{d}\,\mathrm{Tr}\big(P H^{\alpha} P\big)
    = \frac{1}{d}\,\mathrm{Tr}\big(H^{\alpha}\big)\,.
    \label{eq:trace_identity}
\end{align}
This holds for \emph{any} fixed $(H,P)$, independently of whether $H$ is
chaotic: the Pauli conjugation leaves every spectral moment invariant. For the
traceless GUE, whose spectral density approaches the semicircle
$\rho(E) = \tfrac{1}{2\pi}\sqrt{4-E^2}$ on $[-2,2]$, the moments are, in the
large-$d$ limit,
\begin{align}
    \frac{1}{d}\,\mathrm{Tr}\big(H^{\alpha}\big)
    \;\longrightarrow\; \int_{-2}^{2} E^{\alpha}\,\rho(E)\, dE
    = \begin{cases}
        0, & \alpha \text{ odd}, \\[4pt]
        C_{\alpha/2}, & \alpha \text{ even},
    \end{cases}
    \label{eq:semicircle_moments}
\end{align}
with $C_p = \tfrac{1}{p+1}\binom{2p}{p}$ the Catalan numbers. Equation \eqref{eq:trace_identity} is exact at finite $d$; \eqref{eq:semicircle_moments} is its large-$d$ value up to the usual finite-size
corrections to the spectral density.

\subsection{Delocalization of the Pauli operator}

The distribution of the weights $|P_{nm}|^2$ in \eqref{eq:matrix_element}
depends on the eigenvectors of $H$. For a GUE Hamiltonian, the eigenvector
matrix $W$ is Haar distributed, so $P_W = W^\dagger P W$ has the statistics of
a fixed operator expressed in a random basis. The total weight is fixed exactly:
\begin{align}
    \sum_{n,m} |P_{nm}|^2 = \mathrm{Tr}\big(P^\dagger P\big) = d\,,~~ \text{so the mean weight is } \; \overline{|P_{nm}|^2} = \frac{1}{d}\,.
    \label{eq:mean_weight}
\end{align}
Moreover, under Haar averaging the individual weights concentrate about this
mean: the Pauli operator is delocalized over the energy basis rather than
supported on a few matrix elements. This does not imply that each weight equals
$1/d$ exactly; it states only that the weights are spread across the spectrum
with typical magnitude $1/d$.

\begin{figure}[h]
    \hspace{-0.5cm}
    \includegraphics[width=1\linewidth]{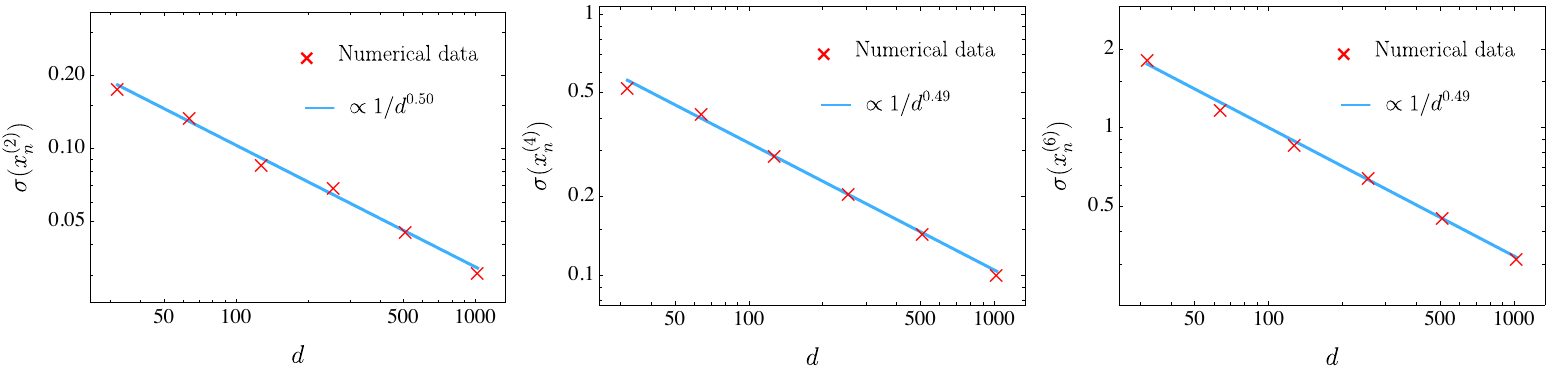}
    \caption{Scaling of the standard deviation of $x_n^{(\alpha)} = \bra{E_n} H_2^{\alpha} \ket{E_n}$, with the dimension $d$ for $\alpha = 2,4,6$. In these three cases, the standard deviation scales as $\sigma(x_n^{(\alpha)}) \propto d^{-\gamma}$, where  $\gamma \approx 1/2$. We consider a fixed pair of $(H,P)$.}
    \label{fig:EigOvpGUE}
\end{figure}

Rearranging $|P_{nm}|^2 = (1/d) + (|P_{nm}|^2 - 1/d)$, and noticing $\sum_m E_m^{\alpha} = \mathrm{Tr}(H^{\alpha})$, Eq.\,\eqref{eq:matrix_element} can be decomposed as
\begin{align}
\bra{E_n} H_2^{\alpha} \ket{E_n}
= \frac{1}{d}\,\mathrm{Tr}\big(H^{\alpha}\big)
      + \sum_{m}\bigg(|P_{nm}|^2 - \tfrac{1}{d}\bigg) E_m^{\alpha} \,,
    \label{eq:fluctuation}
\end{align}
where the first term is the exact spectral moment shared by all eigenstates
[Eq.\,\eqref{eq:trace_identity}] and the second term encodes eigenstate-to-eigenstate fluctuations. Because the weights are delocalized, these fluctuations are suppressed as the Hilbert-space dimension grows, so that for typical eigenstates of a chaotic realization
\begin{align}
    x_n^{(\alpha)} := \bra{E_n} H_2^{\alpha} \ket{E_n}
    \;\approx\; \frac{1}{d}\,\mathrm{Tr}\big(H^{\alpha}\big) \,,
    \label{eq:typicality}
\end{align}
up to finite-size corrections. In Fig.\,\ref{fig:EigOvpGUE}, we numerically demonstrate this suppression by computing the standard deviation $\sigma(x_n^{(\alpha)})$ for $\alpha = 2,4,6$. We observe that the eigenstate-to-eigenstate spread of $\bra{E_n} H_2^{\alpha} \ket{E_n}$ decreases systematically with increasing Hilbert-space dimension $d$. For the cases considered, the numerical fit yields $\sigma(x_n^{(2)}) \propto d^{-\gamma}$ with $\gamma \approx 1/2$, in excellent agreement with the expected $d^{-1/2}$ scaling for GUE matrices. Since this scaling is inferred from finite-size numerics, the fitted exponent for an individual Hamiltonian realization may exhibit small fluctuations around $\gamma = 1/2$. However, by the typicality of GUE ensembles, averaging over a sufficiently large collection of independent Hamiltonian realizations, where each defines a distinct 2PK protocol, is expected to yield a mean exponent converging to $\gamma = 1/2$. Therefore, \eqref{eq:typicality} suggests that in the large $d$ limit, typical eigenstates of $H$ effectively give the same expectation value of $P H^{\alpha} P$, namely the global spectral average $\mathrm{Tr}(H^{\alpha})/d$. In other words, the Pauli kick reproduces, for the leading-order frame potential, the delocalized weight structure \eqref{eq:mean_weight} that an independently drawn Hamiltonian would produce, though the two constructions remain distinct at the level of the spectrum, as we discuss next.

\subsection{Comparison with independent Hamiltonians}

It is instructive to contrast this structure with a protocol built from two
statistically independent chaotic Hamiltonians $H_1, H_2$, as in
Refs.\,\cite{Zhou:2026ubz, Sun:2026wrk}. There, $H_1$ and $H_2$ have independent spectra and independent eigenbases, no fixed operator relates them, and the collapse \eqref{eq:power_collapse} does not apply. The analogue of \eqref{eq:matrix_element} instead involves the overlap of two unrelated
eigenbases,
\begin{align}
    \bra{E_n^{(1)}} H_2^{\alpha} \ket{E_n^{(1)}}
    = \sum_{m} |O_{nm}|^2 \, \big(E_m^{(2)}\big)^{\alpha} ,
    \qquad O_{nm} := \langle E_n^{(1)}|E_m^{(2)}\rangle \,,
    \label{eq:overlap_element}
\end{align}
with a \emph{different} spectrum $\{E_m^{(2)}\}$ and overlap weights
$|O_{nm}|^2$ that likewise satisfy $\sum_m |O_{nm}|^2 = 1$ and, for Haar-random bases, $\overline{|O_{nm}|^2} = 1/d$.

The two constructions therefore share the delocalization property that
underlies \eqref{eq:typicality}, but differ at the level of the spectral
identity. In the 2PK case, Eq.\,\eqref{eq:trace_identity} equates the spectral moments of $H_1$ and $H_2$ \emph{exactly} for the given pair, as an operator identity. For independent Hamiltonians, $d^{-1}\mathrm{Tr}(H_1^{\alpha})$ and
$d^{-1}\mathrm{Tr}(H_2^{\alpha})$ are distinct quantities that coincide only in
distribution, up to sample-to-sample differences. The 2PK protocol thus attains a shared spectrum by construction, rather than by statistical coincidence, while retaining through the delocalization of a generic fixed $P$ in a chaotic eigenbasis—the weight structure that governs the independent case. The reshuffling of a single fixed spectrum by the Pauli operator, rather than any spectral randomness in the independent case, is consistent with the numerical insensitivity (for GUE) to the choice of $P$ established in Sec.\,\ref{sec:pauli_independence}.

\section{Exact first-order frame potential}
\label{sec:k1exact}
 
The design order of the $2$PK protocol is established numerically in the
main text.  Here we show that the \emph{first-order} ($k=1$) frame
potential admits a closed form that is \emph{exact} for any Hamiltonian
with non-degenerate spectrum and any Pauli kick, and that this closed
form makes the mechanism transparent: a single kick fails and two kicks
succeed for a reason that involves only the delocalization of $P$ in the
energy eigenbasis, not the spectrum.
 
We work in the eigenbasis $H\ket{E_a}=E_a\ket{E_a}$ (assumed
non-degenerate, as holds generically for the GUE and SYK models considered) and use the matrix elements and overlap weights
\begin{align}
P_{ab}:=\bra{E_a}P\ket{E_b},
\qquad
w_{ab}:=|P_{ab}|^2\,.
\label{eq:wdef}
\end{align}
As already noted in Sec.\,\ref{sec:matrix_elements}, $w$ inherits two exact properties from the
Pauli kick.  Hermiticity of $P$ gives symmetry, $w_{ab}=w_{ba}$, and
$P^2=\mathds{I}$ gives, via \eqref{eq:weight_norm},
\begin{align}
\sum_b w_{ab}=\bra{E_a}P^2\ket{E_a}=1\,.
\label{eq:doublystochastic}
\end{align}
Thus $w$ is a real symmetric \emph{doubly stochastic} matrix.  Its
eigenvalues $\{\mu_j\}$ are real with a Perron value
\begin{align}
\mu_1=1
\quad(\text{eigenvector}\propto(1,\dots,1)^{\intercal}),
\qquad
|\mu_j|\le 1 \,.
\label{eq:perron}
\end{align}
 
\subsection{Time average}
 
Denoting $U_{2\mathrm{PK}}= e^{-i H t_3}P\, e^{- i H t_2}P\, e^{- i H t_1}$ in the eigenbasis gives the matrix element
\begin{align}
U^{2\mathrm{PK}}_{ab}(t_1,t_2,t_3)
:= \langle E_a|U^{2\mathrm{PK}}|E_b \rangle = e^{- i(E_a t_3+E_b t_1)} \langle E_a| P e^{-i H t_2} P |E_b \rangle  =  e^{- i(E_a t_3+E_b t_1)}\sum_{m}P_{am}P_{mb}\, e^{-i E_m t_2}\,,
\label{eq:Uelement}
\end{align}
in which the two outer times ($t_1$ and $t_3$) attach to the outer labels $a,b$ while the single middle time $t_2$ attaches to the summed intermediate label $m$.  Let $V$ be a second, independent element of the temporal ensemble with times $t'_i$. Using $\mathrm{Tr}(U^\dagger V)=\sum_{a,b} U^{*}_{ab}\,V_{ab}$, we find
\begin{align}
z:=\mathrm{Tr} \big(U^\dagger V\big)
=\sum_{a,b,m,n}
P^{*}_{am}P^{*}_{mb}\,P_{an}P_{nb}\;
e^{i E_a(t_3-t'_3)}\,
e^{i E_b(t_1-t'_1)}\,
e^{i E_m t_2}\,
e^{-i E_n t'_2}\,.
\label{eq:Zexpansion}
\end{align}
The two middle times $t_2$ (from $U$) and $t'_2$ (from $V$) enter
\emph{separately} rather than as a difference; this is the structural
signature of the two kicks and the origin of the two intermediate labels
$m,n$.
 
The first-order frame potential is
$F^{(1)}_{2\mathrm{PK}}=\overline{|z|^2}$, the overline denoting the average over the six times $\{t_i, t'_i\}$ for $i = 1,2,3$, each uniform on $[0,T]$ with $T\to\infty$, so that $\overline{e^{i\omega t}}=\delta_{\omega,0}$. Writing the
conjugate $\overline{z}$ with hat labels $(\hat{a}, \hat{b}, \hat{m}, \hat{n})$ and averaging each time independently forces every phase to vanish separately.  With a non-degenerate spectrum, this yields the \emph{exact} constraints
\begin{align}
\begin{split}
  &t_3,t'_3:\ a=\hat{a};\quad
t_1,t'_1:\ b=\hat{b}; \\
&t_2:\ m=\hat{m};\quad
~~t'_2:\ n=\hat{n}\,.  
\end{split}
\end{align}
Here we have not made any approximation: at $k=1$ each constraint is a single energy \emph{difference}, so no accidental sum-resonances can occur. Substituting and using \eqref{eq:wdef}, we get
\begin{align}
F^{(1)}_{2\mathrm{PK}}
=\sum_{a,b,m,n}w_{am}\,w_{mb}\,w_{an}\,w_{nb}\,.
\end{align}
Performing the $m$ and $n$ sums, $\sum_m w_{am}w_{mb}=(w^2)_{ab}$, gives
the closed form
\begin{align}
F^{(1)}_{2\mathrm{PK}}
=\sum_{a,b}\big[(w^2)_{ab}\big]^2
=\mathrm{Tr}\big(w^4\big)
=\sum_j\mu_j^4\,,
\label{eq:F1_2PK}
\end{align}
valid for \emph{any} $H$ (integrable or chaotic) with non-degenerate spectrum and \emph{any} $P$;
no eigenvector average or ensemble has been used.
 
The identical computation for the one-kick protocol
$U_{1\mathrm{PK}}=e^{-i H t_2}P\,e^{-i H t_1}$, for which $U^{1\mathrm{PK}}_{ab}=e^{-i(E_a t_2+E_b t_1)}P_{ab}$, and for the general $n$-kick protocol of the main text gives
\begin{align}
F^{(1)}_{1\mathrm{PK}}=\mathrm{Tr}\big(w^2\big)\,,~~~
F^{(1)}_{n\mathrm{PK}}=\mathrm{Tr}\big(w^{2n}\big)=\sum_j\mu_j^{2n}\,,
\label{eq:F1_nPK}
\end{align}
consistent with the limiting cases $n=0$ ($\mathrm{Tr}(w^0)=d$, which is the zero-Pauli kick), $n=1$, and $n=2$.
 
\subsection{1PK vs nPK: convergence towards Haar}
 
Because $\mu_1=1$ by Eq.\,\eqref{eq:perron} and the Haar value is $F^{(1)}_{\mathcal{E}_{\mathrm{Haar}}}=1!=1$, Eq.~\eqref{eq:F1_nPK} separates cleanly
into the Haar value plus an excess,
\begin{align}
F^{(1)}_{n\mathrm{PK}}=1+\underbrace{\sum_{j\geq 2}\mu_j^{2n}}_{\text{excess}} .
\label{eq:excess}
\end{align}
Chaos enters only through the delocalization of $P$ over the energy
eigenbasis. With the eigenvector statistics of Sec.\,\ref{sec:matrix_elements}, the mean weight is $\overline{|P_{ab}|^2}=1/d$ [Eq.\,\eqref{eq:mean_weight}] and, for eigenstate-thermalizing matrix elements, $\overline{|P_{ab}|^4}\simeq
2/d^2$ (for GUE), so that
\begin{align}
F^{(1)}_{1\mathrm{PK}}=\mathrm{Tr}\big(w^2\big)=\sum_{a,b}|P_{ab}|^4\;\xrightarrow{\ d\to\infty\ }\;2 \,.
\label{eq:Trw2}
\end{align}
Since $\mathrm{Tr}(w^2)=1+\sum_{j \geq 2}\mu_j^2$, the subleading eigenvalues carry total squared weight $\sum_{j \geq 2}\mu_j^2 \to 1$, spread over $\sim d$ modes; hence a typical subleading eigenvalue scales as $\mu_j \sim d^{-1/2}$.  It follows that, for GUE,
\begin{align}
F^{(1)}_{1\mathrm{PK}}\to 2\,,~~~
F^{(1)}_{2\mathrm{PK}}=1+\sum_{j\geq 2}\mu_j^4=1+O(1/d)\,, \label{anaj}
\end{align}
and more generally $F^{(1)}_{n\mathrm{PK}}=1+O(1/d)$ for all $n\ge2$ because $\sum_{j\geq 2}\mu_j^{2n}\le\sum_{j\geq 2}\mu_j^4$ when $|\mu_j|\le1$ [Eq.\,\eqref{eq:perron}]. A single kick therefore leaves an $O(1)$ excess and cannot form a $1$-design, whereas two kicks reduce the excess to $O(1/d)$ and recover the Haar value, in agreement with the numerics of Fig.\,\ref{fig:FP12PKGUEplot}. Note that for GOE, $\overline{|P_{ab}|^4}\simeq
3/d^2$, resulting in $F^{(1)}_{1\mathrm{PK}} \rightarrow 3$, consistent with Eq.\,\eqref{dearr} with $\Lambda = 3$ as well (see \eqref{eq:S3-kurtosis} for the definition). Consequently, the sum is $\sum_{j \geq 2}\mu_j^2 \to 2$ and the scaling of $\mu_j$ is still $\mu_j \sim d^{-1/2}$ with some constant numerical factor. Hence, the convergence $F^{(1)}_{n\mathrm{PK}}=1+O(1/d)$ for $n \geq 2$ holds for GOE as well.

We emphasize two important features of Eqs.\,\eqref{eq:F1_2PK}--\eqref{eq:excess}. First, the first-order frame potential is a functional of
the overlap matrix $w$ \emph{alone}: the spectrum $\{E_a\}$ has cancelled
identically. The isospectrality of $H$ and $H_2=PHP$ (cf.\ Proposition 1) is therefore immaterial at $k=1$; the nontrivial content of the design property resides at $k\ge2$, where spectral correlations can enter (see next section \eqref{sec:S3}). Second, the second kick acts precisely by promoting $w^2\mapsto w^4$, converting the design-spoiling factor $\mathrm{Tr}(w^2)\to2$ into the benign $\mathrm{Tr}(w^4)\to1$. The convergence rate toward the Haar value is governed by the spectral gap $1-\mu_2$ of the doubly stochastic matrix $w$; chaotic systems exhibit a large gap due to eigenvector delocalization, whereas the gap is expected to shrink as the system approaches integrability. A systematic investigation is left for future work.

\section{Frame potential for $k\ge 2$ and the one-kick obstruction}
\label{sec:S3}

The first-order frame potential of Sec.\,\ref{sec:k1exact} is exact. Here we
establish Eq.\,\eqref{f2pk} of the main text for all $k\ge2$ as a controlled
expansion in $1/d$ at fixed $k$, and, within the same computation, derive
the one-kick result \eqref{dearr} in closed form. The strategy is to perform
the time average and the eigenvector average exactly at the level of the
combinatorics, so that the Haar value $k!$ emerges as a counting of
surviving index pairings. We treat both protocols together, since they differ only in the kernel
Eq.\,\eqref{eq:S3-kernels} below.
 
Two properties of a chaotic Hamiltonian are used.
 
\begin{itemize}
\item[\textbf{(A1)}] \emph{Non-resonance at order $k$.} For index
multisets $\{a_r\}_{r=1}^k$ and $\{\hat{a}_r\}_{r=1}^k$, the equality
$\sum_r E_{a_r}=\sum_r E_{\hat{a}_r}$ holds only if the two multisets
coincide. This fails only on a measure-zero set of spectra, hence holds with probability one in the random-matrix ensembles and is numerically satisfied within symmetry-resolved SYK sectors, and reduces to non-degeneracy at $k=1$.
 
\item[\textbf{(A2)}] \emph{Delocalization.} The matrix elements
$P_{ab}=\bra{E_a}P\ket{E_b}$ obey the statistics established in Sec.\,\ref{sec:k1exact}: they are centered and asymptotically independent at leading order in $1/d$ up to the Hermiticity
constraint $P_{ba}=P_{ab}^{*}$, with
$\overline{|P_{ab}|^{2}}=1/d$ [Eq.\,\eqref{eq:mean_weight}] and higher joint moments given by Wick contractions up to relative $O(1/d)$. In particular, the fourth moment is fixed by the eigenvector kurtosis
\begin{align}
\overline{|P_{ab}|^{4}}=\frac{\Lambda}{d^{2}},
\qquad
\Lambda:=\frac{\overline{|P_{ab}|^{4}}}{\bigl(\,\overline{|P_{ab}|^{2}}\,\bigr)^{2}}
=\begin{cases}
2\,, & \text{GUE }(\beta_{\mathrm{Dyson}}=2)\,,\\[2pt]
3\,, & \text{GOE }(\beta_{\mathrm{Dyson}}=1)\,,
\end{cases}
\label{eq:S3-kurtosis}
\end{align}
\emph{i.e.}, $\Lambda$ is the kurtosis of a complex ($\beta_{\mathrm{Dyson}}=2$) or real
($\beta_{\mathrm{Dyson}}=1$) Gaussian variable, as introduced in Sec.\,\ref{sec:k1exact}.
\end{itemize}
 
\noindent
Assumption (A2) is an eigenvector-thermalization ansatz. It is controlled for the GXE, where the eigenvector matrix $W$ is Haar-distributed by construction; for the SYK and MFIM models, it is an assumption, supported \emph{a posteriori} by the numerics in the main text. Throughout we write $w_{ab}=|P_{ab}|^{2}$ for the doubly stochastic overlap matrix of Sec.\,\ref{sec:k1exact}, and
$\Delta t_j=t_j-t'_j$.

\subsection{Regrouping the overlap: the two kernels}
 
From the matrix element \eqref{eq:Uelement}, it is convenient to isolate the
intermediate factor
\begin{align}
R_{ab}(t):=\bra{E_a}P\,e^{-iHt}P\ket{E_b}
=\sum_m P_{am}P_{mb}\,e^{-iE_m t}\,,
\label{eq:S3-R}
\end{align}
so that $U^{\rm 2PK}_{ab}=e^{-i(E_a t_3+E_b t_1)}R_{ab}(t_2)$. Let $V$ be
a second, independent element of the temporal ensemble with times $t'_i$.
Using $\mathrm{Tr}(U^{\dagger}V)=\sum_{a,b}U^{*}_{ab}V_{ab}$ and collecting the
two outer phases into
\begin{align}
\zeta_a:=e^{iE_a\Delta t_3},
\qquad
\eta_b:=e^{iE_b\Delta t_1}\,,
\end{align}
which are independent and uniform on the unit circle as $T\to\infty$ for
a non-degenerate spectrum, the overlap collapses to the bilinear form
\begin{align}
z:=\mathrm{Tr}\bigl(U^{\dagger}V\bigr)=\sum_{a,b}C_{ab}\,\zeta_a\,\eta_b
=\zeta^{\mathsf T}C\,\eta \,.
\end{align}
The whole distinction between the two protocols resides in the kernel:
\begin{align}
C^{\rm 2PK}_{ab}=R^{*}_{ab}(t_2)\,R_{ab}(t'_2)\,,
\qquad\qquad
C^{\rm 1PK}_{ab}=P^{*}_{ab}P_{ab}=w_{ab}\,,
\label{eq:S3-kernels}
\end{align}
with $\zeta_a=e^{iE_a\Delta t_2}$ in the one-kick case. The most
important structural feature, already visible below \eqref{eq:Zexpansion}, is that the middle times $t_2$ and $t'_2$ belong to $U$ and $V$ \emph{separately} and therefore enter $C^{\rm 2PK}$ separately rather than as a difference; the two-kick kernel is consequently a mean-zero random-phase object. For one kick, the intermediate sum is absent, and the kernel is real, non-negative and frozen. As we now show, this is a distinction between coherent and incoherent index sums, and it is the entire content of Eqs.\,\eqref{f2pk} and \eqref{dearr}.

\subsection{Exact reduction of the time average}
 
Raising $z$ to the $2k$-th power,
\begin{align}
|z|^{2k}=\sum_{\{a_r,b_r\}}\sum_{\{\hat{a}_r,\hat{b}_r\}}
\prod_{r=1}^{k}C_{a_rb_r}C^{*}_{\hat{a}_r\hat{b}_r}\,
e^{\,i\Delta t_3\sum_r(E_{a_r}-E_{\hat{a}_r})}\,
e^{\,i\Delta t_1\sum_r(E_{b_r}-E_{\hat{b}_r})}\,.
\label{eq:S3-z2k}
\end{align}
The outer times are sampled independently, so averaging over them
imposes the two sum rules $\sum_r E_{a_r}=\sum_r E_{\hat{a}_r}$ and
$\sum_r E_{b_r}=\sum_r E_{\hat{b}_r}$. By (A1), the hatted tuples must then be
permutations of the unhatted ones,
\begin{align}
\hat{a}_r=a_{\sigma(r)},\qquad \hat{b}_r=b_{\tau(r)},\qquad \sigma,\tau\in S_k\,,
\label{eq:S3-perms}
\end{align}
uniquely so whenever the entries are pairwise distinct; configurations
with repeated entries constitute a fraction $O(k^{2}/d)$ of the total and
are absorbed into the error term below. Because $t_2$ and $t'_2$ enter
$C^{\rm 2PK}$ separately, the remaining average factorizes into a pair of
complex conjugates, and one obtains the  manifestly positive
representation
\begin{align}
F^{(k)}_{\rm 2PK}
&=\sum_{\sigma,\tau\in S_k}\;{\sum_{\{a_r,b_r\}}}'\;
\bigl|\mathcal A_{\sigma\tau}(a,b)\bigr|^{2}\,
\bigl[1+O(k^{2}/d)\bigr],
\label{eq:S3-master}\\[4pt]
\mathcal A_{\sigma\tau}(a,b)
&:=\overline{\prod_{r=1}^{k}R^{*}_{a_rb_r}(t)\,
R_{a_{\sigma(r)}b_{\tau(r)}}(t)}\,,
\label{eq:S3-A}
\end{align}
the prime denoting a sum over distinct entries and the overline in \eqref{eq:S3-A} the average over the single remaining time. No eigenvector average has been used so far. At $k=1$ only $\sigma=\tau=\mathds{1}$ occurs and $\mathcal A=\overline{|R_{ab}(t)|^{2}} =(w^{2})_{ab}$, so that \eqref{eq:S3-master} reduces to $\mathrm{Tr}(w^{4})$, reproducing the exact result \eqref{eq:F1_2PK}.

\subsection{Collapse of the intermediate sums}
 
Inserting \eqref{eq:S3-R} into \eqref{eq:S3-A} and averaging over $t$,
assumption (A1) once more forces a permutation $\pi\in S_k$ of the
intermediate labels, giving
\begin{align}
\mathcal A_{\sigma\tau}(a,b)=\sum_{\pi\in S_k}\;{\sum_{\{m_r\}}}'\;
\prod_{r=1}^{k}
P^{*}_{a_rm_r}P^{*}_{m_rb_r}
P_{a_{\sigma(r)}m_{\pi(r)}}P_{m_{\pi(r)}b_{\tau(r)}}\,.
\label{eq:S3-Apc}
\end{align}
As before, repeated intermediate labels contribute only an $O(k^2/d)$ fraction and are omitted from the primed sum; \eqref{eq:S3-Apc} is therefore exact up to this relative correction.

By (A2), each $\pi$-term is a sum of $\sim d^{k}$ contributions of
typical modulus $d^{-2k}$ carrying independent phases. Such a sum
survives coherently, at $O(d^{-k})$, only if all $4k$ matrix elements of the kick operator pair into moduli $|P|^{2}$. Pairing the $a$-type factors $P^{*}_{a_rm_r}$ and
$P_{a_{\sigma(u)}m_{\pi(u)}}$ requires $\sigma(u)=r$ and $\pi(u)=r$
simultaneously, hence $\sigma=\pi$; pairing the $b$-type factors
$P^{*}_{m_rb_r}$ and $P_{m_{\pi(u)}b_{\tau(u)}}$ requires $\pi(u)=r$ and $\tau(u)=r$, hence $\tau=\pi$. (Contractions that identify an $a$- or $b$-label with an $m$-label are possible only on a codimension-one
subset of index configurations and are $O(1/d)$ suppressed.) A coherent
contraction therefore exists if and only if
\begin{align}
\sigma=\tau=\pi\,,
\end{align}
in which case, relabeling $m_{\sigma(r)}\to m_r$ in the two unstarred products,
\begin{align}
\mathcal A_{\sigma\sigma}(a,b)
={\sum_{\{m_r\}}}'\prod_{r=1}^{k}w_{a_rm_r}w_{m_rb_r}
=\prod_{r=1}^{k}(w^{2})_{a_rb_r}\bigl[1+O(k^{2}/d)\bigr],
\label{eq:S3-diag}
\end{align}
independently of $\sigma$.
 
For $\sigma\neq\tau$, take $\pi=\sigma$ (the choice that maximizes
coherence; any other $\pi$ is more strongly suppressed) and set
\begin{align}
\nu := k-f(\tau\sigma^{-1}) \ge 2 \,,
\end{align}
where $f(\cdot)$ denotes the number of fixed points of a permutation; the bound holds because no permutation possesses exactly $k-1$ fixed points. Relabeling $s=\sigma(r)$, the $b$-type product becomes
$\prod_s P^{*}_{m_sb_s} P_{m_sb_{\tau\sigma^{-1}(s)}}$, in which exactly $\nu$ factors are mismatched and hence carry a phase that is random in $m_s$. Each mismatched index is summed incoherently and the corresponding sum is suppressed by the usual random-phase (RMT/ETH) estimate $d^{1/2}$ rather than the coherent enhancement $d$, so that $|\mathcal A_{\sigma\tau}|=O(d^{-k-\nu/2})$ and, summing over the $\sim d^{2k}$ configurations of the outer labels,
\begin{align}
{\sum_{\{a_r,b_r\}}}'\bigl|\mathcal A_{\sigma\tau}\bigr|^{2}
=O(d^{-\nu})=O(1/d^{2}),\qquad \sigma\neq\tau \,.
\label{eq:S3-offdiag}
\end{align}

\subsection{The two-kick result}
 
Inserting \eqref{eq:S3-diag} and \eqref{eq:S3-offdiag} into
\eqref{eq:S3-master}, the $(k!)^{2}$ terms of the double permutation sum
collapse to the $k!$ diagonal ones $\sigma=\tau$, each contributing the
same factorized value
${\sum}'_{a,b}\prod_r[(w^{2})_{a_rb_r}]^{2}=[\mathrm{Tr}(w^{4})]^{k}
[1+O(k^{2}/d)]$. With $\mathrm{Tr}(w^{4})=1+\sum_{j\ge2}\mu_j^{4}=1+O(1/d)$ from Sec.\,\ref{sec:k1exact},
\begin{align}
F^{(k)}_{\mathrm{2PK}}
=k!\,\bigl[\mathrm{Tr}(w^{4})\bigr]^{k}\bigl[1+O(1/d)\bigr]
=k!\,\bigl[1+O(1/d)\bigr]\,,
\label{eq:S3-2PKresult}
\end{align}
which is Eq.\,\eqref{f2pk} of the main text. The Haar value is thus \emph{counted}: $k!$ is the number of diagonal pairs $(\sigma,\sigma)$ in
\eqref{eq:S3-master}, every other pairing being suppressed by $1/d^{2}$
through \eqref{eq:S3-offdiag}.
 
The same mechanism suggests the extension, with $w^{2}\to w^{n}$ in \eqref{eq:S3-diag}:
\begin{align}
F^{(k)}_{n\rm PK}
=k!\,\bigl[\mathrm{Tr}(w^{2n})\bigr]^{k}\bigl[1+O(1/d)\bigr],
\qquad n\ge 2,
\label{eq:S3-nPK}
\end{align}
consistent at $k=1$ with the exact result \eqref{eq:F1_nPK}. Since $|\mu_j|\le1$, one has $\mathrm{Tr}(w^{2n})\le\mathrm{Tr}(w^{4})=1+O(1/d)$ for all $n\ge2$, so every $n\ge2$ forms an approximate $k$-design under the same delocalization assumptions. We stress that \eqref{eq:S3-nPK} does \emph{not} extend to $n=1$: substituting
$n=1$ would give $k!\,\Lambda^{k}$, whereas the correct answer, derived next, is $k!\sum_{\rho\in S_k}\Lambda^{f(\rho)}$. The failure of the substitution is precisely the failure of factorization, and it is the subject of the following subsection.

\subsection{The one-kick obstruction: closed form and ensemble dependence}
\label{sec:S3-1PK}

For one kick the kernel is $C^{\rm 1PK}_{ab}=w_{ab}$ by
\eqref{eq:S3-kernels}: it carries no intermediate time, so the step
\eqref{eq:S3-Apc} does not occur and no phase cancellation is available.
Equation \eqref{eq:S3-master} is replaced by
\begin{align}
F^{(k)}_{\rm 1PK}
=\sum_{\sigma,\tau\in S_k}\;{\sum_{\{a_r,b_r\}}}'\;
\prod_{r=1}^{k}w_{a_rb_r}\,w_{a_{\sigma(r)}b_{\tau(r)}} .
\label{eq:S3-1PKmaster}
\end{align}
Since the index sums are unrestricted apart from distinctness, the
summand depends on $\sigma$ and $\tau$ only through
$\rho:=\tau\sigma^{-1}$; relabeling $s=\sigma(r)$ therefore yields an
overall factor $k!$ and reduces \eqref{eq:S3-1PKmaster} to
\begin{align}
F^{(k)}_{\rm 1PK}
=k!\sum_{\rho\in S_k}\;{\sum_{\{a_s,b_s\}}}'\;
\prod_{s=1}^{k}w_{a_sb_s}\,w_{a_sb_{\rho(s)}}\,.
\label{eq:S3-1PKrho}
\end{align}
The ensemble average now factorizes over the cycles of $\rho$:
 
\begin{itemize}
\item A \emph{fixed point} $\rho(s)=s$ collapses the two factors to
$|P_{a_sb_s}|^{4}$; summing over $a_s,b_s$ and using
\eqref{eq:S3-kurtosis},
\begin{align}
\sum_{a,b}\overline{|P_{ab}|^{4}}\simeq d^{2}\cdot\frac{\Lambda}{d^{2}}
=\Lambda \,.
\end{align}
 
\item A \emph{cycle of length $\ell\ge2$} produces a closed loop of
$2\ell$ distinct weights $w=|P|^{2}$, each averaging to $1/d$ by (A2);
with $\sim d^{2\ell}$ index configurations the contribution is
$d^{2\ell}\cdot d^{-2\ell}=1$, independently of the ensemble.
\end{itemize}
 
\noindent
Hence each $\rho$ contributes $\Lambda^{f(\rho)}$, where $f(\rho)$ is its
number of fixed points. The ensemble dependence of the one-kick frame
potential is therefore carried, at leading order in $1/d$, by the single
parameter $\Lambda$: the spectral correlations have dropped out entirely,
exactly as they did at $k=1$. Since the number of permutations of $S_k$
with exactly $j$ fixed points is $\binom{k}{j}\,!(k-j)$, we obtain 
\begin{align}
F^{(k)}_{\rm 1PK}
=k!\sum_{j=0}^{k}\binom{k}{j}\,!(k-j)\,\Lambda^{j}\,,
\label{eq:S3-eq6}
\end{align}
which is Eq.\,\eqref{dearr} of the main text.
 
\paragraph{Closed form.}
We reorganize \eqref{eq:S3-eq6} as a binomial convolution of the
derangement numbers (exponential generating function $e^{-x}/(1-x)$) with
$\{\Lambda^{j}\}$ (generating function $e^{\Lambda x}$), and the product
$e^{(\Lambda-1)x}/(1-x)$ gives the compact closed form
\begin{align}
F^{(k)}_{\rm 1PK}
=(k!)^{2}\sum_{m=0}^{k}\frac{(\Lambda-1)^{m}}{m!}
\;\xrightarrow[k\to\infty]{}\;(k!)^{2}e^{\Lambda-1}\,.
\label{eq:S3-closed}
\end{align}
For the two Wigner--Dyson classes relevant here:
\begin{align}
\text{GUE }(\Lambda=2):&\qquad
F^{(k)}_{\rm 1PK}
=(k!)^{2}\sum_{m=0}^{k}\frac{1}{m!}\;\to\;(k!)^{2}e \,,
\label{eq:S3-GUE}\\
\text{GOE }(\Lambda=3):&\qquad
F^{(k)}_{\rm 1PK}
=(k!)^{2}\sum_{m=0}^{k}\frac{2^{m}}{m!}\;\to\;(k!)^{2}e^{2} .
\label{eq:S3-GOE}
\end{align}
The first values are $F^{(k)}_{\rm 1PK}=2,10,96,\dots$
(GUE) and $3,20,228,\dots$ (GOE), to be compared with the Haar value
$k!=1,2,6,\dots$; the $k=1$ entry is the exact check
$F^{(1)}_{\rm 1PK}=\mathrm{Tr}(w^{2})=\sum_{a,b}|P_{ab}|^{4}
=\Lambda$ of Sec.\,\ref{sec:k1exact}. The GOE obstruction is uniformly larger
than the GUE one, by a factor $e$ in the large-$k$ tail, as expected from
the heavier tails of a real Gaussian.
 
The two protocols differ in exactly one respect. A sign-definite frozen
kernel admits no phase cancellation and retains
$\mathrm{Tr}(w^{2})\to\Lambda$, with the additional structure
\eqref{eq:S3-closed} generated by the fixed points of $\rho$; the second
kick replaces it by a mean-zero random-phase kernel with
$\mathrm{Tr}(w^{4})=1+O(1/d)$, which factorizes and leaves the bare count $k!$.
 
\paragraph{Relation to the two-step quench.}
The derivation above used only the doubly stochastic character of the
overlap matrix and its first two moments. It therefore applies verbatim
to the two-step quench $e^{-iH_2t_2}e^{-iH_1t_1}$ of Ref.\cite{Zhou:2026ubz} upon
replacing $w_{ab}\to A_{am}=|\langle E^{(1)}_a|E^{(2)}_m\rangle|^{2}$,
for which mutually Haar-random eigenbases give the same
$\overline{|A|^{2}}=1/d$ and kurtosis $\Lambda$. Equation
\eqref{eq:S3-eq6} thus governs the two-step quench as well, and the comparison with the three-step protocol at $k=1$ is carried out in Sec.\,\ref{sec:3sp}.

\subsection{Remarks and regime of validity}
 
\paragraph{(i) The role of the spectrum is unimportant.} Assumption (A1) is the only spectral input, and it enters solely to
enforce the index matchings \eqref{eq:S3-perms} and \eqref{eq:S3-Apc}.
The leading frame potential is a functional of the overlap matrix $w$
alone at every order $k$, as was found exactly at $k=1$. The
isospectrality of $H$ and $PHP$ (cf.\ Proposition~1) is therefore
immaterial to \eqref{eq:S3-2PKresult}, and can enter only through
accidental spectral resonances, which violate (A1) on a measure-zero set
and are $O(1/d)$ suppressed for a chaotic spectrum.
 
\paragraph{(ii) Regime of validity.}

Two conditions control the expansion. Repeated-entry configurations,
discarded below \eqref{eq:S3-perms}, are a fraction $O(k^{2}/d)$ of the
total, requiring
\begin{align}
k^{2}\ll d\,.
\label{eq:S3-cond1}
\end{align}
Off-diagonal pairings, of which there are at most $(k!)^{2}$, each
contribute $O(1/d^{2})$ by \eqref{eq:S3-offdiag} and must be negligible
against the diagonal total $k!$, requiring
\begin{align}
k!\ll d^{2}\,.
\label{eq:S3-cond2}
\end{align}
For the numerics of the main text, $d=2^{8}$ (GXE, SYK), $d=2^{10}$ (MFIM) and $k\le5$, these read
$25\ll256$ and $120\ll65536$; the first is the more restrictive and is
satisfied by roughly an order of magnitude. We emphasize that
\eqref{eq:S3-cond1}--\eqref{eq:S3-cond2} bound the accuracy of the
$1/d$ expansion, not the design order itself: the exact statement at
$k=1$, Eq.\,\eqref{eq:F1_2PK}, carries no such restriction.

\subsection{Single-diagonalization diagnostics for design formation} \label{suffcond}

The derivation above suggests two compact probes of design formation, both
obtainable from a single diagonalization of $H$. The first is the normalized
excess of the second frame potential over its factorized value,
\begin{align}
\kappa_{\mathcal{E}} := \frac{F^{(2)}_{\mathcal{E}}}{2\bigl(F^{(1)}_{\mathcal{E}}\bigr)^{2}}-1 \,,
\label{eq:kappadef}
\end{align}
which measures the failure of the overlap $z=\mathrm{Tr}(U^\dagger V)$ to have a
factorized (Gaussian) fourth moment, and vanishes when the ensemble reproduces
the Haar relation $\overline{|z|^{4}}=2\bigl(\overline{|z|^{2}}\bigr)^{2}$. The
second is the excess of the first frame potential itself. Define $\Delta^{(1)}_{\mathcal{E}} = F_{\mathcal{E}}^{(1)} - 1$, and therefore for the 2PK protocol,
\begin{align}
\Delta_{\mathrm{2PK}}^{(1)} := F^{(1)}_{\mathrm{2PK}}-1 = \mathrm{Tr}\bigl(w^{4}\bigr)-1
= \sum_{j\ge2}\mu_j^{4} \,,
\label{eq:Deltadef}
\end{align}
where the second equality is the exact result \eqref{eq:F1_2PK} and the third uses $\mu_1=1$ from \eqref{eq:perron}. Both quantities are functionals of the overlap matrix $w$ alone and require no temporal sampling. As we now show, they are independent, and \emph{both} are needed: $\kappa$ probes the fourth moment of the Gaussian
\emph{shape} of the distribution of $z$, while $\Delta_{\mathrm{2PK}}^{(1)}$ certifies its \emph{scale}.

The RHS of \eqref{eq:Deltadef} can be fixed explicitly. Since $w$ is doubly stochastic, we may split off the Perron mode exactly. Expressing
\begin{align}
w = \frac{J}{d} + \delta\,, \qquad J_{ab}=1\,, \qquad \delta\,\mathds{1}=0\,,
\label{eq:perron-split}
\end{align}
so that $w^{n} = J/d + \delta^{n}$ for all $n\ge 1$ and, using $\mu_{1}=1$,
\begin{align}
\Delta^{(1)}_{\rm 2PK} =  \mathrm{Tr}\left(\delta^{4}\right)\,.
\label{eq:delta-as-tr-delta4}
\end{align}
By the delocalization assumption (A2), the entries $\delta_{ab}=|P_{ab}|^{2}-1/d$ are centered, symmetric ($\delta_{ab}=\delta_{ba}$), and independent at leading order in $1/d$, with variance
\begin{align}
\sigma^{2} = \overline{|P_{ab}|^{4}} - \left(\overline{|P_{ab}|^{2}}\right)^{2}
\;=\; \frac{\Lambda-1}{d^{2}}\,,
\label{eq:delta-variance}
\end{align}
where we used $\overline{|P_{ab}|^{2}}=1/d$ from \eqref{eq:mean_weight} and the kurtosis relation, introduced in \eqref{eq:S3-kurtosis}. Expanding $\operatorname{Tr}(\delta^{4})=\sum_{a,b,c,d}\delta_{ab}\delta_{bc}\delta_{cd}\delta_{da}$,
the pairings that survive at leading order are $a=c$ and $b=d$ separately: each leaves three free indices and contributes $d^{3}\sigma^{4}$. The doubly paired term $a=c$ \emph{and} $b=d$, the diagonal
entries $w_{aa}$, and the traceless constraint $\delta\,\mathds{1}=0$ are each suppressed by a further
power of $1/d$. Hence
\begin{align}
\Delta^{(1)}_{\rm 2PK} \;=\; 2 d^{3}\sigma^{4}\left[1+O(1/d)\right]
\;=\; \frac{2(\Lambda-1)^{2}}{d}\,\left[1+O(1/d)\right]\,,
\label{eq:delta1-coefficient}
\end{align}
\emph{i.e.}, $\Delta^{(1)}_{\rm 2PK}=2/d$ for the GUE ($\Lambda=2$) and $8/d$ for the GOE ($\Lambda=3$).

\paragraph{The two protocols.}
For one kick, \eqref{eq:S3-eq6} at $k=1,2$ gives
$F^{(1)}_{\mathrm{1PK}}=\Lambda$ and
$F^{(2)}_{\mathrm{1PK}}=2(1+\Lambda^{2})$, whence
\begin{align}
\kappa_{\mathrm{1PK}}=\frac{1}{\Lambda^{2}}
=\frac{1}{\bigl[\mathrm{Tr}(w^{2})\bigr]^{2}}=O(1),
\label{eq:kappa1PK}
\end{align}
\emph{i.e.}, $1/4$ for the GUE and $1/9$ for the GOE, whereas for two kicks
\eqref{eq:S3-offdiag} and \eqref{eq:S3-2PKresult} give
\begin{align}
\kappa_{\mathrm{2PK}}=O(1/d).
\label{eq:kappa2PK}
\end{align}
The obstruction is therefore in the fourth moment: a single kick leaves an 
$O(1)$ excess, which the second kick removes.

\paragraph{$\kappa$ is scale free.}
It is important to delimit what \eqref{eq:kappa2PK} certifies. The diagnostic \eqref{eq:kappadef} is invariant under the rescaling
$F^{(k)}\mapsto\lambda^{k}F^{(k)}$ for
any $\lambda>0$, since $\lambda^{2}$ cancels between numerator and denominator. Consequently, inserting the factorized form \eqref{eq:S3-2PKresult},
$F^{(k)}_{\mathrm{2PK}}
=k!\,\bigl[\mathrm{Tr}(w^{4})\bigr]^{k}\bigl[1+O(1/d)\bigr]$, into
\eqref{eq:kappadef} yields
\begin{align}
\kappa_{\mathrm{2PK}}
=\frac{2\bigl[\mathrm{Tr}(w^{4})\bigr]^{2}}{2\bigl[\mathrm{Tr}(w^{4})\bigr]^{2}}-1+O(1/d)
=O(1/d)
\qquad\text{for any value of}~~\mathrm{Tr}(w^{4})\,,
\label{eq:kappacancel}
\end{align}
so that $\kappa$ is blind to $\Delta_{\mathrm{2PK}}^{(1)}$ by construction. The origin of this
insensitivity is that the two statements contained in \eqref{eq:S3-2PKresult} rest on different hypotheses. The factorization over the $k!$ diagonal permutation pairings, which is what $\kappa$ tests, follows from the non-resonance assumption (A1) together with the doubly stochastic structure of $w$ and the phase-random character of the two-kick kernel \eqref{eq:S3-kernels}. That $\mathrm{Tr}(w^{4})\to1$, by contrast, is a separate consequence of the delocalization assumption (A2). An ensemble may therefore satisfy the first
without the second: any ensemble obeying \eqref{eq:S3-2PKresult} with
$\mathrm{Tr}(w^{4})\to1+a$, $a=O(1)$, has
\begin{align}
F_{\mathcal{E}}^{(k)}=k!\,(1+a)^{k},\qquad \kappa=O(1/d)\,,
\end{align}
\emph{i.e.}, $z$ is asymptotically a centered complex Gaussian, but of variance $1+a$ rather than unity, and the ensemble is \emph{not} an approximate $k$-design. Since the Haar ensemble requires $z$ to be a centered complex Gaussian of
unit variance, and since $\kappa$ fixes only the shape and
$\Delta^{(1)}_{\mathrm{2PK}}$ only the variance, both are necessary
conditions for design formation:
\begin{align}
\mathcal{E}\ \text{forms an approximate }k\text{-design}
\qquad\Longrightarrow\qquad
\Delta_{\mathcal{E}}^{(1)} \to0\ \ \text{and}\ \ \kappa\to0 \,.
\label{eq:criterion}
\end{align}
The reverse implication holds under (A1)--(A2), which give the factorization
$F^{(k)} = k!\,[F^{(1)}]^{k}$~\eqref{eq:S3-2PKresult}; it is not necessarily valid in
general. The Clifford group has
$\Delta^{(1)} = \kappa = 0$ exactly, being an exact 3-design, yet fails to form a 4-design. For the 1PK protocol both criteria fail, since $F^{(1)}_{\mathrm{1PK}}=\Lambda\neq1$ in addition to
\eqref{eq:kappa1PK}; for the 2PK protocol both hold under (A1) and (A2).

\paragraph{$\Delta_{\mathrm{2PK}}^{(1)}$ and the spectral gap of $w$.}
The remaining criterion admits a sharp characterization in terms of the spectrum
of the doubly stochastic matrix $w$. Let $\mu_2:=\max_{j\ge2}|\mu_j|$ denote the
largest subleading eigenvalue, so that $1-\mu_2$ is the spectral gap. Using
$\sum_{j\ge2}\mu_j^{2}=\mathrm{Tr}(w^{2})-1$ in \eqref{eq:Deltadef}, we obtain the two-sided bound, exact at finite $d$,
\begin{align}
\mu_2^{4}\;\le\;\Delta_{\mathrm{2PK}}^{(1)}\;\le\;\mu_2^{2}\,\bigl[\mathrm{Tr}(w^{2})-1\bigr]\,.
\label{eq:gapbound}
\end{align}
The lower bound is the single term $j=2$ of \eqref{eq:Deltadef}; the upper bound
follows from $\mu_j^{4}\le\mu_2^{2}\mu_j^{2}$ for all $j\ge2$. Two consequences
follow. First, $\Delta_{\mathrm{2PK}}^{(1)}\to0$ \emph{implies} $\mu_2\to0$: design formation necessarily requires the gap of $w$ to open. Second, the converse holds whenever $\mathrm{Tr}(w^{2})$ remains bounded as $d\to\infty$. This is the case under (A2), where $\mathrm{Tr}(w^{2})\to\Lambda$, so that
\begin{align}
\Delta_{\mathrm{2PK}}^{(1)}\;\longrightarrow\;0
\qquad\Longleftrightarrow\qquad
\mu_2\;\longrightarrow\;0 \,.
\end{align}
The design property is thus equivalent to the opening of the spectral gap of
$w$, which is in turn a direct statement about the delocalization of the fixed
Pauli operator in the energy eigenbasis. Chaotic systems exhibit a large gap for
precisely this reason; the gap is expected to close as integrability is approached, and $\Delta_{\mathrm{2PK}}^{(1)}$ therefore provides a sharp probe of the chaos-to-integrability crossover, obtainable from a single diagonalization of $H$ with no temporal sampling.

We note finally that, because the leading factorized contribution cancels
identically in \eqref{eq:kappacancel}, $\kappa_{\mathrm{2PK}}$ is governed
entirely by the off-diagonal permutation pairings \eqref{eq:S3-offdiag}, which are
suppressed by $1/d^{2}$ relative to the diagonal ones. Its precise scaling is
therefore a considerably finer quantity than that of $\Delta_{\mathrm{2PK}}^{(1)}$, and resolving it numerically would require correspondingly higher statistics. A detailed study of $\kappa_{\mathrm{2PK}}$ beyond leading order, and of the behavior of
\eqref{eq:gapbound} across a chaos-to-integrability transition, is left for
future work.

\section{Insensitivity to the choice of Pauli string in GUE}
\label{sec:pauli_independence}

\begin{figure}[t]
    \centering
    \includegraphics[width=0.4\linewidth]{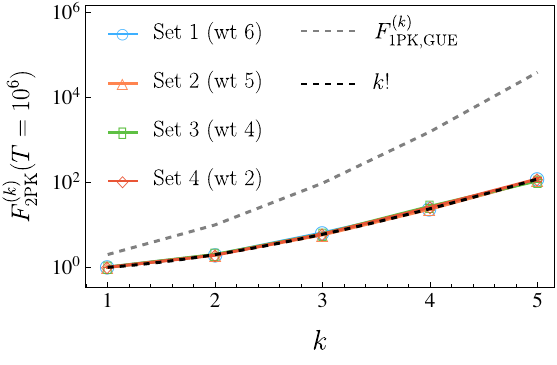}
    \caption{Frame potential $F^{(k)}_{\mathrm{2PK}}(T)$ for the GUE at $d = 2^{8}$, evaluated for four independently drawn fixed Pauli strings (markers) of different weights on a single fixed Hamiltonian realization, compared with the Haar value $k!$ (dashed line). The four datasets coincide within error bars, demonstrating that the long-time frame potential is insensitive to the choice of Pauli string. Error bars denote the standard error over $10^{4}$ temporal realizations; the long-time limit is taken with $T = 10^{6}$. The four curves are mutually consistent at every $k$ ($\chi^2/\mathrm{dof} \le
    1.14$) and jointly consistent with $k!$ ($\chi^2/\mathrm{dof} = 0.67$).}
    \label{fig:pauli_independence}
\end{figure}

In the 2PK protocol, the Hamiltonian $H$ and the Pauli operator $P$ are both
held fixed, with randomness entering only through the sampled evolution times.
Since $P$ is fixed rather than averaged over, the frame potential is in
principle a function of the particular string chosen. Here we verify that the
emergence of the approximate unitary design is independent of this choice,
confirming that the Haar-like behavior originates from the chaotic dynamics
rather than from any special property of the kick operator. However, we emphasize that the insensitivity of Pauli strings holds for the GXE (and most likely for all-to-all models like SYK) but will surely fail for local spin models, as evident from Table \ref{tab:3sp}.

We fix a single realization of the traceless GUE Hamiltonian at $d = 2^{8}$
and, keeping this realization fixed, evaluate the 2PK frame potential
$F^{(k)}_{\mathrm{2PK}}(T)$ for four independently drawn Hermitian Pauli
strings of Pauli weight $6$, $5$, $4$, and $2$ respectively. The weight 6 case is the same as Fig.\,\ref{fig:FP12PKGUEplot}. Each string is drawn uniformly from the non-identity Pauli strings $\mathcal{P}_n \backslash \{\mathds{I}^{\otimes n}\}$. The long-time limit is taken with $T = 10^{6}$, and each frame potential is estimated from $10^{4}$ independent temporal realizations, with the same conventions as the main text. The results are shown in
Fig.\,\ref{fig:pauli_independence}, where the four curves are seen to collapse onto the Haar value $k!$.

The four datasets are statistically indistinguishable. At each order $k$, the
four values are mutually consistent, with a reduced chi-squared
$\chi^2/\mathrm{dof} \le 1.14$ (computed across the four strings at fixed $k$,
$\mathrm{dof} = 3$), and the relative spread across strings, \emph{i.e.}, the standard deviation divided by the mean, is below $2\%$ at $k = 1, 2$ and remains below $7\%$ for $k \le 5$, comparable to the statistical uncertainty of an individual run. Notably, this includes a string of Pauli weight two, indicating
that even a low-weight kick is sufficient for the formation of the design in GXE. This is because the eigenbasis carries no notion of locality; Pauli weight is not a meaningful parameter. We therefore find no detectable dependence of the long-time frame potential on the choice of Pauli string, as expected from the unitary invariance. For the GXE, the design is thus a property of the chaotic Hamiltonian dynamics alone; in local models it is a joint property of the pair $(H,P)$, quantified by $c$ and $\mathrm{Tr}(w^4)$ in Sec.\,\ref{supp:MFIM}.

\section{The finite-size scaling for the 2PK protocol in GUE}
\label{sec:finitesize}

\begin{figure}[t]
\centering
\includegraphics[width=0.31\linewidth]{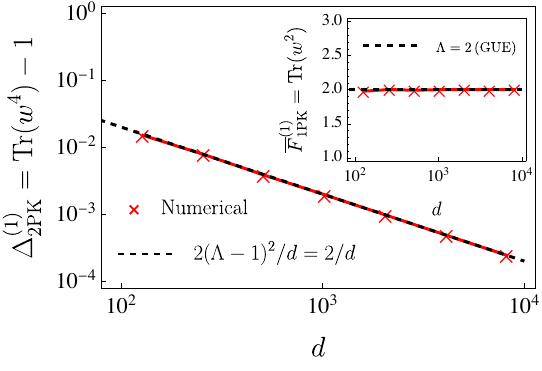}
\hspace{-1ex}
\includegraphics[width=0.33\linewidth]{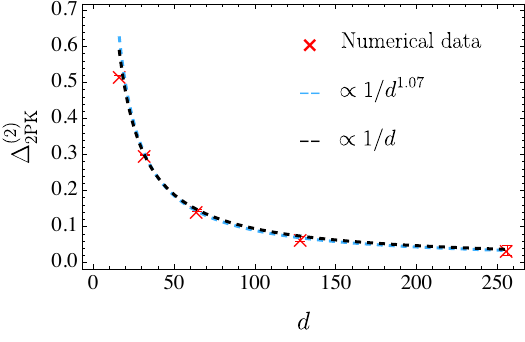}
\hspace{-1ex}
\includegraphics[width=0.33\linewidth]{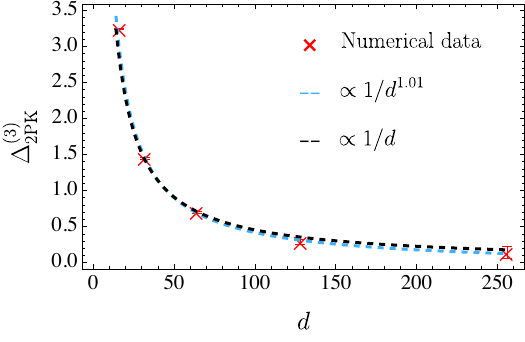}
\caption{Finite-size behavior of $\Delta^{(k)}_{\mathrm{2PK}}$ for $k=1$ (left panel), $k=2$ (middle panel) and $k=3$ (right panel) for the GUE Hamiltonian. Here $d$ denotes the Hilbert-space dimension (matrix size). The left panel is obtained using \eqref{eq:Deltadef} without temporal sampling, and fitted against the analytic result \eqref{eq:delta1-coefficient}. The inset shows $\mathrm{Tr}(w^2)$, which converges to the GUE kurtosis $\Lambda = 2$~[Eq.\,\eqref{eq:S3-kurtosis}]. For the middle and right figures, numerical data are generated using the temporal ensemble with the long-time limit $T=10^6$. In these cases, fitting is done from $d = 2^5$ to $d = 2^8$. The results are averaged over $\{5\times10^7, 10^7, 5\times10^6, 5\times10^5, 10^5\}$ independent temporal realizations for increasing system sizes.}
\label{fig:finitesize}
\end{figure}

To examine the approach of the 2PK protocol toward an exact unitary design, we study the finite-size scaling of the long-time frame potential. The left panel of Fig.\,\ref{fig:finitesize} shows $\Delta^{(k)}_{\mathrm{2PK}}$, evaluated from Eq.\,\eqref{eq:Deltadef} without temporal sampling. The numerical results converge to the analytic prediction \eqref{eq:delta1-coefficient} for $k=1$, $\Delta^{(1)}_{\rm 2PK}=2(\Lambda-1)^{2}/d = 2/d$ for the GUE kurtosis $\Lambda=2$. The middle and right panels show $\Delta^{(2)}_{\mathrm{2PK}}$ and $\Delta^{(3)}_{\mathrm{2PK}}$, respectively, obtained from temporal sampling as a function of the Hilbert-space dimension $d$. The long-time average is evaluated over a time window $T=10^6$ and averaged over a large number of independent temporal realizations, with the number of samples chosen to ensure statistical convergence at each system size. In both cases, the numerical data exhibit a clear approach toward the Haar value as $d$ increases. Equivalently, fitting for $k=2$ and $k = 3$ with
\begin{align}
\Delta^{(k)}_{\mathrm{2PK}} = F^{(k)}_{\mathrm{2PK}}-k! \sim d^{-\gamma}\,,
\end{align}
where $\gamma$ characterizes the leading finite-size correction, yields $\gamma \approx 1.07$ (for $k = 2$) and $\gamma \approx 1.01$ (for $k = 3$), in excellent agreement with the natural $1/d$ scaling, consistent with the left panel. The systematic decrease of $\Delta^{(k)}_{\mathrm{2PK}}$ with increasing Hilbert-space dimension is therefore consistent with the 2PK protocol approaching an increasingly accurate unitary $k$-design in the large Hilbert-space dimension limit using only a single chaotic Hamiltonian interspersed with two Pauli kicks.

\section{2PK protocol on Mixed Field Ising models}
\label{supp:MFIM}

\begin{figure}[h]
    \hspace{-0.5cm}
    \includegraphics[width=0.8\linewidth]{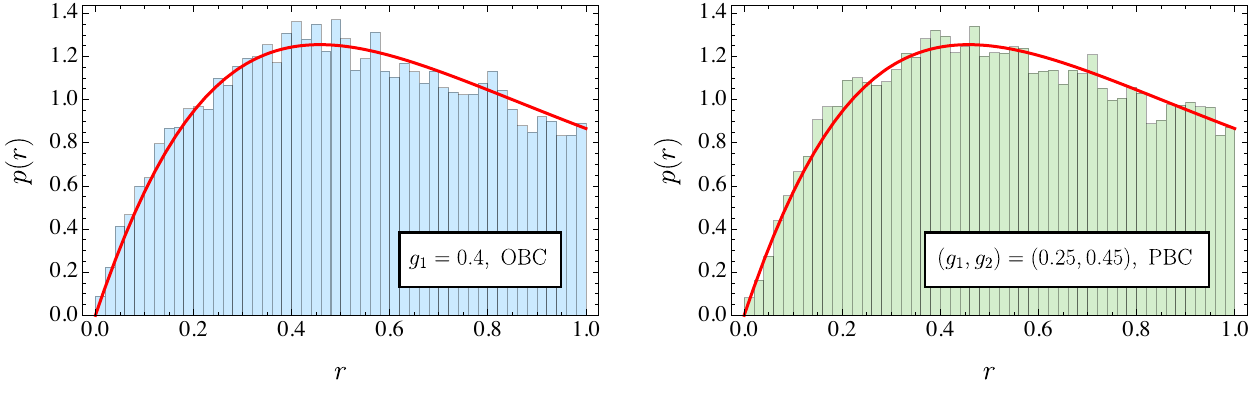}
    \caption{Histograms of the $r$ values in the MFIM model with $g_1 = 0.4$ in OBC (left panel) and $(g_1, g_2) = (0.25, 0.45)$ in PBC (right panel). No disorder is present in the model, so the result is from a single diagonalization. The system size is $L =14$, so the Hilbert space dimension is $d = 2^L = 16384$. In both cases, the statistics follow the GOE result, shown by the solid red line.}
    \label{fig:MFIMHistPlotL14}
\end{figure}

In this section, we provide details on the applicability of the 2PK protocol to local spin-chain models. Specifically, we consider the Mixed Field Ising models (MFIM) used in the main text,
\begin{align}
    H_{\mathrm{MFIM}} = \sum_i (Z_i Z_{i+1} + h_x X_i + h_z  Z_i) + \sum_i g_i X_i\,.
\end{align}
We study two different realizations of this model: an open boundary condition (OBC) system with $g_1=0.4$, and a periodic boundary condition (PBC) system with $(g_1,g_2)=(0.25,0.45)$. The remaining parameters are fixed to $(h_x,h_z)=(-1.05,0.5)$ \cite{Banuls:2010zki} for both cases. Therefore, the model is completely fixed by these parameters, and no randomness exists. The distributions of the nearest-neighbor level-spacing ratios \cite{Oganesyan:2007wpd, Atas2013distribution} are shown in Fig.\,\ref{fig:MFIMHistPlotL14}, along with the GOE prediction (solid red line) for comparison. The excellent agreement with the GOE statistics confirms that both models exhibit quantum chaotic behavior. 

To provide the details of the mechanism underlying the 2PK protocol, we begin
with a generic local spin chain of $L$ sites, written in the Pauli basis as
\begin{align}
    H = \sum_{S} h_S\, S\,,
    \label{eq:Hpauli}
\end{align}
where $S \in \{\mathds{1},X,Y,Z\}^{\otimes L}$ is an $L$-qubit Pauli string and
$h_S \in \mathds{R}$ is its weight ($h_S^2$ is the \emph{Frobenius weight}); the coefficients are real because both $H$
and $S$ are Hermitian. Any local Hamiltonian can be cast into this form, with $h_S \neq 0$ only for strings of bounded support.

\subsection{Pauli conjugation and the effective Hamiltonian}

Any two Pauli strings either commute or anti-commute, since on each site the
constituent Pauli matrices do, and the two strings anti-commute precisely when
the number of anti-commuting sites is odd. For the fixed kick $P \in \mathcal{P}_L\backslash\{\mathds{1}\}$ we may therefore define the sign
\begin{align}
    P S = \varepsilon_S\, S P \,,
    \qquad \varepsilon_S \in \{+1,-1\}\,,
\end{align}
and, using $P^2 = \mathds{1}$,
\begin{align}
    P S P = \varepsilon_S\, S P P = \varepsilon_S\, S\,.
    \label{eq:PSP}
\end{align}
The second effective Hamiltonian generated by the protocol
(cf.\ Proposition 1) is thus obtained from $H$ by flipping the sign of a
subset of its Pauli components,
\begin{align}
    H_2 \;\equiv\; P H P \;=\; \sum_S \varepsilon_S\, h_S\, S\,.
    \label{eq:H2}
\end{align}
Equation~\eqref{eq:H2} makes the isospectrality of $H$ and $H_2$ manifest at
the level of the operator content: the two Hamiltonians consist of the same
Pauli strings with the same magnitudes, differing only in a pattern of signs of the terms.

\subsection{The overlap ratio \texorpdfstring{$c$}{c}}

The natural measure of how much the kick disturbs the Hamiltonian is the normalized Hilbert--Schmidt overlap of $H$ with its conjugate,
\begin{align}
c = \frac{\mathrm{Tr} \left(H\,PHP\right)}{\mathrm{Tr}\left(H^2\right)}\,.\label{eq:cdef}
\end{align}
Inserting \eqref{eq:Hpauli} and \eqref{eq:PSP} and using the orthogonality of Pauli strings, $\mathrm{Tr}(SS') = d\,\delta_{SS'}$ with $d=2^L$, gives
\begin{align}
    \mathrm{Tr}\left(H\,PHP\right)
    = \sum_{S,S'} h_S h_{S'} \varepsilon_{S'} \mathrm{Tr}(SS')
    = d \sum_S \varepsilon_S h_S^2 \,,
    \qquad
    \mathrm{Tr}\left(H^2\right) = d \sum_S h_S^2 \,,
\end{align}
so that the Hilbert-space dimension cancels and
\begin{align}
c = \frac{\sum_S \varepsilon_S h_S^2}{\sum_S h_S^2}= \frac{\sum_{S \in \mathcal{C}} h_S^2 \;-\; \sum_{S \in \mathcal{A}} h_S^2}{\sum_S h_S^2} = 1 - 2f \,.\label{eq:c1m2f}
\end{align}
Here $\mathcal{C} = \{S : [P,S]=0\}$ and $\mathcal{A} = \{S : \{P,S\}=0\}$ are
the disjoint sets of commuting and anti-commuting strings, and
\begin{align}
    f = \frac{\sum_{S\in\mathcal{A}} h_S^2}{\sum_S h_S^2}\,,
\end{align}
is the fraction of the Frobenius weight of $H$ that the kick flips. Thus $c$
has a direct operational meaning: it measures how much of the Hamiltonian the
Pauli insertion leaves invariant. Equivalently, in the energy eigenbasis
$H\ket{E_a} = E_a \ket{E_a}$, using $P_{ba} = P_{ab}^{*}$ and
$w_{ab} = |P_{ab}|^2$,
\begin{align}
    \mathrm{Tr}(HPHP) = \sum_a \langle E_a|H PHP|E_a \rangle = \sum_{ab} E_a E_b |P_{ab}|^2 = \sum_{ab} E_a w_{ab} E_b\,.
\end{align}
Consequently, $c$ can be expressed as
\begin{align}
c = \frac{\sum_{a,b} E_a\, w_{ab}\, E_b}{\sum_a E_a^2} = \frac{E^{\mathsf T} w\, E}{\|E\|^2} \,,
\label{eq:cenergy}
\end{align}
\emph{i.e.}, $c$ is the autocorrelation of the energy vector under one action of the doubly stochastic matrix $w$ introduced in Sec.\,\ref{sec:k1exact}.

\subsection{Geometric interpretation and the two degenerate limits}

\begin{figure}[h]
    \hspace{-0.5cm}
    \includegraphics[width=1\linewidth]{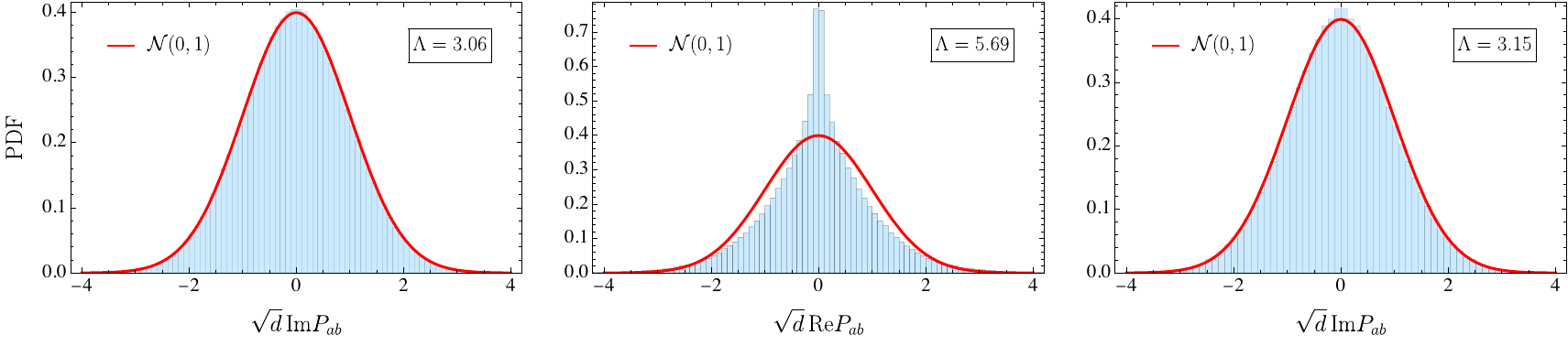}
    \caption{Distribution of off-diagonal Pauli matrix elements $P_{ab} = \langle E_a|P|E_b \rangle$ in the energy eigenbasis. Histograms of the rescaled off-diagonal matrix elements $\sqrt{d}\, \mathrm{Re}\,P_{ab}$ or $\sqrt{d}\,\mathrm{Im}\,P_{ab}$ for the Pauli strings $P_1$ (left: Case 3 and right: Case 5) and $P_2$ (middle: Case 4) in the chaotic MFIM (cf.\ Table \ref{tab:3sp}). Since the Hamiltonian is real symmetric, its eigenvectors can be chosen real. Consequently, if the Pauli string contains an odd (even) number of $Y$ operators, the matrix elements are purely imaginary (real) in the energy eigenbasis, and we therefore plot $\mathrm{Im}\,P_{ab}$ ($\mathrm{Re}\,P_{ab}$). The red curve denotes the standard normal distribution $\mathcal{N}(0,1)$. The left and right panels correspond to $\Lambda \approx 3$, consistent with the Gaussian behavior in GOE, whereas the middle panel, with $\Lambda \gg 3$, shows a clear deviation from Gaussianity. Here $\Lambda$ is evaluated from the off-diagonal elements only; it coincides with $\mathrm{Tr}(w^2)$ when $P$ is delocalized in the energy basis (Cases 3, 5) but not when $P$ is nearly diagonal (Case 4), where the diagonal entries carry appreciable weight. The system parameters are the same as those listed in Table \ref{tab:3sp}.}
    \label{fig:Gauss}
\end{figure}

Because $P$ is unitary, $\|PHP\| = \|H\|$ exactly, and \eqref{eq:cdef} is
the cosine of the angle $\theta$ between two operators of equal norm,
\begin{align}
    c= \frac{\mathrm{Tr} \left(H\,PHP\right)}{\mathrm{Tr}\left(H^2\right)} = \frac{\langle H, PHP\rangle}{||H|| \,||PHP||} = \cos\theta \,,
    \qquad \theta = \angle\big(H,\, PHP\big) \,.
\end{align}
where $\langle A, B \rangle = \mathrm{Tr}(A^{\dagger} B)$ and $||A|| = \sqrt{\langle A, A\rangle}$. The two extremes are degenerate and must be avoided. For $c \to +1$ one has
$\theta \to 0$, \emph{i.e.}, $PHP = H$ and $[P,H]=0$; then $w = \mathds{1}$,
$\mathrm{Tr}(w^{2n}) = d$, and
\begin{align}
U_{\rm 2PK} = e^{-iH(t_1+t_2+t_3)} \,. \label{tr1}
\end{align}
For $c \to -1$ one has $\theta \to \pi$, \emph{i.e.}, $PHP = -H$, and
\begin{align}
U_{\rm 2PK} = e^{-iHt_3}e^{+iHt_2}e^{-iHt_1} = e^{-iH(t_1-t_2+t_3)} \,. \label{tr2}
\end{align}
In both limits, the three independently sampled times collapse into a single
effective time, the temporal ensemble degenerates to the one-parameter orbit
of a single Hamiltonian, and no design can form. The favorable regime is
therefore $|c| \ll 1$: the two effective Hamiltonians are then orthogonal in
operator space, $\langle H, PHP\rangle \simeq 0$, and by \eqref{eq:c1m2f} the
kick flips essentially exactly half of the Hamiltonian's weight. This is the
precise sense in which the requirement $[P, H]\neq 0$ quoted in the main text
must be sharpened for the protocol to succeed: it is not enough that $P$ fail
to commute with $H$; it must fail to commute with an $O(1)$ fraction of it. Therefore, the condition on $|c| \ll 1$ is purely geometric and has to be viewed as a necessary but not a sufficient condition for the design formation. While $c$ determines whether the kick generates a nontrivial operator rotation, Gaussianity of $P_{ab}$ probes whether this rotation resembles the random basis transformation required for design formation. This is shown in Fig.\,\ref{fig:Gauss}. It will be interesting to examine the higher moments of the form $\mathrm{Tr}(H^m P H^nP)$, for generic $m$ and $n$, which might provide more details toward design formation.

\subsection{Locality and the Pauli weight of the kick}

For the Gaussian ensembles, the choice of $P$ is immaterial, and the reason is
structural. The eigenvector matrix $W$ is Haar distributed, so $P_W = W^\dagger P W$ is the fixed operator $P$ expressed in a uniformly random basis, and its statistics can depend only on those properties of $P$ preserved by unitary conjugation, namely its spectrum. Since every non-identity Hermitian Pauli string satisfies $P^2 = \mathds{1}$ and $\mathrm{Tr} P = 0$, all such strings have eigenvalues $\pm1$ with equal multiplicity $d/2$ and are therefore spectrally identical: the distribution of $w_{ab} = |P_{ab}|^2$, and hence of $\mathrm{Tr}(w^4)$, is statistically the same for every $P$. This is the origin of the insensitivity established in Sec.\,\ref{sec:pauli_independence}. The argument uses only the Haar distribution of $W$ and does not survive the restriction to a local Hamiltonian, whose eigenvectors are organized by the locality of $H$ rather than being uniformly random. 
There, the relevant quantity is the overlap ratio \eqref{eq:cdef}: a kick of
Pauli weight $\omega$ can anticommute only with those terms of $H$ whose
support it overlaps, a fraction $1-\binom{L-\omega}{q}/\binom{L}{q}\approx
q\omega/L$ of the $q$-body terms for $q\omega\ll L$. Since $f$ is bounded by
this fraction, $f\to0$ and hence $c\to1$ as $\omega/L\to0$, so low-weight kicks
drive the protocol into the trivial limit $PHP\simeq H$. Large weight is necessary but not sufficient, since the signs $\varepsilon_S$ may still cancel: for a bond $Z_iZ_j$ and a product kick, $\{Z_iZ_j,P\}=0$ requires \emph{exactly one} of $P_i,P_j \in \{X,Y\}$, so any uniform string produces two sign flips per bond and leaves the entire interaction invariant. We therefore draw $P$ uniformly from $\{\mathds{I}, X,Y,Z\}^{\otimes L}\backslash \{\mathds{I}^{\otimes L}\}$ and retain those with $|c| \lesssim 0.2$, which for the models considered here typically requires $\omega \gtrsim L/2$.

\subsection{Comparison with the three-step protocol (3SP) for local Hamiltonians}
\label{sec:3sp}

The three-step protocol (3SP) of Ref.\,\cite{Zhou:2026ubz} generates the temporal ensemble $U_{\rm 3SP}=e^{-iH_3t_3}e^{-iH_2t_2}e^{-iH_1t_1}$ from three statistically independent chaotic Hamiltonians. Repeating the time average of Sec.\,\ref{sec:k1exact} for this
ensemble, we write
\begin{align}
A_{am}=\big|\braket{E^{(1)}_a|E^{(2)}_m}\big|^{2}\,, \qquad
B_{mp}=\big|\braket{E^{(2)}_m|E^{(3)}_p}\big|^{2}\,,
\end{align}
both of which are real doubly stochastic matrices. The four independent time variables force all four index matchings, giving the closed form
\begin{align}
    F^{(1)}_{\rm 3SP}=\sum_{a,p}\big[(AB)_{ap}\big]^{2}=\|AB\|^{2}\,.
    \label{eq:F13SP}
\end{align}
The quantity $\mathrm{Tr}\big(AA^{\mathsf T}\big)=\|A\|^{2}$ plays for 3SP the role that $\mathrm{Tr}(w^{2})$ plays for 1PK, approaching $\Lambda$ when the two eigenbases are mutually random and $d$ when they coincide; correspondingly $\|AB\|^{2}$ is the analogue
of $\mathrm{Tr}(w^{4})=\|w^{2}\|^{2}$. Both therefore approach unity in the same way, up to $O(1/d)$, when the relevant eigenbases are mutually random.
In the local setting, the two protocols are separated by whether such a rotation is achievable at all, as quantified by 
$c$ in Table \ref{tab:3sp}.

To compute the frame potentials concretely, we consider the MFIM model discussed in the main text, with some variations. We rewrite them for convenience and denote them $H_1$ and $H_2$:
\begin{align}
    H_1 &=  \sum_{i} (Z_i Z_{i+1} -1.05  X_i + 0.5  Z_i) + 0.4 X_1\,, ~~ \mathrm{(OBC)}\,, \label{hamh1} \\
    H_2 &= \sum_{i} (Z_i Z_{i+1} -1.05  X_i + 0.5  Z_i) + \sum_i g_i X_i\,, ~~~ g_i \sim W \mathcal{N}(0,1)\,,~ W = 0.2~~ \mathrm{(PBC)}\,. \label{hamh2}
\end{align}

\begin{figure}[t]
\centering
\includegraphics[width=0.31\linewidth]{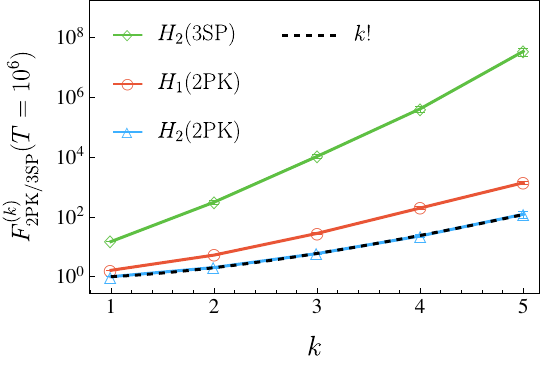}
\hspace{5ex}
\includegraphics[width=0.31\linewidth]{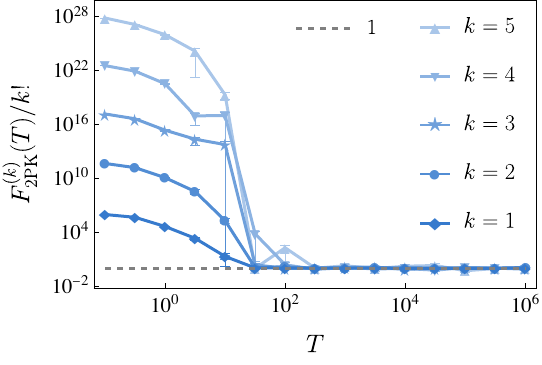}
\caption{(Left panel) The frame potential for the Hamiltonians $H_1$ \eqref{hamh1} and $H_2$ \eqref{hamh2} under the 2PK and 3SP protocols. For the 2PK protocol, the Pauli operators are chosen as $P_2$ for $H_1$ (red) and $P_1$ for $H_2$ (blue). For 3SP (green), three independent realizations are taken and kept fixed. The parameters are $L = 10$, $T = 10^6$, and $5 \times 10^3$ temporal realizations are taken. (Right panel) The 2PK Frame potential normalized by $k!$ for different $T$, which converges to unity shown by the gray dashed line. This implies that the curves stabilize for $T \geq 10^4$ and justifies the long-term limit $T = 10^6$.}
\label{fig:FP2PK3SPMFIMplot}
\end{figure}

The Pauli operators chosen are $P_1 = ZXYZYXYXYY$ (high weight) and $P_2 = IIIIIZIZZI$ (low weight). Table \ref{tab:3sp} lists the
normalized Hilbert--Schmidt overlap $\cos\theta$ between the two effective
Hamiltonians of each protocol. Clearly, for GUE both the 2PK and 3SP protocols lead to the approximate design. The situation changes for local Hamiltonians. Cases 3--5 involve deterministic Hamiltonians, so the random-Hamiltonian ensemble interpretation of 3SP is not applicable. However, 2PK can be well applicable here, and Case 3 and Case 5 form an approximate design, while Case 4 does not. However, the most interesting cases are 6--7. The chaotic properties of $H_2$ are established in \cite{Camargo:2025zxr}. Here, the 2PK protocol significantly outperforms the 3SP, where the setup is the same, leading to the approximate design within 2PK but not within 3SP. Fig.\,\ref{fig:FP2PK3SPMFIMplot} (left panel) shows the results. The long time limit for all MFIM models is taken as $T = 10^6$, which is validated in Fig.\,\ref{fig:FP2PK3SPMFIMplot} (right panel).

\begin{table}[h]
\centering
\begin{tabular}{llllllrr}
\hline
Cases~ & ~~~~~~~~~~~~~~~~~Model & Protocol & ~~~~2nd eff. Ham. & $~~\mathrm{Tr}(w^4)$ & $~~c = \cos\theta$ & $\theta$ & ~Design? \\
\hline
Case 1~ & GUE ~[Fig.\,\ref{fig:FP12PKGUEplot}] [Fig.\,\ref{fig:pauli_independence}] & 2PK & $PHP$, for any $P$ & $\approx 1.008$   & $~~~O(10^{-4})$ & $\approx 90.00^{\circ}$ & Yes \\
Case 2~ & GUE ~(Ref.\,\cite{Zhou:2026ubz}) & 3SP & Independent GUE  & ~~~~~\textemdash        & $~~~O(10^{-4})$ & $\approx 90.00^{\circ}$ & Yes \\
\hline
Case 3~ & MFIM ($g_1 = 0.4$, OBC) [Fig.\ref{fig:FP2PKallSYKMFIMplot}] & 2PK & $PHP$, $P = P_1$  & $~~~1.009$ & $~~~-0.102$ & $95.86^{\circ}$ & Yes \\
Case 4~ & MFIM ($g_1 = 0.4$, OBC) [Fig.\ref{fig:FP2PK3SPMFIMplot}] & 2PK & $PHP$, $P = P_2$  & $~~~1.638$  & $~~~+0.697$ & $45.80^{\circ}$ & No\\
Case 5~ & MFIM [$(g_1, g_2) = (0.25, 0.45)$, PBC] [Fig.\,\ref{fig:FP2PKallSYKMFIMplot}] & 2PK & $PHP$, $P = P_1$  & $~~~1.011$                  & $~~~-0.188$ & $~100.82^{\circ}$ & Yes \\
Case 6~ & MFIM [$g_i = W \mathcal{N}(0,1)$ [$W = 0.2$], PBC] [Fig.\,\ref{fig:FP2PK3SPMFIMplot}] & 2PK & $PHP$, $P = P_1$     & $\approx 1.011$ & $\approx -0.167$ & $\approx 99.66^{\circ}$ & Yes\\
Case 7~ & MFIM [$g_i = W \mathcal{N}(0,1)$ [$W = 0.2$], PBC] [Fig.\,\ref{fig:FP2PK3SPMFIMplot}] & 3SP & Ind. disorder real. & ~~~~~\textemdash  & $\approx +0.984$ & $\approx~\, 9.85^{\circ}$ & No \\
\hline
\end{tabular}
\caption{Operator-space angle between the two Hamiltonians of each
protocol, $\cos\theta=\mathrm{Tr}(H_1H_2)/(\|H_1\| \|H_2\|)$, evaluated at $L=10$ (MFIM) and $d=2^8$ (GUE). Values $|\cos\theta| \simeq 1$ signal nearly parallel or antiparallel Hamiltonians, for which the sampled times collapse and no design can form, while $\cos\theta \simeq 0$ indicates orthogonal operator directions. Cases 3--5 are fully deterministic and reproducible exactly; for the GUE (Cases 1, 2) and the disordered MFIM (Cases 6, 7) we average $c$ and $\theta$ (the average is: $\langle \theta \rangle = \langle \arccos (c) \rangle \neq \arccos \langle c \rangle$) independently over $500$ realizations, and the $\approx$ notation indicates that a single draw will fluctuate about these means. The Pauli
operators are $P_1 = ZXYZYXYXYY$ and $P_2 = IIIIIZIZZI$; for the GUE the result is independent of the choice of $P$, in accordance with
Sec.\,\ref{sec:pauli_independence}. Conjugation by a Pauli string preserves the Frobenius norm exactly and therefore acts as a pure rotation, whereas two
Hamiltonians drawn from the same local family share most of their operator
content and remain nearly parallel. We use a single draw for Case 6 and three
independent draws for Case 7 to implement the 2PK and 3SP protocols. In all 2PK cases, the corresponding $\mathrm{Tr}(w^4)$ values are shown, which are close to unity, suggesting the approximate design formation (cf.\ Eq.\,\eqref{eq:S3-2PKresult}).}
\label{tab:3sp}
\end{table}

The origin of the difference between 3SP and 2PK is that the two protocols generate their second Hamiltonian by different means. In 3SP, the couplings are redrawn, and the achievable operator-space angle is limited by the operator content common to the family. In 2PK, the couplings are not redrawn at all: conjugation flips their signs, $PHP = \sum_S \varepsilon_S h_S S$, which preserves the norm identically and therefore acts as a pure rotation, reaching $\theta \simeq \pi/2$ whenever the kick anticommutes with \emph{half} the Frobenius weight of $H$. The isospectrality of $H$ and $PHP$,
noted as a caveat in Proposition 1, is thus not a limitation of the protocol
in the local setting, since the leading frame potential depends only on the overlap matrix. It is instead a by-product of unitary conjugation, which is also what makes the kick a norm-preserving rotation in operator space. This rotation is the mechanism by which 2PK outperforms multi-Hamiltonian quenches for local Hamiltonians.

\section{Proof of the finite-temperature frame-potential bound}
\label{sec:SM-bound}

We prove the inequality $F^{(k, \beta)}_{\mathcal{E}} \geq F^{(k, \beta)}_{\mathcal{E}_{\mathrm{Haar}}}$, from which \eqref{ineq} and \eqref{lowerbound} follow to leading order in $1/d$ stated in the main text. Our input is that the 2PK protocol forms an approximate $k$-design in the large-$d$ limit, which we verified numerically up to $k = 5$. We show that \emph{(i)} any ensemble forming an approximate unitary $k$-design evaluates the finite-temperature frame potential to
$k!\,(\mathcal{Z}(2\beta)/d)^{k}$ up to $O(1/d)$; \emph{(ii)} for an \emph{arbitrary} ensemble, this expression is a lower bound, saturated precisely by $k$-designs; and \emph{(iii)} the bound is itself no smaller than the Haar value $k!$. Throughout our derivation, we take $\rho_\beta=e^{-\beta H}$, $\mathcal{Z}(\beta)=\mathrm{Tr}\,e^{-\beta H}$, and
$|a\rangle$ are the eigenstates of $H$ with energies $E_a$, so
$(\rho_\beta)_{ab}=e^{-\beta E_a}\delta_{ab}$.

\emph{Derivation of finite-temperature result.} Let $U,V$ be independent elements of an ensemble $\mathcal{E}$ and set $G=U^\dagger V$. The integrand given by $|\mathrm{Tr}(\rho_\beta G)|^{2k}
=[\mathrm{Tr}(\rho_\beta G)]^{k}[\overline{\mathrm{Tr}(\rho_\beta G)}]^{k}$ is a balanced polynomial of degree $(k,k)$ in the entries of $G$. A unitary $k$-design reproduces, by definition, the Haar average of every such polynomial. Moreover, by left-invariance of the Haar measure, $G=U^\dagger V$ is Haar-distributed whenever $V$ is. Hence, for an approximate $k$-design
\begin{align}
F^{(k,\beta)}_{\mathcal{E}}
=\big\langle |\mathrm{Tr}(\rho_\beta G)|^{2k}\big\rangle_{G\sim\mathrm{Haar}}
\,[1+O(1/d)]\,,
\label{eq:SM-reduce}
\end{align}
the $O(1/d)$ measuring the departure of $\mathcal{E}$ from an exact design. The
2PK ensemble is such a design.

In the energy eigenbasis $\mathrm{Tr}(\rho_\beta G)=\sum_a e^{-\beta E_a}G_{aa}$. For a Haar-random $G$ one has $\langle G_{aa}\rangle=0$ and, exactly, $\langle G_{aa}\,\overline{G_{bb}}\rangle=\delta_{ab}/d$. To leading order in
$1/d$ the $\{G_{aa}\}$ are jointly complex Gaussian, so
$z\equiv\mathrm{Tr}(\rho_\beta G)$ is a mean-zero complex Gaussian with variance
\begin{align}
\sigma^2= \overline{|z|^2}
=\frac1d\sum_a e^{-2\beta E_a}=\frac{\mathcal{Z}(2\beta)}{d}\,.
\label{eq:SM-var}
\end{align}
Using the complex-Gaussian moment $\langle|z|^{2k}\rangle=k!\,\sigma^{2k}$, we have
\begin{align}
F^{(k,\beta)}_{\mathrm{2PK}}
=k!\left(\frac{\mathcal{Z}(2\beta)}{d}\right)^{k}[1+O(1/d)]\,,
\label{eq:SM-value}
\end{align}
which is Eq.\,\eqref{ineq}. Equivalently, the exact Haar average is the
Weingarten sum
\begin{align}
\big\langle |\mathrm{Tr}(\rho_\beta G)|^{2k}\big\rangle_{\mathrm{Haar}}
=k! \sum_{\sigma\in S_k}\mathrm{Wg}(\sigma,d)\prod_{c\,\in\,\mathrm{cyc}(\sigma)}\mathrm{Tr}\big(\rho_\beta^{\,2|c|}\big)\,,
\label{eq:SM-wg}
\end{align}
where $\mathrm{cyc}(\sigma)$ are the cycles of $\sigma$ and $|c|$ their lengths; its leading (identity) term, with $\mathrm{Wg}(\mathrm{id},d)=d^{-k}[1+O(1/d^2)]$ and $\mathrm{Tr}(\rho_\beta^2)=\mathcal{Z}(2\beta)$, reproduces \eqref{ineq}. For $k=1$, \eqref{eq:SM-wg} is exact, $F^{(1,\beta)}=\mathcal{Z}(2\beta)/d$.

\emph{Lower bound for an arbitrary ensemble.}
We express the frame potential as an operator expectation
\begin{align}
F^{(k,\beta)}_{\mathcal{E}}
=\mathrm{Tr}\Big[\big(\rho_\beta^{\otimes k}\otimes\bar\rho_\beta^{\otimes k}\big)\,
\mathcal{M}^{(k)}_{\mathcal{E}}\Big]\,,~~~~
\mathcal{M}^{(k)}_{\mathcal{E}}
=\big\langle W^{\otimes k}\otimes\bar W^{\otimes k}\big\rangle\,,
\end{align}
with $W=U^\dagger V$. Since $U,V$ are independent,
$\mathcal{M}^{(k)}_{\mathcal{E}}=\Phi^\dagger\Phi$ with
$\Phi=\langle V^{\otimes k}\otimes\bar V^{\otimes k}\rangle$, while for the
Haar ensemble $\Phi=\Pi$, the orthogonal projector onto the permutation commutant, so $\mathcal{M}^{(k)}_{\mathcal{E}_{\mathrm{Haar}}}=\Pi$. Every twirl leaves the permutation operators invariant, whence $\Phi \Pi=\Pi\Phi=\Pi$; therefore, for any $|x\rangle$, we have
\begin{align}
\|\Phi|x\rangle\|^2
=\|\Pi|x\rangle\|^2+\|\Phi(\mathds{1}-\Pi)|x\rangle\|^2\ \geq \langle x|\Pi|x\rangle\,,
\end{align}
\emph{i.e.}, $\mathcal{M}^{(k)}_{\mathcal{E}}-\mathcal{M}^{(k)}_{\mathcal{E}_{\mathrm{Haar}}}
=\Phi^\dagger\Phi-\Pi\succeq0$. Since also
$\rho_\beta^{\otimes k}\otimes\bar\rho_\beta^{\otimes k}\succeq0$ (because
$\rho_\beta=e^{-\beta H}\succeq0$), and the trace of a product of two positive semidefinite operators is nonnegative, we find
\begin{align}
F^{(k,\beta)}_{\mathcal{E}}-F^{(k,\beta)}_{\mathcal{E}_{\mathrm{Haar}}}
=\mathrm{Tr}\Big[\big(\rho_\beta^{\otimes k}\otimes\bar\rho_\beta^{\otimes k}\big)
\big(\mathcal{M}^{(k)}_{\mathcal{E}}-\Pi\big)\Big]\ \geq 0\,. \label{eq:positivity}
\end{align}
Thus $F^{(k,\beta)}_{\mathcal{E}}\ge F^{(k,\beta)}_{\mathcal{E}_{\mathrm{Haar}}} =k!\,(\mathcal{Z}(2\beta)/d)^{k}[1+O(1/d)]$, with equality iff $\mathcal{E}$ is a unitary $k$-design (positivity of $\rho_\beta$ makes the bounding operator full rank, so saturation forces $\mathcal{M}^{(k)}_{\mathcal{E}}=\Pi$). This proves the lower bound \eqref{lowerbound} of the main text; the 2PK protocol saturates it up to the $O(1/d)$ corrections in \eqref{ineq}.

\emph{Positivity and the infinite-temperature limit.}
Finally, the $d$ positive numbers $e^{-2\beta E_a}$ satisfy the
arithmetic-mean\,--\,geometric-mean inequality,
\begin{align}
\frac{\mathcal{Z}(2\beta)}{d}
=\frac1d\sum_a e^{-2\beta E_a}
\ \ge\ \Big(\prod_a e^{-2\beta E_a}\Big)^{1/d}
=e^{-2\beta\,\mathrm{Tr}(H)/d}\,. \label{eq:AMGM}
\end{align}
For a traceless Hamiltonian, $\mathrm{Tr}(H)=0$ (as for the Majorana and Spin SYK
models, and for the traceless part of the GUE used throughout), the right-hand side of \eqref{eq:AMGM} equals unity exactly, so that
\begin{align}
\Big(\frac{\mathcal{Z}(2\beta)}{d}\Big)^{k}\ \ge\ 1 ,
\label{eq:ZoverD}
\end{align}
with equality if and only if $\beta=0$, where arithmetic mean = geometric mean. For ensembles that are traceless only on average, such as the full GUE, a
single realization has $\mathrm{Tr}(H)=O(1)$, hence $\mathrm{Tr}(H)/d=O(1/d)$, and
the right-hand side of \eqref{eq:AMGM} is instead $1+O(1/d)$; restricting to the
traceless part removes this correction.

Collecting the ingredients, we obtain
\begin{align}
F^{(k,\beta)}_{\mathcal{E}} \geq  F^{(k,\beta)}_{\mathcal{E}_{\mathrm{Haar}}} = k! \Big(\frac{\mathcal{Z}(2\beta)}{d}\Big)^{k}\big[1+O(1/d)\big]
 \geq k!\,\big[1+O(1/d)\big] \,,
\label{eq:lowerbound-SM}
\end{align}
which is Eq.\,\eqref{lowerbound} of the main text. We can further examine
where the $1/d$ correction enters. The first inequality is \emph{exact} at finite $d$: it follows from the operator inequality \eqref{eq:positivity}, and is
saturated if and only if $\mathcal{E}$ is a unitary $k$-design. The second
inequality \eqref{eq:ZoverD} is likewise exact for traceless $H$, and is saturated
if and only if $\beta=0$. The $O(1/d)$ arises solely from evaluating the Haar value in closed form: the diagonal entries $\{G_{aa}\}$ are jointly complex Gaussian only to leading order in $1/d$, and the moment identity $\langle|z|^{2k}\rangle=k!\,\sigma^{2k}$ inherits the corresponding correction. In particular, at $\beta=0$ the Haar average is exact for $k \leq d$, so that $F^{(k,0)}_{\mathcal{E}}\ge k!$ holds with no correction; for $\beta>0$ the bound $k!$ is attained up to the $O(1/d)$ accuracy of \eqref{eq:lowerbound-SM}.

\section{Results with finite-temperature Wightman inner product}
\label{sec:Wightman}

The finite-temperature frame potential of the main text, Eq.~\eqref{eq:F2PKbeta},
inserts the full Gibbs weight $\rho_\beta=e^{-\beta H}$ on one side of the trace. A natural alternative is the \emph{symmetric}, or Wightman, regularization, in which $\rho_\beta^{1/2}=e^{-\beta H/2}$ is inserted symmetrically
\begin{align}
F^{(k,\beta)}_{\mathrm{2PK},\mathrm{W}}
:=\mathds{E}_{U,V\in\mathcal{E}_{\mathrm{2PK}}}
\Big|\mathrm{Tr}\big(\rho_\beta^{1/2}\,U^\dagger\,\rho_\beta^{1/2}\,V\big)\Big|^{2k}\,,
\label{eq:FW}
\end{align}
corresponding to the thermal (Wightman) inner product
$\langle A,B\rangle_\beta=\mathrm{Tr}(\rho_\beta^{1/2}A^\dagger\rho_\beta^{1/2}B)$ commonly employed for real-time thermal and out-of-time-order correlators
\cite{Maldacena:2015waa, Parker:2018yvk}.

\emph{Analytic result.} For an approximate $k$-design the ensemble average again
reduces to a Haar average (as in the proof of Eq.\,\eqref{ineq}), and one finds
\begin{align}
F^{(k,\beta)}_{\mathrm{2PK},\mathrm{W}}
= k!\left(\frac{\mathcal{Z}(\beta)}{d}\right)^{2k}\big[1+O(1/d)\big]\,,
\label{eq:FWresult}
\end{align}
to be compared with the one-sided result
$F^{(k,\beta)}_{\mathrm{2PK}}=k!\,(\mathcal{Z}(2\beta)/d)^{k}$. The symmetric
insertion thus replaces the R\'enyi-2 partition function (purity)
$\mathcal{Z}(2\beta)=\mathrm{Tr}\,\rho_\beta^{2}$ by the squared partition
function $\mathcal{Z}(\beta)^2=(\mathrm{Tr}\,\rho_\beta)^{2}$.

\begin{figure}[t]
    \hspace{-0.5cm}
\includegraphics[width=0.7\linewidth]{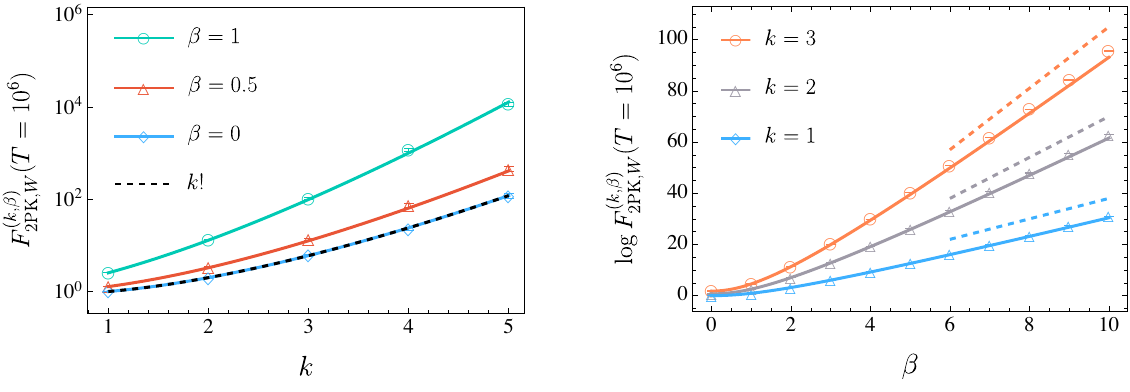}
    \caption{\textbf{Left panel:} Finite-temperature frame potential \eqref{eq:FWresult} for the GUE under the 2PK protocol for different $\beta$ in the Wightman definition \eqref{eq:FW}. \textbf{Right panel:} Corresponding results with $\beta$ for different $k$. The parameters are same as Fig.\,\ref{fig:FP2PKGUEwithbetaplot}.}
    \label{fig:Wightman}
\end{figure}

\emph{Derivation.} Let $Q=\rho_\beta^{1/2}U^\dagger\rho_\beta^{1/2}$. Absorbing it, the integrand becomes
$|\mathrm{Tr}(QV)|^{2k}$, which we average over $V$ and $U$ successively.

For fixed $U$, the trace is linear in the entries of $V$,
$\mathrm{Tr}(QV)=\sum_{a,b}Q_{ba}V_{ab}$, so for Haar $V$ it is a mean-zero
complex Gaussian. Its variance follows from the second moment of the Haar unitary
entries, $\mathds{E}_V[V_{ab}V^{*}_{cd}]=\frac{1}{d}\,\delta_{ac}\delta_{bd}$:
\begin{align}
\mathds{E}_V\big|\mathrm{Tr}(QV)\big|^2
=\sum_{a,b,c,d}Q_{ba}\,Q^{*}_{dc}\;
\mathds{E}_V\big[V_{ab}\,V^{*}_{cd}\big]
=\frac1d\sum_{a,b}\big|Q_{ab}\big|^2
=\frac{\mathrm{Tr}(QQ^\dagger)}{d}\,.
\end{align}
Therefore, the asymptotic limit of the complex-Gaussian moment $\mathds{E}|z|^{2k}=k!\,(\mathds{E}|z|^2)^k$ gives
\begin{align}
\mathds{E}_V\big|\mathrm{Tr}(QV)\big|^{2k}
= k!\left(\frac{\mathrm{Tr}(QQ^\dagger)}{d}\right)^{k} [1+O(1/d)]\,.
\label{eq:Wstep1}
\end{align}
Using $Q^\dagger=\rho_\beta^{1/2}U\rho_\beta^{1/2}$, the identity
$\rho_\beta^{1/2}\rho_\beta^{1/2}=\rho_\beta$, and cyclicity of the trace, we have
\begin{align}
\mathrm{Tr}(QQ^\dagger)
= \mathrm{Tr}\big(\rho_\beta^{1/2}U^\dagger\rho_\beta\,U\rho_\beta^{1/2}\big)
= \mathrm{Tr}\big(\rho_\beta\,U^\dagger\rho_\beta\,U\big).
\label{eq:Wstep2}
\end{align}
Averaging \eqref{eq:Wstep2} over $U$ with the unitary $1$-design twirl
$\mathds{E}_U[U^\dagger\rho_\beta U]=(\mathrm{Tr}\,\rho_\beta/d)\,\mathds{1}$,
\begin{align}
\mathds{E}_U\,\mathrm{Tr}\big(\rho_\beta U^\dagger\rho_\beta U\big)
= \mathrm{Tr}\Big(\rho_\beta\,\frac{\mathrm{Tr}\,\rho_\beta}{d}\,\mathds{1}\Big)
= \frac{(\mathrm{Tr}\,\rho_\beta)^2}{d}
= \frac{\mathcal{Z}(\beta)^2}{d}.
\label{eq:Wstep3}
\end{align}
Since $\mathrm{Tr}(\rho_\beta U^\dagger\rho_\beta U)$ concentrates on its mean up to $O(1/d)$, one has $\mathds{E}_U(\cdot)^k=(\mathds{E}_U\,\cdot)^k[1+O(1/d)]$;
combining \eqref{eq:Wstep1}--\eqref{eq:Wstep3} then yields
\begin{align}
F^{(k,\beta)}_{\mathrm{2PK},\mathrm{W}}
= k!\left(\frac{\mathcal{Z}(\beta)^2}{d^{2}}\right)^{k}\big[1+O(1/d)\big]
= k!\left(\frac{\mathcal{Z}(\beta)}{d}\right)^{2k}\big[1+O(1/d)\big]\,,
\end{align}
which is Eq.\,\eqref{eq:FWresult}.

\emph{Properties.} The Wightman frame potential shares the qualitative structure
of the one-sided definition. \emph{(i)} At $\beta=0$, $\mathcal{Z}(0)=d$ and $F=k!$.\emph{(ii)} For a traceless Hamiltonian, the arithmetic-mean--geometric-mean inequality
gives $\mathcal{Z}(\beta)/d\ge1$, hence
$F^{(k,\beta)}_{\mathrm{2PK},\mathrm{W}}\ge k!$, with equality at $\beta=0$.
\emph{(iii)} In the low-temperature limit $\mathcal{Z}(\beta)\to e^{-\beta E_0}$, so
\begin{align}
F^{(k,\beta)}_{\mathrm{2PK},\mathrm{W}}
\ \sim\ \frac{k!}{d^{2k}}\,e^{2k\beta|E_0|}\,,\qquad(\beta\to\infty)\,,
\end{align}
\emph{i.e.}, the same growth exponent $\alpha_k=2k|E_0|$ as \eqref{gs}. The ground-state scaling is therefore \emph{independent of the choice of thermal
regularization}; only the prefactor ($d^{-2k}$ vs.\ $d^{-k}$) and the
intermediate-$\beta$ profile differ. Moreover, by the Cauchy--Schwarz inequality
$\mathcal{Z}(\beta)^2\le d\,\mathcal{Z}(2\beta)$, so
$F^{(k,\beta)}_{\mathrm{2PK},\mathrm{W}}\le F^{(k,\beta)}_{\mathrm{2PK}}$: the symmetric insertion produces a milder thermal enhancement at the same $\beta$.

\emph{Numerical verification.} Figure \ref{fig:Wightman} confirms
\eqref{eq:FWresult} for the GUE under the 2PK protocol. The numerically
evaluated Wightman frame potential agrees with $k!\,(\mathcal{Z}(\beta)/d)^{2k}$
across $\beta$ and $k$, and its large-$\beta$ slope approaches $\alpha_k=2k|E_0|=4k$ for the semicircle edge $E_0=-2$, consistent with the one-sided result of the main text.

\end{document}